\documentclass[JHEP,11pt]{article}
\usepackage{verbatim}
\usepackage{jheppub}
\usepackage[T1]{fontenc}
\usepackage{slashed}
\usepackage[utf8]{inputenc}
\usepackage{array}
\usepackage{wrapfig}
\usepackage{multirow}
\usepackage{tabu}
\usepackage{amsmath,amssymb,amsthm,amsfonts}
\usepackage{enumitem}
\usepackage{graphicx}
\usepackage{placeins}
\usepackage{longtable}
\usepackage{float}
\usepackage{caption,subcaption}
\usepackage[dvipsnames]{xcolor}
\usepackage{tikz,tikz-3dplot}
\usepackage[compat=1.0.0]{tikz-feynman}
\usetikzlibrary{calc,arrows.meta,shapes.geometric}
\usepackage{soul}
\usepackage[usestackEOL]{stackengine}
\usepackage{cleveref}

\def\I{\mathcal{I}}

\def\r{\boldsymbol{r}}
\def\x{\boldsymbol{x}}
\def\v{\boldsymbol{v}}
\def\w{\boldsymbol{w}}

\def\s{\boldsymbol{s}}
\def\t{\boldsymbol{t}}
\def\eps{\epsilon}

\newcommand{\leftrarrows}{\mathrel{\raise.75ex\hbox{\oalign{%
  $\scriptstyle\leftarrow$\cr
  \vrule width0pt height.5ex$\hfil\scriptstyle\relbar$\cr}}}}
\newcommand{\lrightarrows}{\mathrel{\raise.75ex\hbox{\oalign{%
  $\scriptstyle\relbar$\hfil\cr
  $\scriptstyle\vrule width0pt height.5ex\smash\rightarrow$\cr}}}}
\newcommand{\Rrelbar}{\mathrel{\raise.75ex\hbox{\oalign{%
  $\scriptstyle\relbar$\cr
  \vrule width0pt height.5ex$\scriptstyle\relbar$}}}}

\newcommand{\GlauberVertexHard}[1]{%
\begin{tikzpicture}[#1]%
\draw[fill, thick, White] (0,0) circle (5pt); \draw (0,-0.2) edge [line width = 0.8mm, GreenYellow] (0,0.2) node [] {}; \draw[ultra thick, Blue] (0,0) circle (5pt);
\end{tikzpicture}%
}
\newcommand{\GlauberVertexJetA}[1]{%
\begin{tikzpicture}[#1]%
\draw[fill, thick, White] (0,0) circle (5pt); \draw (0,-0.2) edge [line width = 0.8mm, GreenYellow] (0,0.2) node [] {}; \draw[ultra thick, Green] (0,0) circle (5pt);
\end{tikzpicture}%
}
\newcommand{\GlauberVertexJetB}[1]{%
\begin{tikzpicture}[#1]%
\draw[fill, thick, White] (0,0) circle (5pt); \draw (0,-0.2) edge [line width = 0.8mm, GreenYellow] (0,0.2) node [] {}; \draw[ultra thick, teal] (0,0) circle (5pt);
\end{tikzpicture}%
}
\newcommand{\GlauberTextHard}{\raisebox{\dimexpr 2.2ex-\height}{\GlauberVertexHard{}}}
\newcommand{\GlauberTextJetA}{\raisebox{\dimexpr 2.2ex-\height}{\GlauberVertexJetA{}}}
\newcommand{\GlauberTextJetB}{\raisebox{\dimexpr 2.2ex-\height}{\GlauberVertexJetB{}}}
\newcommand\greencheckmark[1][]{%
  \tikz[scale=0.4,#1]{\fill(0,.35) -- (.25,0) -- (1,.7) -- (.25,.15) -- cycle;}%
}
\newcommand\crossmark[1][]{%
  \tikz[scale=0.4,#1]{
    \fill(0,0)--(0.1,0) .. controls (0.5,0.4) .. (1,0.7)--(0.9,0.7) ..  controls (0.5,0.5) ..(0,0.1) --cycle;
    \fill(1,0.1)--(0.9,0.1) .. controls (0.5,0.3) .. (0,0.7)--(0.1,0.7) .. controls (0.5,0.4) ..(1,0.2) --cycle;
  }%
}

\makeatletter
\def\leftrightarrowsfill@{\arrowfill@\leftrarrows\Rrelbar\lrightarrows}
\newcommand{\xleftrightarrows}[2][]{\ext@arrow 3399\leftrightarrowsfill@{#1}{#2}}
\makeatother

\preprint{IPPP/24/47}

\title{\boldmath Dissecting polytopes: Landau singularities and asymptotic expansions in $2\to 2$ scattering}

\author[a]{Einan Gardi,} 
\author[a]{Franz Herzog,}
\author[b]{Stephen Jones,}
\author[c]{Yao Ma}

\affiliation[a]{Higgs Centre for Theoretical Physics, School of Physics and Astronomy,\\The University of Edinburgh, Edinburgh EH9 3FD, Scotland, UK}
\affiliation[b]{Institute for Particle Physics Phenomenology, Durham University, Durham DH1 3LE, UK}
\affiliation[c]{Institute for Theoretical Physics, ETH Zürich, 8093 Zürich, Switzerland}

\emailAdd{einan.gardi@ed.ac.uk}
\emailAdd{fherzog@ed.ac.uk}
\emailAdd{stephen.jones@durham.ac.uk}
\emailAdd{yaomay@phys.ethz.ch}

\abstract{
Parametric representations of Feynman integrals have a key property: many, frequently all, of the Landau singularities appear as endpoint divergences. This leads to a geometric interpretation of the singularities as faces of Newton polytopes, which facilitates algorithmic evaluation by sector decomposition and asymptotic expansion by the method of regions. Here we identify cases where some singularities appear instead as pinches in parametric space for general kinematics, and we then extend the applicability of sector decomposition and the method of regions algorithms to such integrals, by dissecting the Newton polytope on the singular locus. We focus on $2\to 2$ massless scattering, where we show that pinches in parameter space occur starting from three loops in particular nonplanar graphs due to cancellation between terms of opposite sign in the second Symanzik polynomial. While the affected integrals cannot be evaluated by standard sector decomposition, we show how they can be computed by first linearising the graph polynomial and then splitting the integration domain at the singularity, so as to turn it into an endpoint divergence. Furthermore, we demonstrate that obtaining the correct asymptotic expansion of such integrals by the method of regions requires the introduction of new regions, which can be systematically identified as facets of the dissected polytope. In certain instances, these hidden regions exclusively govern the leading power behaviour of the integral. In momentum space, we find that in the on-shell expansion for wide-angle scattering the new regions are characterised by having two or more connected hard subgraphs, while in the Regge limit they are characterised by Glauber modes.
}

\begin{document}
\maketitle
\flushbottom

\section{Introduction}
\label{section-introduction}

Evaluating Feynman integrals is an essential and central ingredient of Quantum Field Theory in a vast range of applications. Despite impressive progress over many years~\cite{Smirnov:2004ym,Henn:2014qga,Weinzierl:2022eaz}, it remains notoriously difficult. While analytic approaches to Feynman integrals, wherever available, provide highly valuable insights and eventually facilitate fast and flexible evaluation, numerical approaches~\cite{SmnTtyk09FIESTA,SmnSmnTtyk11FIESTA2,Smn14FIESTA3,Smn16FIESTA4,Smn22FIESTA5,KnkUeda10,Carter:2010hi,Borowka:2015mxa,Heinrich:2023til,pySecDec17,Brsk20,BrskMchTld23,LiuMa23AMFlow,Hdg21DiffExp,AmdlBcnDvtRanaVcn23SeaSyde} remain important, in particular when the complexity increases, in terms of the loop order, the number of kinematic scales, or the relevant function space. For similar reasons, also asymptotic expansions of Feynman integrals in ratios of external kinematic parameters become increasingly important. Specifically, the \emph{Method of Regions} (MoR)~\cite{BnkSmn97,Smirnov:2002pj,PakSmn11,Jtz11,JtzSmnSmn12,SmnvSmnSmv19,GrdHzgJnsMaSchlk22,Ma23} provides a general strategy to perform such expansions in dimensional regularisation\footnote{Alternatives, which avoid dimensional regularisation, also exist, see e.g.~\cite{Plenter:2020lop}.}, exploiting the fact that dimensional regularisation regularises both infrared and ultraviolet divergences. This method has been used to study a variety of kinematic limits and it provides insight on the infrared singularity structure of integrals and amplitudes. It also sets the basis for formulating effective field theories which capture the relevant degrees of freedom in certain limits, such as Heavy Quark Effective Theory (HQET)~\cite{IsgWise89,IsgWise90,Nbt98book,Mnh00book,Grz04book} or Soft-Collinear Effective Theory (SCET)~\cite{BchBrgFrl15book, BurFlmLk00, BurPjlSwt02-1, BurPjlSwt02-2, BnkChpkDhlFdm02, RstStw16}.  

The tremendous challenge presented by Feynman integrals is largely due to their complex analytic structure, dictated by their singularities (see e.g.~\cite{EdenLdshfOlvPkhn02book,Stm95book,ClsSprStm04,Stmg18,Agarwal:2021ais}).
Given any Feynman integral, a necessary condition for the emergence of singularities --- be it infrared divergences which manifest themselves as poles in the dimensional regulator, or branch points at particular configurations of the external kinematic parameters such as kinematic thresholds ---  is given by the Landau equations~\cite{Lnd59,Bjorken:1959fd,Nakanishi}. In momentum representation, the Landau equations imply that the integration contours of some loop momenta must be \emph{pinched} by singularities of the integrand. The solutions of the Landau equations then identify manifolds in the loop momentum space where the Feynman integral possibly diverges, typically characterised by momenta that are collinear, soft, etc. These manifolds, often referred to as \emph{pinch surfaces} in literature, have been studied extensively over the past decades~\cite{ClmNtn65,Stm78I,ClsStm81,Stm96lectures}, providing the basis for establishing factorisation properties of scattering amplitudes and cross sections.

Landau singularities and the associated pinch surfaces also dictate the properties of asymptotic expansions of Feynman integrals. Indeed, whenever the resulting function is non-analytic in the limit of interest, one expects that a straightforward expansion (a Taylor expansion of the integrand) about this limit should not commute with the integration over the loop momenta. Instead, one must carefully consider how the integration variables behave in the limit considered. 
In the MoR, one asserts that the integration domain can be divided into several distinct regions, such that the Feynman integral can be expressed as a sum over integrals expanded in these regions. In each such region, there is a distinct way in which the loop momenta behave in the limit, which in turn dictates how the integrand should be expanded (so that the expansion converges inside the region). Thus, each such region is governed by the vicinity of a particular pinch surface associated with a solution of the Landau equations~\cite{GrdHzgJnsMaSchlk22}, which determines the leading behaviour of the integrand. 
In practice, once the expansion is performed in each region, one simply performs the loop momentum integration over the entire domain~\cite{Smirnov:2002pj,Jtz11}, with the expectation that the extension beyond the domain of convergence would cancel amongst the various regions in the sum,\footnote{This remarkable property has been elucidated in some detail in ref.~\cite{Jtz11}, specifying conditions under which it is guaranteed to hold.} owing to the properties of dimensional regularisation. While no general proof of the MoR is available, this strategy has been utilised extensively, and shown to work in every instance, provided that certain issues are handled with sufficient care (e.g.~that all integrals are regularised). A remaining challenge is the need to identify a complete set of regions in loop momentum space at the outset, and so far there is no algorithmic method for doing so.

In addition to the momentum representation, Feynman integrals can also be expressed in parameter space using e.g. Feynman or Lee-Pomeransky representation~\cite{LeePmrsk13}. In these representations, the singularity structure is encoded in the zeros of the second Symanzik polynomial~$\mathcal{F}$. One advantage of these representations is that many (sometimes all) of the singularities manifest themselves as \emph{endpoint divergences}. These are fully characterised by the scalings of the parameters as they approach the endpoint of the integration domain (typically zero or infinity). The set of exponents defining this scaling for a particular Landau singularity are referred to as \emph{weight vectors}, or \emph{region vectors} in the context of the MoR. These vectors are entirely independent of the kinematic coefficients appearing in $\mathcal{F}$, and it is therefore possible to determine them geometrically~\cite{KnkUeda10,MzrTln22,AkHmHlmMzr22,BrskMchTld23,FvlMzrTln23prl,FvlMzrTln23pld}.
To this end one first maps each monomial of the Symanzik polynomials\footnote{Such a geometric construction can be based on the monomials of the product of the two Symanzik polynomials, ${\cal U}\cal{F}$, which is natural when working with Feynman parameterisation, or the monomials of their sum, ${\cal U}+\cal{F}$, which appears in the Lee-Pomeransky representation. See section~\ref{section-preliminaries} below for further details.}, corresponding to a Feynman graph with $N$ edges, to an $N$-dimensional vector whose entries are the powers of the parameters in that monomial.
Regarding each of these vectors as a point in an $N$-dimensional space, one then constructs a Newton polytope by forming the convex hull of all these points. One then finds that the normal vectors to the faces of the polytope are precisely the weight vectors characterising the endpoint divergences of the integral. This geometric construction has many applications. In particular, it facilitates the formulation of efficient algorithms to determine Landau singularities~\cite{Brown:2009ta,Panzer:2014caa,FvlMzrTln23pld,FvlMzrTln23prl,Dlapa:2023cvx,Helmer:2024wax}, the study of the infrared singularity structure of Feynman integrals in dimensional regularisation~\cite{AkHmHlmMzr22}, the numerical computation Feynman integrals using \emph{sector decomposition} \cite{SmnTtyk09FIESTA,SmnSmnTtyk11FIESTA2,Smn14FIESTA3,Smn16FIESTA4,Smn22FIESTA5,KnkUeda10,Carter:2010hi,Borowka:2015mxa,pySecDec17,Heinrich:2023til}, or the closely related tropical Monte Carlo integration~\cite{Brsk20,BrskMchTld23}, and the asymptotic expansion of Feynman integrals by the MoR directly in parameter space, as implemented in computer codes such as Asy2~\cite{JtzSmnSmn12} (as part of the program FIESTA~\cite{SmnTtyk09FIESTA,SmnSmnTtyk11FIESTA2,Smn14FIESTA3,Smn16FIESTA4,Smn22FIESTA5}), ASPIRE~\cite{AnthnrySkrRmn19} and pySecDec~\cite{pySecDec17,HrchJnsSlk22}.

Our recent study~\cite{GrdHzgJnsMaSchlk22} focused on the application of the geometric MoR to the~\emph{on-shell expansion} of massless particle scattering in general kinematics. Here one begins with integrals defined for off-shell external legs and expands them about the on-shell limit. It was shown there that region vectors correspond to pinch surfaces defined by solutions of the Landau equations in momentum space. In particular, this paper proposed that in the on-shell expansion, the region vectors identified using the Newton polytope stand in one-to-one correspondence with regions in momentum space, and that they involve just three types of modes: hard, collinear and soft. On this basis, a purely graph-theoretical algorithm was formulated to determine all these regions, which makes no reference to the parametric representation. A rigorous proof for this proposition was later given in ref.~\cite{Ma23}, with extensions to other types of expansions derived (see also ref.~\cite{Herzog:2023sgb}). These provide promising avenues to a more complete theoretical understanding of the underpinnings of the MoR and infrared divergences.

To summarise, a key advantage of formulating Feynman integrals in parameter space is the fact many of the singularities appear as endpoint divergences. This stands in contrast to momentum space, where (infrared) singularities in Minkowski space typically manifest themselves as pinches of the loop momenta. As a consequence, parametric representations have a geometric interpretation and they lend themselves naturally to algorithmic methods, providing highly-valued tools as well as a more general theoretical understanding of asymptotic expansion of Feynman integrals and the related singularities.

The present work is motivated by the observation that some singularities do in fact appear as pinches in parameter space. From the Landau equations, this occurs when both ${\cal F}$ and $\partial {\cal F}/\partial \alpha_i$ vanish away from the integration boundary. While this situation has been considered since the early days (see e.g.~\cite{EdenLdshfOlvPkhn02book}), its consequences for evaluating Feynman integrals using sector decomposition and for applying the MoR in parameter space have not yet received much attention.\footnote{Notable exceptions in the application of the MoR on integrals involving internal massive propagators near threshold or in the forward limit~\cite{JtzSmnSmn12,AnthnrySkrRmn19} will be discussed below (see section~\ref{section-motivating_examples}).} As we will see, existing sector decomposition algorithms fail when a pinch singularity is present. Similarly, algorithms based on identifying regions as facets of the Newton polytope, fail to capture the complete set of regions when expanded about a limit in which a pinch is present. The goal of this study is to overcome these challenges, so as to extend the applicability of sector decomposition and MoR algorithms to these types of integrals.

In contrast to endpoint divergences, pinch singularities appear in the Feynman parameterisation within the interior of the domain of integration. Such solutions of the Landau equations are sensitive to the signs of the monomials of the Symanzik $\mathcal{F}$ polynomial, and hence cannot be identified from the Newton polytope alone. Specifically, they arise when terms of opposite sign in $\mathcal{F}$ and in its derivatives cancel. The simplest example is the one-loop self-energy integral in the threshold limit (see section~\ref{section-motivating_examples}). In this case, $\mathcal{F}_{\text {bub}}=(q^2/4)(\alpha_1-\alpha_2)^2+y(\alpha_1+\alpha_2)^2$, where $\alpha_{1,2}$ are the Feynman parameters associated with the two propagators and $y\to 0 $ at threshold.
Upon expanding the integral in powers of $y$ one encounters a pinch singularity at $\alpha_1=\alpha_2$. To handle this, one can split the domain of integration at the singularity into the two sectors, $\alpha_1>\alpha_2$ and $\alpha_2>\alpha_1$. One then employs a change of integration variables, in particular in the sector $\alpha_1>\alpha_2$ one replaces $\alpha_1$ by $\tilde{\alpha}_1 = \alpha_1-\alpha_2>0$, after which the threshold singularity manifests itself as an endpoint singularity at $\tilde{\alpha}_1\to 0$ in the new variables $(\tilde{\alpha}_1,\alpha_2)$. By summing both sectors one covers the entire integration domain of the original integral, with each sector free of pinch singularities, rendering it amenable to the standard geometric MoR algorithm.
This example illustrates both the problem and the solution devised by Asy2~\cite{JtzSmnSmn12} and ASPIRE~\cite{AnthnrySkrRmn19}. These tools can resolve a pinch singularity in cases where there is \emph{a linear shift} of the parameters that turns the singularity into an endpoint divergence. It turns out, however, that this is merely the simplest cancellation structure occurring, while multi-loop integrals possess far more.

To see why the problem of pinches in parameter space might be a rather general feature of Minkowski-space Feynman integrals, recall that according to the graph-theoretical definition of the Symanzik ${\cal F}$ polynomial (see eq.~(\ref{UFterm_general_expression}) below) all Mandelstam invariants corresponding to channels on which the graph has cuts, enter with the same sign. However, a Feynman integral can only be defined after momentum conservation has been imposed, i.e.~all momentum-conservation relations between the Mandelstam invariants have been employed. This inevitably implies that certain monomials proportional to a given Mandelstam invariant will finally appear with opposite signs.

\begin{figure}[t]
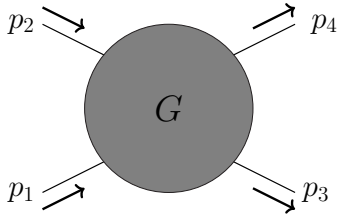

\centering
\include{figs/generic_2to2_forward_scattering}
\vspace{-2em}\caption{General massless two-to-two scattering, $1+2\to 3+4$. The Mandelstam invariants $s=(p_1+ p_2)^2$,\,$t=(p_1- p_3)^2$,\, $u=(p_1- p_4)^2$ admit $s+t+u=0$.}
\label{figure-generic_2to2_forward_scattering}
\end{figure}

It is easiest to demonstrate this for $2\to 2$ massless scattering (see figure~\ref{figure-generic_2to2_forward_scattering}), on which we focus in this paper. In this case, momentum conservation implies $s+t+u=0$. For planar graphs only two of the three invariants appear at the outset (before momentum conservation is used) so for every such integral there is a choice of kinematic regime in which all terms in ${\cal F}$ have the same sign -- referred to here as the \emph{same-sign regime} -- and any other kinematic region can be reached via analytic continuation. Thus, pinch singularities cannot occur in planar graphs for $2\to 2$ massless scattering.
However, in the nonplanar case ${\cal F}$ is bound to contain all three invariants, $s$, $t$, and $u$, so upon using momentum conservation to eliminate say $u$, the terms in ${\cal F}$ proportional to $s$ and to $t$ would each contain monomials with opposite signs. This exemplifies the situation where an integral does not have a \emph{same-sign regime}, and it opens up the possibility that a nontrivial cancellation in ${\cal F}$ would occur in \emph{general kinematics}, leading to a pinch and potentially, an infrared divergence of the integral to which the Newton polytope is blind.

The requirements for a pinch which would prohibit the computation of the integral by sector decomposition, are far more strict still: not only ${\cal F}$ but also its derivatives must vanish. Furthermore, this should occur within the domain of integration. These requirements are not easily satisfied, but we find that starting from three loops there are indeed massless $2\to 2$ graphs that display such a pinch (even in general kinematics), and in such cases, application of existing sector decomposition and MoR algorithms is doomed to fail. Although not analysed in the context of these algorithms, we point out that graphs admitting pinches in parameter space have been identified in the literature as interesting, and their asymptotic expansions have been analysed by multiple authors~\cite{Pkhn63I,Pkhn63II,IslLdsfTlr63,Hld63,Mdst63,Tktpls63,Tktpls63nonplanar,Hld64,Lam69,Ldsf74,Cls77book,Pkhn80book}, both in wide-angle and forward kinematics.

The simplest example of a massless $2\to 2$ Feynman graph featuring a pinch in parameter space is the 8-propagator graph $G_{\bullet\bullet}$ (see figure~\ref{possibly_hidden_singular_graph_dot_dot}), a topology sometimes referred to as the \emph{crown}. After eliminating $u$, the terms proportional to $s$ in the Symanzik ${\cal F}$ polynomial take the form $(\alpha_a\alpha_b - \alpha_c\alpha_d)(\alpha_e\alpha_f - \alpha_g\alpha_h)$, and similarly for the terms proportional to $t$, albeit with a different combination of Feynman parameters.
In this case, all the $\partial \mathcal{F}/\partial \alpha_i$ (and thus $\mathcal{F}$) vanish upon satisfying three independent nonlinear relations between the $\alpha_i$, which are reached within the domain of integration. This is therefore a genuine singularity of the integral, and it occurs for any values of $s$ and $t$. 
Clearly, the singularity cannot be mapped to an endpoint of the integration domain via a linear shift. However, through a suitable rescaling of the parameters, we are able to linearise the polynomial near the singular locus and subsequently split the integration domain into sectors, in order to finally map the singularity onto the endpoint of the integration domain in each sector. In terms of the associated geometry, this operation amounts to dissecting the Newton polytope into 24 separate polytopes, each of which is amenable to applying the standard sector decomposition algorithm. Ultimately, this allowed us to evaluate the integral numerically with high precision, and check the result against the analytic computation~\cite{Henn:2020lye,Bargiela:2021wuy}. 

We proceed by using the same technique of dissecting the polytope to perform asymptotic expansions of integrals in this class. To verify the results and gain another perspective, we also perform asymptotic expansions in momentum space\footnote{To this end the work of Botts and Sterman~\cite{BottsStm89} provides an excellent starting point.} and compare them to analytic results where available. We consider two cases: wide-angle scattering and the Regge limit. In the context of wide-angle scattering we study the on-shell expansion. Specifically, we start with a four-point integral with massless propagators and one off-shell external leg, with the remaining three external legs strictly on shell, and then expand in the mass (virtuality) of the off-shell leg. Since the expansion is performed about the on-shell limit, where a pinch is present, the expansion is sensitive to the pinch surface. Indeed, we find that this singularity generates a new region, which is not visible using the original Newton polytope. We refer to this as a \emph{hidden region} and we show that it appears as a facet only after dissecting the polytope as explained above. This hidden region has a clear physical interpretation: it features two distinct hard vertices, providing the simplest example\footnote{These contributions are known to be suppressed in gauge theory~\cite{BottsStm89SudakovEffects,BottsStm89,LbtdStm85}, but provide important examples for our study of scalar integrals.} of \emph{Landshoff scattering}~\cite{Hld64,Ldsf74,BottsStm89}. This stands in contrast to facet regions, which have just one hard subgraph in wide-angle scattering~\cite{GrdHzgJnsMaSchlk22,Ma23}.
Turning to the Regge limit of $2\to 2$ scattering, we similarly show that a hidden region emerges upon dissecting the polytope. In this case the new region involves a loop momentum in the \emph{Glauber mode}. For the particular case of the 8-propagator graph $G_{\bullet\bullet}$ in figure~\ref{figure-possibly_hidden_singular_graphs} in both the wide-angle on-shell expansion and the Regge-limit expansion, the hidden regions alone govern the leading asymptotic behaviour: all other regions, which are visible as facets of the original Newton polytope, are power suppressed compared to the hidden one!

This paper is organised as follows. In section~\ref{section-preliminaries} we set up our notation, briefly recalling the Feynman and Lee-Pomeransky parametric representations of Feynman integrals, the Landau equations and the geometric approach to sector decomposition and to the MoR. In section~\ref{section-hidden_landau_singularities} we have two main objectives. The first is to use the Landau equations in order to set up an algorithmic search for graphs that are prone to having a pinch in parameter space, which we apply to $2\to 2$ massless scattering graphs up to four loops. In the rest of the paper we analyse the three-loop examples we identified. The second objective is to devise a method of applying sector decomposition to integrals with a pinch by dissecting the polytope as outlined above. We end this section with a comparison between the numerical computation and analytic results.
Next, in section~\ref{section-revisit_onshell_expansion_wideangle_scattering} we turn to study the MoR in the aforementioned class of three-loop examples, focusing on the on-shell expansion for wide-angle scattering. We perform a power counting analysis in both parameter space and momentum space, demonstrating the presence of a hidden region. We also provide a numerical analysis of the same graph near the on-shell limit. We demonstrate that this class of integrals provides a departure from the graphical rules found in refs.~\cite{GrdHzgJnsMaSchlk22}, where only endpoint singularities were considered, in that they involve more than one connected hard subgraph.
In section~\ref{section-high_energy_expansion_forward_scattering} we perform a similar analysis in the Regge limit, starting with the strictly on-shell integrals. In this case we are able to verify our conclusions regarding MoR expansion, by comparing it to the expansion of the analytic result in powers of $-t/s$. Finally, in section~\ref{section-conclusion_outlook} we summarise our conclusion and provide an outlook. 

\section{Preliminaries}
\label{section-preliminaries}

In this section, we review the concepts and definitions necessary for our subsequent discussions of Landau singularities and the MoR. We begin in section~\ref{section-parameter_representations} by introducing the Feynman the Lee-Pomeransky representations of Feynman integrals. In the context of the Feynman representation, we recall in section~\ref{section-Landau_equations_parameter_space} the Landau equations, which we later on use for identifying Landau singularities. 
In section~\ref{section-newton_polytopes_sector_decomposition} we describe the geometric approach to sector decomposition.
Finally, using the Lee-Pomeransky representation, we briefly recall in section~\ref{section-Newton_polytope_approach} how Newton polytopes can be used to systematically identify regions suitable for the asymptotic expansion of Feynman integrals.

\subsection{Parametric representations}
\label{section-parameter_representations}

Throughout this paper, we will use $G$ to denote a Feynman graph, $\I$ to denote a Feynman integral, and $N$ and $L$ to denote the numbers of propagators and loops of $G$, respectively. We start from the Feynman representation of $\I(G)\equiv \I(\s)$, which reads:
\begin{eqnarray}
\I(\s)=\frac{\Gamma(\nu-LD/2)}{\prod_{e\in G}\Gamma(\nu_e)} \int_0^\infty \left(\prod_{e\in G} \mathrm{d} \alpha_e\, \frac{\alpha_e^{\nu_e}}{\alpha_e}\right) \delta \left(\sum_{e\in G}\alpha_e -1\right) \frac{\left[\mathcal{U}(\boldsymbol{\alpha})\right]^{\nu-(L+1)D/2}}{\left[\mathcal{F}(\boldsymbol{\alpha};\s)\right]^{\nu-LD/2}}\,,
\label{eq:feynman_representation}
\end{eqnarray}
where $D$ is the dimension of space-time, $\nu_e$ is the exponent of the denominator associated to the propagator of edge $e$, and $\nu\equiv \sum_{e\in G}\nu_e$. We have used $\boldsymbol{\alpha}$ to denote the set of Feynman parameters $\alpha_1,\dots,\alpha_N$, and $\s$ to denote the set of Mandelstam variables of $\mathcal{I}$, which consist of Lorentz invariants formed with the external momenta $\{p_i\}$ as well as internal squared masses $\{m_e^2\}$, if present.

The polynomials $\mathcal{U}(\boldsymbol{\alpha})$ and $\mathcal{F}(\boldsymbol{\alpha};\s)$, which are called the first and second Symanzik polynomials respectively, are given by:
\begin{eqnarray}
\label{UFterm_general_expression}
\mathcal{U}(\boldsymbol{\alpha})=\sum_{T^1}^{}\prod_{e\notin T^1}^{}\alpha_e,\qquad \mathcal{F}(\boldsymbol{\alpha};\s)=\sum_{T^2}^{} (-s_{T^2}^{}) \prod_{e\notin T^2}^{}\alpha_e +\mathcal{U}(\boldsymbol{\alpha})\sum_{e}^{}m_e^2 \alpha_e - i\varepsilon\ , 
\end{eqnarray}
where $\varepsilon>0$ represents the Feynman prescription. The notations $T^1$ and $T^2$ denote a \emph{spanning tree} and a \emph{spanning 2-tree} of the graph $G$, respectively. The symbol $s_{T^2}^{}$ is the square of the total momentum flowing between the components of the spanning 2-tree $T^2$.

An alternative representation of Feynman integrals, which will be particularly useful for applying the MoR in what follows, is the Lee-Pomeransky representation:
\begin{equation}
\I(\s) = \frac{\Gamma(D/2)}{\Gamma\left((L+1)D/2-\sum_{e\in G}\nu_e\right) \prod_{e\in G}\Gamma(\nu_e)}\cdot \int_0^\infty \left( \prod_{e\in G} \frac{ \mathrm{d} x_e}{x_e}\right)\cdot \left(\prod_{e\in G} x_e^{\nu_e} \right) \cdot \Big( \mathcal{P}(\x;\s) \Big)^{-D/2}.
\label{eq:lee_pomeransky_definition}
\end{equation}
Similar to above, we have used $\x$ to denote the set of Lee-Pomeransky parameters $x_1,\dots,x_N$, with $N$ the number of edges (enumerated by $e$) of the Feynman graph~$G$; $\nu_{e}$ is the exponent of the denominator associated with the propagator $e$. The Lee-Pomeransky polynomial, $\mathcal{P}(\x;\s)$, is equal to the sum over the Symanzik polynomials:
\begin{eqnarray}
\label{lee_pomeransky_integrand_definition}
    \mathcal{P}(\x;\s)\equiv \mathcal{U}(\x)+\mathcal{F}(\x;\s)\,.
\end{eqnarray}
Note that the Feynman and Lee-Pomeransky parameters can be related via:
\begin{eqnarray}
\label{eq:Feynman_LP_relation}
\alpha_e=\frac{x_e}{x_1+x_2+\dots+x_N}.
\end{eqnarray}
Using this relation the Lee-Pomeransky representation, eq.~(\ref{eq:lee_pomeransky_definition}), can be translated back into the Feynman parameterisation, eq.~(\ref{eq:feynman_representation}) as follows. One first inserts 
\[
1=\int_{-\infty}^\infty dX \delta \left(\sum_{e\in G} x_e -X\right) 
\]
into eq.~(\ref{eq:lee_pomeransky_definition}), and then changes the variables $x_e= X\alpha_e$ according to (\ref{eq:Feynman_LP_relation}). At this point one uses the homogeneity properties of the Symanzik polynomials in (\ref{UFterm_general_expression}), to find that the $\mathcal{U}$ polynomial scales as $X^L$ while the $\mathcal{F}$ polynomial scales as $X^{L+1}$, which makes it possible to integrate over $X$ and obtain a ratio of gamma functions. The result is exactly (\ref{eq:feynman_representation}).

\subsection{Landau equations in parameter space}
\label{section-Landau_equations_parameter_space}

In the Feynman representation, a singularity can occur when either  $\mathcal{U}$ or $\mathcal{F}$ vanish. Such singularities can occur at the endpoint of the domain of integration, $\alpha_i\to 0$, forcing every monomial to vanish independently. These are often referred to as \emph{endpoint singularities} in the literature.
Additionally, since the $s_{T^2}$ in ${\cal F}$ in (\ref{UFterm_general_expression}) may have different signs, $\mathcal{F}=0$ singularities may arise due to cancellations between different terms. Such singularities may sometimes be avoided by analytic continuation, namely by deforming the integration contour of the relevant $\alpha_k$-parameters into the complex plane by introducing a small imaginary part $i\varepsilon_k$:
\begin{equation}
\label{cont_def_}
 \alpha_k\to \alpha_k-i\varepsilon_k (\boldsymbol{\alpha}) \,. 
\end{equation}
Under such a change of variables the $\mathcal{F}$ polynomial transforms as follows
\begin{equation}
\mathcal{F}(\boldsymbol{\alpha};\s)\to \mathcal{F}(\boldsymbol{\alpha};\s) 
-i \sum_k \varepsilon_k(\boldsymbol{\alpha}) \frac{\partial \mathcal{F}(\boldsymbol{\alpha};\s)}{\partial \alpha_k}+\mathcal{O}(\varepsilon^2)\,.
\end{equation}
Thus, provided that for some $k$ (for which $\alpha_k\neq 0$) one has $\frac{\partial \mathcal{F}(\boldsymbol{\alpha};\s)}{\partial \alpha_k}\neq 0$, the contour-deformed  integral avoids the singularity. This provides a definition of the integral by analytic continuation (note that according to the Feynman prescription in eq.~(\ref{UFterm_general_expression}), one should choose $\varepsilon_k(\boldsymbol{\alpha})$ in eq.~(\ref {cont_def_}) to have the same sign as $\frac{\partial \mathcal{F}(\boldsymbol{\alpha};\s)}{\partial \alpha_k}$ in the neighbourhood where $\mathcal{F}$ vanishes). 
However, if both ${\cal F}=0$ and $\partial \mathcal{F}/\partial \alpha_k=0$ for all $k$ for which $\alpha_k\neq 0$, the contour becomes \emph{pinched}. In this case the singularity cannot be avoided, and would have an imprint on the analytic structure of $\I(\s)$. Singularities of this type are referred to as \emph{pinch singularities}. 

In conclusion, on a singular surface, therefore, each $\alpha_k$ must either vanish or have its contour pinched. The necessary condition for a singularity to occur are summarised by the Landau equations, which can be expressed as~\cite{Lnd59,EdenLdshfOlvPkhn02book}
\begin{eqnarray}
\mathcal{F}=0,\qquad \alpha_k \, \frac{\partial\mathcal{F}}{\partial \alpha_k} =0\quad \text{for each }k\in\{1,\dots,N\}.
\label{eq:Landau_equation_parameter_representation}
\end{eqnarray}
We remark that due to the homogeneity of $\mathcal{F}$ we also have, by Euler's theorem, 
\begin{equation}
\sum_{k=1}^N \alpha_k \,\frac{\partial \mathcal{F}}{\partial \alpha_k} \,=(L+1)\mathcal{F}\,,
\end{equation}
where $L$ is the number of loops in $G$. Thus, the condition $\mathcal{F}=0$ is automatically fulfilled when $\alpha_k\partial\mathcal{F}/\partial \alpha_k=0$ for all $k$.

Recently, progress has been made in solving the Landau equations algorithmically~\cite{FvlMzrTln23pld,FvlMzrTln23prl,Dlapa:2023cvx,JiangLiuXuYang24,Helmer:2024wax,CrHtCrrGrx24}, which is a profoundly difficult problem in the general setting~\cite{Brown:2009ta,Panzer:2014caa}. Solutions can be characterised by manifolds of any dimension, residing in the large space of kinematic invariants times the $N$-dimensional \emph{complex} parameter space $\{\alpha_k\}$. Our interest here instead is limited to singularities which may obstruct the computation of the integral using sector decomposition or introduce regions in an asymptotic expansion by the MoR. Specifically, these restrictions allow us to focus on solutions to eq.~(\ref{eq:Landau_equation_parameter_representation}) that appear within the integration domain, $\alpha_k\geqslant 0$.

\subsection{Newton polytopes for sector decomposition}
\label{section-newton_polytopes_sector_decomposition}

One of the central objects of study in this work is the Newton polytope associated to Feynman integrals either in Feynman or Lee-Pomeransky representation. The Newton polytope can be used to study the behaviour of integrands in the limit that sets of parameters become small or large simultaneously. The behaviour of the integrand near the boundaries of parameter space can give rise to ultraviolet or infrared singularities which appear as poles in the dimensional regulator after integration.
Aside from a potential overall ultraviolet divergence in the $\Gamma$-function prefactors, the Newton polytope fully captures the singular structure of integrals in the same-sign regime, where all terms in ${\cal F}$ have the same sign. Away from the same-sign regime, or for integrals without a same-sign regime, additional singularities can be present, for example at thresholds where intermediate particles go on shell.
Here, we briefly recap how Newton polytopes can be used to study and blow-up, or resolve, singularities at the integration boundary using sector decomposition.\footnote{We refer the interested reader to refs.~\cite{Bogner:2007cr,KnkUeda10,Kaneko:2010kj,Schlenk:2016cwf,Schlenk:2016epj,Weinzierl:2022eaz} for a more in-depth treatment.} Then, in section~\ref{section-hidden_landau_singularities}, we describe a situation in which the Newton polytope alone is insufficient to resolve the singularities of certain integrals that do not have a same-sign regime.

The Newton polytope corresponding to the integral $\mathcal{I}(\s)$ in Feynman parameter space is obtained as follows.
Starting from the Feynman representation given in eq.~\eqref{eq:feynman_representation} the Dirac $\delta$-distribution is first integrated out after using the Cheng-Wu theorem to replace it by $\delta(1-\alpha_N)$.
Next, the product of the Symanzik polynomials, $\mathcal{U}$ and $\mathcal{F}$, given in eq.~\eqref{UFterm_general_expression}, is constructed
\begin{align}
\mathcal{U}(\boldsymbol{\alpha}) \cdot
\mathcal{F}(\boldsymbol{\alpha};\s) = \sum_{i=1}^m b_i(\s)\, \alpha_1^{r_i,1} \cdots \alpha_{N-1}^{r_i,{N-1}} \equiv  \sum_{i=1}^m b_i(\s)\boldsymbol{\alpha}^{\boldsymbol{r}_i},
\end{align}
where $\boldsymbol{r}_i$ are $(N-1)$ dimensional exponent vectors.
The Newton polytope, $\Delta(\mathcal{U} \cdot \mathcal{F})$, is then defined as the convex hull of the vertices given by $\boldsymbol{r}_i$,
\begin{align}
\label{eq:convex_hull_UtimesF}
\Delta(\mathcal{U} \cdot \mathcal{F}) &\overset{\text{def}}{=} \mathrm{convHull}(\boldsymbol{r}_1,\ldots,\boldsymbol{r}_m) = \left\{ \sum_{i=1}^m a_i \boldsymbol{r}_i\ \Big|\ a_i \geqslant 0 \land \sum_{i=1}^m a_i = 1 \right\}\\
&= \bigcap_{f\in F} \left\{ \boldsymbol{\rho}\in\mathbb{R}^{N-1} \mid \boldsymbol{\rho}\cdot\boldsymbol{n}_f + a_f \geqslant 0 \right\},\quad a_f \in \mathbb{Z}\quad \forall\ f.
\label{eq:newton_polytope_UtimesF}
\end{align}
In the first line, the Newton polytope is described by the vertices, $\boldsymbol{r}_i$, while in the second line, the polytope is instead defined by inward pointing vectors, $\boldsymbol{n}_f$, normal to the facets of the polytope $f \in F$.

Next we describe how Newton polytope geometry is used to perform sector decomposition of the integral. Let $S_j$ denote the set of facets incident on the extremal vertex, $\boldsymbol{r}_j$; they define a local change of variables,
\begin{align}
\alpha_i = \prod_{f \in S_j} y_f^{\langle \boldsymbol{n}_f,\boldsymbol{e}_i \rangle},
\label{eq:sd_variable_change}
\end{align}
where $\boldsymbol{e}_i$ are standard basis vectors in $\mathbb{R}^{N-1}$.
If more than $N-1$ facets are incident on an extremal vertex then a further subdivision, known as triangulation, is required. This means that each extremal vertex leads to one or more (if triangulation is required) sectors.

After applying the change of variables for each sector, the integral can now be written as the sum,
\begin{align}
\label{eq:sector_decomposed_integral}
\mathcal{I}(\s) = \sum_{\sigma} |\sigma| \int_0^1 \left( \prod_{f \in \sigma} \frac{\mathrm{d} y_f}{y_f} y_f^{\langle \boldsymbol{n}_f, \boldsymbol{\nu} \rangle - (t_{\mathcal{U}} + t_{\mathcal{F}}) a_f} \right) \, \left(\tilde{\mathcal{U}}_\sigma (\boldsymbol{y}_f)\right)^{t_\mathcal{U}} \left(\tilde{\mathcal{F}}_\sigma(\boldsymbol{y}_f ;\s)\right)^{t_\mathcal{F}}.
\end{align}
Here $\sigma$ indexes the simplicial cones associated with the extremal vertices, $|\sigma|$ is the Jacobian determinant associated to the change of variables in each sector, $\boldsymbol{\nu}$ is a vector of propagator exponents built from $\nu_e$, and $t_\mathcal{U} = \nu-(L+1)D/2$ with $t_\mathcal{F} = -\nu+LD/2$.
The polynomials $\tilde{\mathcal{U}}_\sigma$ and $\tilde{\mathcal{F}}_\sigma$ can be
obtained by substituting the change of variables of eq.~\eqref{eq:sd_variable_change} and factoring out an overall monomial, giving
\begin{align}
\mathcal{U}(\boldsymbol{y}_f) = \left(\prod_{f \in \sigma} y_f^{-a_f} \right) \tilde{\mathcal{U}}_\sigma(\boldsymbol{y}_f), \qquad \qquad 
\mathcal{F}(\boldsymbol{y}_f;\s) = \left( \prod_{f \in \sigma} y_f^{-a_f} \right) \tilde{\mathcal{F}}_\sigma(\boldsymbol{y}_f ;\s)\,.
\end{align}
The polynomials $\tilde{\mathcal{U}}_\sigma$ and $\tilde{\mathcal{F}}_\sigma$ are guaranteed to each contain a constant term.
The singularities appearing at the boundaries of integration now appear as an overall factor in eq.~\eqref{eq:sector_decomposed_integral} and the remaining polynomials, $\tilde{\mathcal{U}}_\sigma$ and $\tilde{\mathcal{F}}_\sigma$, are guaranteed to be finite for any subset of $y_f \rightarrow 0$.

Recently, there has been a resurgence of interest in using geometric concepts to understand the singularity structure of Feynman integrals, including both dimensionally regulated and kinematic singularities.
For example, a similar geometrical construction has been used to obtain a general formula for predicting the leading singularities in the $\epsilon$ expansion of generic Feynman integrals~\cite{AkHmHlmMzr22}. 
In refs.~\cite{FvlMzrTln23prl,FvlMzrTln23pld}, a geometrical construction was used to find solutions of the Landau equations depending on the kinematic invariants.
In another development, concepts from tropical geometry, including \emph{generalised permutahedra}, were used to devise a procedure for efficiently integrating finite Feynman integrals~\cite{Brsk20,BrskMchTld23}.
Also in the context of tropical geometry, very recent work has focused on studying subtraction formulae using the facets of the Newton polytopes~\cite{Svtr24}.
Symmetries of Feynman integrals have also been investigated using the symmetries of Newton polytopes~\cite{dlCrz24}.

\subsection{Newton polytopes for the MoR}
\label{section-Newton_polytope_approach}

In parameter space, the scaleful regions contributing to the MoR expansion can be found algorithmically using Newton polytopes.
The approach is believed to capture all relevant regions, provided the terms of the $\mathcal{F}$ or, equivalently, $\mathcal{P}$ polynomials do not cancel each other in the positive hyperoctant~\cite{JtzSmnSmn12, Smirnov:2002pj, Smirnov:2004ym} leading to a pinch in the limit about which we expand.
Here we briefly review the Newton polytope approach to the MoR in our notation, assuming such cancellations do not occur. Then, in the remainder of this paper, we examine the situation where such cancellations do occur and are relevant for both the evaluation and expansion of Feynman integrals.

The MoR begins by assuming that a subset of invariants $\t \subset \s$ are small compared to the rest, and that we are interested in the asymptotic expansion of eq.~(\ref{eq:lee_pomeransky_definition}) around the kinematic limit in which they vanish.
To this end, it is convenient to introduce an $|\s|$-dimensional scaling vector $\w=(w_1,w_2,w_3,\dots)$, with $w_i=1$ if $s_i\in \t$ and $w_i=0$ otherwise, and introduce a parameter $\lambda\ll 1$ such that $s_i\sim \lambda^{w_i} Q^2$, with $Q^2$ some fixed nonzero scale. 
A specific class of expansions we would be interested in, following \cite{GrdHzgJnsMaSchlk22}, is the on-shell expansion in massless scattering, where the set $\t$ consists exclusively of the virtualities $p_i^2$ of some external legs. A more detailed definition is given in eq.~(\ref{eq:wideangle_onshell_kinematics}) below. We will also discuss the Regge limit in on-shell $2\to 2$ massless scattering where the set $\t$ consists exclusively of the Mandelstam invariant $t$, while $s$ remains large.

The Newton polytope of interest for applying the MoR can be obtained using the Lee-Pomeransky polynomial\footnote{Alternatively, we can obtain the region vectors using the \emph{product} $\mathcal{U}(\boldsymbol{\alpha}) \cdot \mathcal{F}(\boldsymbol{\alpha};\s)$ rather than the \emph{sum} appearing in $\mathcal{P}(\boldsymbol{x};\s)$. As shown in ref.~\cite{SmnvSmnSmv19}, the resulting region vectors are in one-to-one correspondence and differ only by a constant shift, see also refs.~\cite{HrchJnsSlk22,GrdHzgJnsMaSchlk22}.}, defined in eq.~\eqref{lee_pomeransky_integrand_definition}, with the Mandelstam invariants rescaled as $\s\to \lambda^{\boldsymbol{w}}\s$,
\begin{align}
&\mathcal{P}(\boldsymbol{x};\lambda^{\w} \s) = \sum_{i=1}^m c_i(\s)\, x_1^{r_{i,1}} \dots x_N^{r_{i,N}}\, \lambda^{r_{i,N+1}},
\label{LP_polynomial_rescaling_momenta}
\end{align}
where, for Feynman integrals, the powers $r_{i,j} \in \{0,1,2\}$. For each $i$, we define the $(N+1)$-dimensional exponent vector $\boldsymbol{r}_i\equiv (r_{i,1}, \ldots, r_{i,N}; r_{i,N+1}) \equiv (\hat{\r}_i; r_{i,N+1})$. The Newton polytope $\Delta(\mathcal{P})$ is then the \emph{convex hull} of the exponent vectors,
\begin{align}
\label{eq:convex_hull_UplusF}
\Delta(\mathcal{P}) &\overset{\text{def}}{=} \mathrm{convHull}(\boldsymbol{r}_1,\ldots,\boldsymbol{r}_m) = \left\{ \sum_i^m a_i \boldsymbol{r}_i\ \Big|\ a_i \geqslant 0 \land \sum_i^m a_i = 1 \right\}\\
&= \bigcap_{f\in F} \left\{ \boldsymbol{\rho}\in\mathbb{R}^{N+1} \mid  \boldsymbol{\rho}\cdot\boldsymbol{v}_f + a_f \geqslant 0 \right\},\quad a_f \in \mathbb{Z}\quad \forall\ f,
\label{eq:newton_polytope_UplusF}
\end{align}
where $\boldsymbol{v}_f$ are inward pointing vectors normal to the facets of the polytope $f \in F$.
The Newton polytope $\Delta(\mathcal{P})$ is an $(N+1)$-dimensional polytope, the first $N$ dimensions correspond to the Lee-Pomeransky parameters $x_1, \ldots, x_N$, which are integrated over, while the $(N+1)$-th dimension corresponds to the expansion parameter $\lambda$.

The \emph{lower facets} of $\Delta(\mathcal{P})$ are defined to be those facets which have an inward pointing normal vector $\boldsymbol{v}_f$ with a positive $(N+1)$-th component, $v_{N+1} > 0$.
Let us call the set of lower facets $F_+$, with $F_+ \subset F$.
The lower facets capture the behaviour of $\mathcal{P}$ where $\lambda \rightarrow 0$ and describe the regions relevant for the MoR.
The integrand of each region can by obtained by rescaling the integrand eq.~\eqref{eq:lee_pomeransky_definition} as $\s\to \lambda^{\boldsymbol{w}}\s$ and $\x \to \lambda^{\boldsymbol{u}_R} \x$, $\lambda \to \lambda^{v_{R,N+1}}$, where $\boldsymbol{v}_R = (\boldsymbol{u}_R, v_{R,N+1})$ and $R \in F_+$.

Given a scaling vector $\boldsymbol{u}_R$ for some region $R$, the scaling rule of each Lee-Pomeransky parameter $x_e\in \x$ can be directly translated into the scaling of rule of the corresponding edge momentum $l_e$ as follows~\cite{GrdHzgJnsMaSchlk22,Engel:2022kde}:
\begin{equation}
\label{x_scaling_rule}
x_e \sim \frac{1}{l_e^2} \sim \lambda^{u_{R,e}}.
\end{equation}
This relates regions in parameter space to regions in momentum space. Specifically, in the on-shell expansion scaling vectors $\boldsymbol{u}_R$ involve just three types of entries: $u_{R,e}\in \{0,-1,-2\}$, which correspond respectively to the following three momentum modes: \emph{hard}, \emph{collinear} and \emph{soft}~\cite{GrdHzgJnsMaSchlk22,Ma23}. Note that in momentum space the virtuality does not always uniquely fix the scaling of individual components of the loop momentum (for example a jet loop momentum can be collinear to different external vectors). In the on-shell expansion, upon imposing momentum conservation this mapping is uniquely determined.

As described in ref.~\cite{GrdHzgJnsMaSchlk22}, the lower facets of the Newton polytope capture the solutions of the Landau equations that are relevant for infrared regions, as well as the hard region.
However, the Newton polytope is completely insensitive to the coefficients $c_i(\s)$ appearing in eq.~\eqref{LP_polynomial_rescaling_momenta}.
If we allow the coefficients to differ in sign, the terms of $\mathcal{P}$ may cancel each other, potentially leading to further \emph{pinch solutions} of the Landau equations not captured by the facet analysis presented above.
These solutions can give rise to additional scaleful regions that must be included in the MoR analysis. We refer to the regions overlooked by the Newton polytope procedure as \emph{hidden regions}.

\section{Computing integrals with pinched contours in parametric representation}
\label{section-hidden_landau_singularities}

As already mentioned, our prime motivation for this work stems from the MoR, and specifically from the geometric approach to the MoR. In contrast to momentum space, the geometric approach offers an algorithmic procedure of identifying regions as (lower) facets of the Newton polytope defined by the Symanzik graph polynomials~\cite{PakSmn11, HrchJnsSlk22}. 
In the Lee-Pomeransky representation this polytope is defined as the convex hull of all monomials in $\mathcal{P}(\x;\s)$ of eq.~(\ref{lee_pomeransky_integrand_definition}) in the space defined by the \emph{powers} of the parameters $x_i$. Importantly, the coefficients of these monomials are not encoded in the polytope. For integrals defined in Euclidean space, all monomials in $\mathcal{P}(\x;\s)$ have the same sign, and then the information contained in the polytope is sufficient to identify all regions. In Minkowski space, the situation is different in that the signs of distinct monomials appearing in the ${\cal P}$ polynomial may play an important role in dictating the singularity structure of the integral: additional singularities may emerge due to particular cancellations between monomials of opposite sign. In certain cases, these singularities may be essential for identifying MoR regions. Indeed, it is well known that some regions, which are necessary for obtaining the correct asymptotic expansion of certain Feynman integrals~\cite{JtzSmnSmn12,AnthnrySkrRmn19}, do not correspond to facets of the polytope associated with $\mathcal{P}(\x;\s)$. In this work we refer to such regions as \emph{hidden regions}.

A key step in developing the geometric approach to the MoR was taken in ref.~\cite{JtzSmnSmn12}, which was the first to propose a method to systematically identify such hidden regions. Ref.~\cite{JtzSmnSmn12} presented a Mathematica code, {\tt{asy2.m}}, which is designed to identify hidden regions associated with a \emph{linear} cancellation between monomials.
In such cases it separates the integration domain at the singularity locus, and maps each sector back onto the standard integration domain by a suitable shift of the integration parameters. The newly defined polynomial in each such sector, which is free of the aforementioned cancellation, gives rise to a new Newton polytope whose facets correspond to regions. This technique has been demonstrated~\cite{JtzSmnSmn12,SmnvSmnSmv19,AnthnrySkrRmn19} to identify all regions in certain classes of integrals. 
We proceed by briefly recalling the method and then reviewing known examples where it applies. 

\subsection{Motivating examples}
\label{section-motivating_examples}

The method to resolve hidden regions, known as \texttt{PreResolve} in \texttt{asy2}, works by selecting a variable appearing in a positive monomial, $\alpha_i$, and a variable appearing in a negative monomial, $\alpha_j$, then making a linear change of variables,
\begin{subequations}
    \begin{align}
    \alpha_i &\rightarrow m_{ii}\, \alpha_i + m_{ij}\, \alpha_j, \\
    \alpha_j &\rightarrow m_{ji}\, \alpha_i + m_{jj}\, \alpha_j
\end{align}
\end{subequations}
with the parameters $m$ chosen (if possible) to eliminate a negative monomial.
The above replacements are tried on all pairs of variables appearing in opposite sign monomials and iterated provided the number of negative monomials is not increased by the substitution.
The resolution procedure is considered successful if, after a finite number of iterations, the modified $\mathcal{F}$ polynomial has only positive monomials.
A similar procedure was introduced in the paper of Ananthanarayan et al.~\cite{AnthnrySkrRmn19} which works by applying the above procedure to the Gr\"{o}bner basis of the $\mathcal{F}$ polynomial and its first order derivatives, rather than directly on $\mathcal{F}$ itself.
In the remainder of this section we will consider some examples where the above procedures work and discuss the interpretation of the output.
In sections~\ref{section-revisit_onshell_expansion_wideangle_scattering} and \ref{section-high_energy_expansion_forward_scattering}, we will discuss in detail situations in which the above procedures fail.

The first example for which \texttt{PreResolve} works is given by a one-loop massive bubble close to threshold, 
\begin{align}
\label{Threshold_example}
{\cal I}_{\text {bub}}(q^2,y)&=\int \frac{d^Dk}{i\pi^{D/2}} \frac{1}{(k^2-m^2)\big((k-q)^2-m^2\big)}\nonumber\\
 &=\Gamma(\eps) \int d\alpha_1 d\alpha_2 \frac{\delta(1-\alpha_1-\alpha_2)  (\alpha_1+\alpha_2)^{-2+2\eps}}{\Big(\mathcal{F}_{\text {bub}}(\alpha_1,\alpha_2;q^2,y)\Big)^{\eps}}
\end{align}
with $D=4-2\epsilon$ and 
\[
\mathcal{F}_{\text {bub}}(\alpha_1,\alpha_2;q^2,y)=(q^2/4)(\alpha_1-\alpha_2)^2+y(\alpha_1+\alpha_2)^2\,\qquad\quad \text{and}\quad\quad   y=m^2-q^2/4.
\]
In the threshold limit $y\to 0$, the integrand develops a singularity due to the vanishing of the term $(q^2/4)(\alpha_1-\alpha_2)^2$ along the line $\alpha_1=\alpha_2$. To understand the nature of this singularity it is instructive to consider the Landau equations:
\begin{equation}
\alpha_1=0\quad  \text{or}\quad   \frac{d\mathcal{F}_{\text {bub}}}{d\alpha_1}=0=(q^2/2)(\alpha_1-\alpha_2)+2y(\alpha_1+\alpha_2) \,,
\end{equation}
\begin{equation}
\alpha_2=0  \quad\text{or} \quad \frac{d\mathcal{F}_{\text {bub}}}{d\alpha_2}=0=(q^2/2)(\alpha_2-\alpha_1)+2y(\alpha_1+\alpha_2) \,.
\end{equation}
At $y=0$ the equations 
\begin{equation}
\label{derivative_condition}
\frac{d\mathcal{F}_{\text {bub}}}{d\alpha_1}=\frac{d\mathcal{F}_{\text {bub}}}{d\alpha_2}=0
\end{equation}
are satisfied on the line $\alpha_1=\alpha_2$, identifying this as a pinch singularity. The $\delta$ constraint fixes its location at $\alpha_1=\frac12 =\alpha_2$. Note that for nonzero $y$ the only solution is $\alpha_1=0=\alpha_2$ which is an endpoint singularity rather than a pinch, which is, however, not supported by the $\delta$-constraint, and is therefore discarded. 

The pinch can also be described from the perspective of an expansion in $y$ as follows. When $y$ approaches zero, the solution of the condition $\mathcal{F}_{\text {bub}}(\alpha_1,\alpha_2;q^2,y)=0$ is: 
\begin{equation}
\alpha_2=1-\alpha_1,\qquad \alpha_1 \rightarrow \frac12 \pm \frac{ i \sqrt{y}}{ \sqrt{q^2}}. 
\end{equation}
Thus, for nonzero $y$ there are two complex roots in $\alpha_1$, at which $\mathcal{F}_{\text {bub}}$ vanishes. In the limit $y\to0$ these roots merge at $\alpha_1=1/2=\alpha_2$, pinching the contour on the real axis from above and below. This is the very point where the derivative conditions~(\ref{derivative_condition}) are satisfied. 

The program \texttt{asy2} has a feature to resolve such singularities, and to find the underlying regions within the MoR approach by sector decomposing over the two sectors: 
\begin{enumerate}
    \item[1)]$\alpha_1<\alpha_2$ : $(\alpha_1,\alpha_2)\to (\alpha_1/2,\alpha_1/2+\alpha_2)$
    \item[2)]$\alpha_1>\alpha_2$ : $(\alpha_1,\alpha_2)\to (\alpha_1+\alpha_2/2,\alpha_2/2)$
\end{enumerate}
In each of these sectors the boundaries are mapped back onto the Feynman simplex ($1=\alpha_1+\alpha_2$) via the indicated mappings, and as a result the line singularity is mapped onto the boundary. For example, in sector 1 the resulting integral is
\begin{align}
{\cal I}_{\text {bub}}^{(1)}(q^2,y)&=\frac{\Gamma(\eps)}{2} \int d\alpha_1 d\alpha_2 \frac{\delta(1-\alpha_1-\alpha_2)  (\alpha_1+\alpha_2)^{-2+2\eps}}{\Big((q^2/4)(\alpha_2)^2+y(\alpha_1+\alpha_2)^2\Big)^{\eps}}\,,
\end{align}
where no further singularities occur in the domain of integration. Thus, in the new formulation all regions can be identified using the Newton polytope approach.

Another example of this kind, also discussed in ref.~\cite{JtzSmnSmn12}, is the following pentagon integral for a forward scattering kinematic configuration
\begin{align}
\label{Forward_DY_example}
\begin{split}
{\cal I}_{\text{penta}}(Q^2,m^2)&=\int \frac{d^Dk}{i\pi^{D/2}} \frac{1}{(k^2-m^2)(k^2-2p\cdot k)(k^2+2p\cdot k)(k^2-2q\cdot k)(k^2+2k\cdot q)}\\
&=-\Gamma(3+\eps) \int d\alpha_1\dots d\alpha_5\, \delta(1-\alpha_1\dots-\alpha_5)\,\times \\ &\hspace*{100pt}
\Big(\mathcal{U}_{\text{penta}} 
(\{\boldsymbol{\alpha}\})\Big)^{1+2\eps} \,
\Big(\mathcal{F}_{\text{penta}}
(\{\boldsymbol{\alpha}\}; Q^2, m^2)\Big)^{-3-\eps}
\end{split}
\end{align} 
where
\begin{subequations}
\begin{eqnarray}
\mathcal{U}_{\text{penta}} (\{\boldsymbol{\alpha}\}) &=&\alpha_1+\alpha_2+\alpha_3+\alpha_4 +\alpha_5\,,\\
\mathcal{F}_{\text{penta}}
(\{\boldsymbol{\alpha}\}; Q^2, m^2) 
&=&\alpha_1\mathcal{U}_{\text{penta}}(\{\boldsymbol{\alpha}\})  m^2+(\alpha_2-\alpha_3)(\alpha_4-\alpha_5)Q^2 \,.
\end{eqnarray}
\end{subequations}
The Landau conditions are
\begin{subequations}
\begin{eqnarray}
\label{eq:LEpenta1}
&& \alpha_1=0\quad  \text{or}\quad   \frac{d\mathcal{F}_{\text{penta}}}{d\alpha_1}=0= (\alpha_1+\mathcal{U}_{\text{penta}})m^2\,,\\   
\label{eq:LEpenta2}
&& \alpha_2=0\quad  \text{or}\quad   \frac{d\mathcal{F}_{\text{penta}}}{d\alpha_2}=0= (\alpha_1)m^2+(\alpha_4-\alpha_5)Q^2\,,\\   
\label{eq:LEpenta3}
&& \alpha_3=0\quad  \text{or}\quad   \frac{d\mathcal{F}_{\text{penta}}}{d\alpha_3}=0= (\alpha_1)m^2-(\alpha_4-\alpha_5)Q^2\,,\\   
\label{eq:LEpenta4}
&& \alpha_4=0\quad  \text{or}\quad   \frac{d\mathcal{F}_{\text{penta}}}{d\alpha_4}=0= (\alpha_1)m^2+(\alpha_2-\alpha_3)Q^2\,,\\   
\label{eq:LEpenta5}
&&  \alpha_5=0\quad  \text{or}\quad   \frac{d\mathcal{F}_{\text{penta}}}{d\alpha_5}=0= (\alpha_1)m^2-(\alpha_2-\alpha_3)Q^2\,.
\end{eqnarray}
\end{subequations}
To satisfy \eqref{eq:LEpenta1} we must require $\alpha_1=0$ since there exists no solution to $\frac{d\mathcal{F}_{\text{penta}}}{d\alpha_1}=0$. Alternatively, we can consider the limit $m^2/Q^2\to0$ in which $\frac{d\mathcal{F}_{\text{penta}}}{d\alpha_1}=0$ is automatically satisfied. In either of these cases eqs.~(\ref{eq:LEpenta2})-(\ref{eq:LEpenta5}) allow for a pinch singularity located at 
\begin{equation}
\label{eq:pentapinch}
\alpha_4-\alpha_5=0=\alpha_2-\alpha_3\,. 
\end{equation}
Note that other singularities are also possible, e.g. $\alpha_2=0=\alpha_3$ solves the Landau equations for any value of $\alpha_4$ and $\alpha_5$, since the corresponding derivatives of $\mathcal{F}_{\text{penta}}$ vanish. This singularity should, however, be interpreted as an endpoint singularity in the sense that each monomial vanishes independently and no cancellation takes place. Note also that if we had only $\alpha_2=\alpha_3$ then $\mathcal{F}_{\text{penta}}$ would still vanish, but $\alpha_4,\alpha_5$ could be used to avoid the singularity by deforming the contour into the complex plain. Thus, the only pinch singularity appears subject to the two conditions, $\alpha_2=\alpha_3$ and $\alpha_4=\alpha_5$, assuming either $\alpha_1\to0$ or $m^2\to0$. Given that the singularity is linear in the difference of two parameters, a similar resolution strategy as described above can map these singularities to the boundary, allowing the Newton polytope approach to be used sector by sector to reveal the regions. 

In the next sections we will see that more general situations of cancellation structures may occur, involving higher degree polynomials, e.g. $\alpha_i\alpha_j = \alpha_k\alpha_l$ for some $i,j,k,l$, similarly leading to pinch singularities, which obstruct both the evaluation of the integral and its expansion with available strategies.

\subsection{Searching for pinch singularities in parameter space}
\label{section-finding_diagrams_prone_to_hidden_landau_singularities}

Given a Feynman integral, the faces of the corresponding Newton polytope suffice to identify all the Landau singularities as long as there exists an analytic continuation to a same-sign regime where $\mathcal{F}$ is positive definite. When there is no same-sign regime, Landau singularities may arise from the cancellation of $\mathcal{F}$ terms with opposite signs. To understand such cancellation structures, let us define the polynomials $\mathcal{F}_+$ and $\mathcal{F}_-$ at a given kinematic point as the positive and negative monomials of $\mathcal{F}$, respectively, with $\mathcal{F}=\mathcal{F}_+ + \mathcal{F}_-$. The second Landau equation then becomes
\begin{eqnarray}
\label{eq:Landau_equation_positive_negative_F}
\alpha_e \frac{\partial\mathcal{F}_+}{\partial \alpha_e} + \alpha_e \frac{\partial\mathcal{F}_-}{\partial \alpha_e} =0,\quad\text{for } e\in\{1,\dots,N\}.
\end{eqnarray}
Since $\alpha_e$ are all non-negative, the sign of a given monomial in $\mathcal{F}$ is directly linked to that of the Mandelstam invariant it contains. Importantly, before examining $\mathcal{F}$ using the Landau equations, one must impose all constraints relating kinematic invariants based on momentum conservation. In particular, the separation of the monomials in ${\cal F}$ into $\mathcal{F}_+$ and $\mathcal{F}_-$ must be done after all such relations are taken into account, so ${\cal F}$ is written in terms of a minimal set of \emph{independent} kinematic variables. 
Having done that, one may distinguish between two types of cancellation scenarios which may arise. One scenario arises when terms with the same invariant cancel between them, independently of the kinematics. The second scenario is when the cancellation involves terms associated with more than one invariant, leading to a potential Landau singularity at a particular kinematic configuration. In this paper, we focus exclusively on the first type of cancellation, which occur for general kinematics (under certain restrictions, e.g. on-shell conditions) potentially giving rise to infrared divergences.
With this in mind, we first treat the coefficient of each independent invariant $s_{ij}\in \boldsymbol{s}\setminus \boldsymbol{t}$ as a separate ${\cal F}$ polynomial, which we denote by ${\cal F}^{(s_{ij})}$ and search for cancellations among its terms, such that (\ref{eq:Landau_equation_positive_negative_F}) admits a nontrivial solution. If such a solution exists for all independent invariants in ${\cal F} (\boldsymbol{\alpha},\boldsymbol{s})$, we investigate if a compatible solution exists. If it does, we have identified a pinch singularity.

Our objective is to identify a necessary condition for the existence of pinch singularities in parameter space within the domain of integration. This situation is characterised by a solution of the Landau equations~(\ref{eq:Landau_equation_parameter_representation}) for positive $\alpha_i$, leading to the following {\bf cancellation requirement:} \emph{there exists at least one Feynman parameter $\alpha_i \neq 0$ with $\frac{\partial\mathcal{F}}{\partial \alpha_i}$ containing non-vanishing terms of both signs that cancel on the solution.}  This requirement is satisfied in the motivating examples of eqs.~(\ref{Threshold_example}) and (\ref{Forward_DY_example}) above.

This requirement allows for the possibility that a set of parameters $\{\alpha_j\}$ strictly vanish. The pinch then corresponds to a solution of the Landau equations which is also present in a sub-topology of the Feynman graph under consideration. When some first derivatives $\frac{\partial\mathcal{F}}{\partial \alpha_{j}}$ are either positive definite or negative definite, the Landau equations imply that each monomial in $\alpha_{j} \frac{\partial\mathcal{F}}{\partial \alpha_{j}}$ must strictly vanish. The cancellation requirement is then that after all such parameters $\{\alpha_{j}\}$ are set to zero, there is still some $\alpha_i\neq 0$ such that $\frac{\partial\mathcal{F}}{\partial \alpha_i}$, which vanishes on the solution, receives contributions from both $\mathcal{F}_+$ and $\mathcal{F}_-$.

Note that this implies that the form of an $\mathcal{F}(\boldsymbol{\alpha};\s)$ polynomial that is prone to having pinch singularities is rather constrained. For example, the following expressions
\begin{eqnarray}
\label{eq:cancellation_incompatible_with_Landau_equations}
    \mathcal{F}(\boldsymbol{\alpha};\s)= (\alpha_a-\alpha_b) f(\boldsymbol{\alpha};\s), \quad\text{ or }\quad \mathcal{F}(\boldsymbol{\alpha};\s)= (\alpha_a\alpha_b - \alpha_c\alpha_d) f(\boldsymbol{\alpha};\s),
\end{eqnarray}
with $f(\boldsymbol{\alpha};\s)$ positive or negative definite, do not satisfy the cancellation requirement. To see this, let us consider the first equation of (\ref{eq:cancellation_incompatible_with_Landau_equations}) as an example. If the cancellation requirement is to be fulfilled, there must exist some dependence on $\alpha_e$ in $f(\boldsymbol{\alpha};\s)$ with $e\neq a, b$, such that $\alpha_e\frac{\partial \mathcal{F}_+}{\partial\alpha_e} =\alpha_a\alpha_e \frac{\partial f}{\partial \alpha_e}$ and $\alpha_e\frac{\partial \mathcal{F}_-}{\partial\alpha_e} =-\alpha_b\alpha_e \frac{\partial f}{\partial \alpha_e}$ are non-vanishing and cancel on the solution. Now we show that in fact these terms vanish independently. To this end, consider the Landau equations~(\ref{eq:Landau_equation_parameter_representation}) with respect to $\alpha_a$ and $\alpha_b$, which imply that each monomial in $\alpha_{a} \frac{\partial\mathcal{F}}{\partial \alpha_{a}}$ and $\alpha_{b} \frac{\partial\mathcal{F}}{\partial \alpha_{b}}$ strictly vanishes on the solution. We note, however, that all the monomials of $\alpha_e \frac{\partial \mathcal{F}}{\partial\alpha_e}$ are included in $\mathcal{F} = \alpha_{a} \frac{\partial\mathcal{F}}{\partial \alpha_{a}} + \alpha_{b} \frac{\partial\mathcal{F}}{\partial \alpha_{b}}$, all of which must vanish on the solution. Hence there is no way that the cancellation requirement could be satisfied. We conclude that having $\mathcal{F}$ terms of distinct signs is not enough. 

As pointed out in the past by several authors~\cite{Pkhn63II,Tktpls63nonplanar,Hld64,Lam69,Ldsf74}, a typical expression of $\mathcal{F}(\boldsymbol{\alpha};\s)$ for $2\to 2$ scattering that features a pinch singularity includes terms proportional to 
\begin{equation}
\label{cancel_pattern}
(\alpha_a\alpha_b - \alpha_c\alpha_d)(\alpha_e\alpha_f - \alpha_g\alpha_h).
\end{equation}
We now present an algorithm which searches for integrals potentially satisfying the cancellation requirement above. We will see that diagrams with cancellation patterns such as (\ref{cancel_pattern}) are identified by the algorithm.

\vspace*{10pt}

\noindent {\bf{Algorithm to identify potential pinch singularities in ${\cal F}^{(s_{ij})}$}}

\vspace*{10pt}
\noindent
Based on the cancellation requirement stated above, we now construct an algorithm which identifies all the graphs that potentially contain a pinch in parameter space, under the assumption that $\mathcal{F}=\mathcal{F}^{(s_{ij})}(\{\alpha_k\})$, i.e. $\mathcal{F}$ is proportional to a single Mandelstam invariant $s_{ij}$, and hence a function of the Feynman parameters alone. We then follow three simple steps: 
\begin{itemize}
    \item [] Step 1. Compute $\mathcal{F}_+$ and $\mathcal{F}_-$. If either of them vanishes, exit the algorithm outputting that there are no pinch Landau singularities. Otherwise go to Step 2.
    \item [] Step 2. Compute $\partial\mathcal{F}_+/\partial \alpha_i$ and $\partial\mathcal{F}_-/\partial \alpha_i$ for all the $\alpha_i$ that $\mathcal{F}$ depends on. If none of these derivatives vanish, exit the algorithm outputting that there are pinch singularities (which may or may not be within the integration domain). Otherwise go to Step 3.
    \item [] Step 3. Identify the $i$ for which~$\partial\mathcal{F}_+/\partial \alpha_i = 0$ or $\partial\mathcal{F}_-/\partial \alpha_i = 0$, replace $\mathcal{F}$ by $\mathcal{F}|_{\alpha_i=0}$, and return to Step 1.
\end{itemize}

\vspace*{10pt}

\noindent {\bf{Application to massless on-shell $2\to 2$ wide-angle scattering}}

\vspace*{10pt}
\noindent
Here we focus on singularities in $2\to 2$ massless wide-angle scattering. In order to apply the above algorithm to this case we must consider the interplay of the three channels $s$, $t$ and~$u$ represented in the ${\cal F}$ polynomial, which satisfy the momentum conservation and on-shell conditions, i.e. $s+t+u=0$.
Given that we have, for example, eliminated $u=-s-t$ from ${\cal F}$, then the complete ${\cal F}$ polynomial takes the form\footnote{Here and below we use the subscript $u$ to indicate that we eliminated it from ${\cal F}$, setting $u=-s-t$.}:
\begin{equation}
    {\cal F}_{u}(\boldsymbol{\alpha};s,t)\equiv {\cal F}^{(s)}(\boldsymbol{\alpha};s)+{\cal F}^{(t)}(\boldsymbol{\alpha};t) \,.
\end{equation} 
We now apply the above algorithm separately on ${\cal F}^{(s)}(\boldsymbol{\alpha};s)$ and on ${\cal F}^{(t)}(\boldsymbol{\alpha};t)$. If we cannot exclude a pinch in either of the two, we go ahead and look for a solution of the \emph{combined set} of Landau equations: 
\begin{equation}
\frac{\partial {\cal F}^{(s)}(\boldsymbol{\alpha};s)}{\partial \alpha_i} =0,\qquad 
\frac{\partial {\cal F}^{(t)}(\boldsymbol{\alpha};t)}{\partial \alpha_i} =0, \qquad \forall i
\end{equation}
If such a solution exists, we have a true candidate for a pinch singularity.

A rather obvious potential caveat could be that while the polynomial ${\cal F}_u$, obtained by eliminating $u$ in favour of $-s-t$ seems to satisfy the conditions for a pinch, upon considering the same graph using an alternative set of independent invariants, say by $t=-s-u$, the resulting polynomial 
\begin{equation}
{\cal F}_{t}(\boldsymbol{\alpha};s,u)\equiv{\cal F}^{(s)}(\boldsymbol{\alpha};s)+{\cal F}^{(u)}(\boldsymbol{\alpha};u) 
\end{equation}
has no pinch solution. Such a scenario is indeed realised for planar graphs. Of course the conclusion is that this graph has no pinch (in fact, in the case of planar graphs ${\cal F}$ has a same-sign regime). Such false candidate graphs for a pinch can be easily eliminated upon running the algorithm on all sets of independent invariants, i.e. on ${\cal F}_{t}(\boldsymbol{\alpha};s,u)$, on ${\cal F}_{u}(\boldsymbol{\alpha};s,t)$ and on ${\cal F}_{s}(\boldsymbol{\alpha};t,u)$. If a pinch singularity is excluded in any of them, then there is no such solution.

\begin{figure}[t]
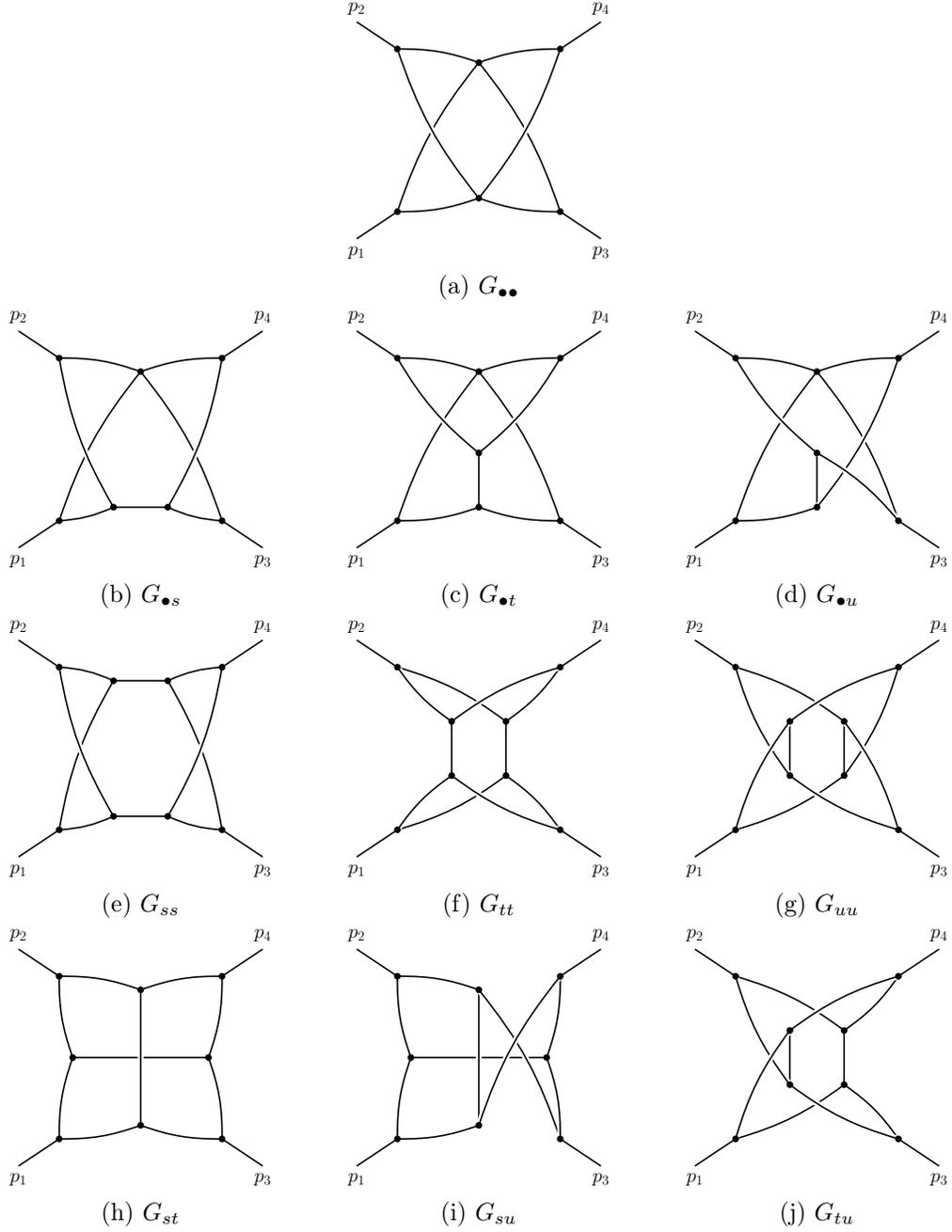

\centering
\begin{subfigure}[b]{0.25\textwidth}
\centering
\include{figs/possibly_hidden_singular_graph_dot_dot}
\vspace{-3em}\caption{$G_{\bullet \bullet}$}
\label{possibly_hidden_singular_graph_dot_dot}
\end{subfigure}
\\
\begin{subfigure}[b]{0.25\textwidth}
\centering
\include{figs/possibly_hidden_singular_graph_dot_s}
\vspace{-3em}\caption{$G_{\bullet s}$}
\label{possibly_hidden_singular_graph_dot_s}
\end{subfigure}
\qquad
\begin{subfigure}[b]{0.25\textwidth}
\centering
\include{figs/possibly_hidden_singular_graph_dot_t}
\vspace{-3em}\caption{$G_{\bullet t}$}
\label{possibly_hidden_singular_graph_dot_t}
\end{subfigure}
\qquad
\begin{subfigure}[b]{0.25\textwidth}
\centering
\include{figs/possibly_hidden_singular_graph_dot_u}
\vspace{-3em}\caption{$G_{\bullet u}$}
\label{possibly_hidden_singular_graph_dot_u}
\end{subfigure}
\\
\begin{subfigure}[b]{0.25\textwidth}
\centering
\include{figs/possibly_hidden_singular_graph_s_s}
\vspace{-3em}\caption{$G_{s s}$}
\label{possibly_hidden_singular_graph_s_s}
\end{subfigure}
\qquad
\begin{subfigure}[b]{0.25\textwidth}
\centering
\include{figs/possibly_hidden_singular_graph_t_t}
\vspace{-3em}\caption{$G_{t t}$}
\label{possibly_hidden_singular_graph_t_t}
\end{subfigure}
\qquad
\begin{subfigure}[b]{0.25\textwidth}
\centering
\include{figs/possibly_hidden_singular_graph_u_u}
\vspace{-3em}\caption{$G_{u u}$}
\label{possibly_hidden_singular_graph_u_u}
\end{subfigure}
\\
\begin{subfigure}[b]{0.25\textwidth}
\centering
\include{figs/possibly_hidden_singular_graph_s_t}
\vspace{-3em}\caption{$G_{s t}$}
\label{possibly_hidden_singular_graph_s_t}
\end{subfigure}
\qquad
\begin{subfigure}[b]{0.25\textwidth}
\centering
\include{figs/possibly_hidden_singular_graph_s_u}
\vspace{-3em}\caption{$G_{s u}$}
\label{possibly_hidden_singular_graph_s_u}
\end{subfigure}
\qquad
\begin{subfigure}[b]{0.25\textwidth}
\centering
\include{figs/possibly_hidden_singular_graph_t_u}
\vspace{-3em}\caption{$G_{t u}$}
\label{possibly_hidden_singular_graph_t_u}
\end{subfigure}
\caption{All the massless four-point three-loop graphs with a possible pinch Landau singularity in parameter space.}
\label{figure-possibly_hidden_singular_graphs}
\end{figure}

By running this algorithm through all massless $2\to 2$ graphs with three- and four-point vertices and up to four loops, we find that for \emph{both} wide-angle scattering kinematics and nearly-forward scattering kinematics, there are no one-loop or two-loop graphs exhibiting pinch Landau singularities. At the three-loop order, there are 10 graphs, shown in figure~\ref{figure-possibly_hidden_singular_graphs}, which potentially contain pinch Landau singularities. We have used the symbols $\bullet$, $s$, $t$, and~$u$ to denote different internal scattering topologies in these graphs: $\bullet$ represents scattering at a four-point vertex, while $s$, $t$, and $u$ represent scattering in the $s$, $t$, and $u$ channels, respectively.
In section~\ref{section-identifying_pinch_dissecting_polytope} we show how knowledge of the pinch singularity facilitates the numerical evaluation of the integral, by enabling us to identify a polytope dissection that maps the singularity to the boundary of the domain of integration.
In sections~\ref{section-revisit_onshell_expansion_wideangle_scattering} and~\ref{section-high_energy_expansion_forward_scattering} we discuss how the pinch singularity leads to new regions, referred to as \emph{hidden regions}, that were not visible as a facet of the Newton polytope of the integral in wide-angle scattering and forward scattering kinematics, respectively. 

\begin{figure}[t]
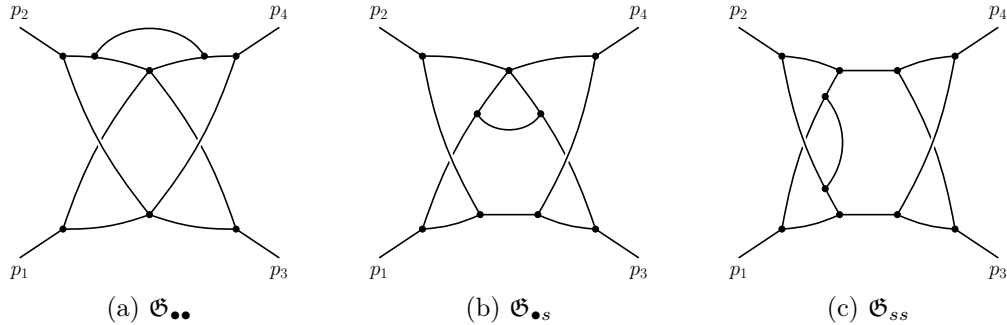

\centering
\begin{subfigure}[b]{0.25\textwidth}
\centering
\include{figs/four_loop_possibly_hidden_singular_graph1}
\vspace{-3em}\caption{$\mathfrak{G}_{\bullet \bullet}$}
\label{four_loop_possibly_hidden_singular_graph1}
\end{subfigure}
\qquad
\begin{subfigure}[b]{0.25\textwidth}
\centering
\include{figs/four_loop_possibly_hidden_singular_graph2}
\vspace{-3em}\caption{$\mathfrak{G}_{\bullet s}$}
\label{four_loop_possibly_hidden_singular_graph2}
\end{subfigure}
\qquad
\begin{subfigure}[b]{0.25\textwidth}
\centering
\include{figs/four_loop_possibly_hidden_singular_graph3}
\vspace{-3em}\caption{$\mathfrak{G}_{ss}$}
\label{four_loop_possibly_hidden_singular_graph3}
\end{subfigure}
\caption{Example four-loop graphs with possible pinch Landau singularities in parameter space, due to one appearing in a three-loop subgraph. They are labelled $\mathfrak{G}_{\bullet \bullet}$, $\mathfrak{G}_{\bullet s}$, and $\mathfrak{G}_{ss}$, because they contain $G_{\bullet\bullet}$, $G_{\bullet s}$, and $G_{ss}$ of figure~\ref{figure-possibly_hidden_singular_graphs} respectively as subgraphs.}
\label{figure-four_loop_possibly_hidden_singular_graphs}
\end{figure}

\begin{figure}[t]
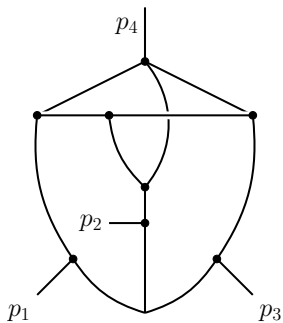

\centering
\include{figs/four_loop_possibly_hidden_singular_graph_85774}
\vspace{-3em}
\caption{A four-loop graph example $\mathfrak{G}'$ that was identified by the algorithm as possibly having a pinch Landau singularity in parameter space, but which does not contain any of the three-loop structures in figure~\ref{figure-possibly_hidden_singular_graphs} as a
subgraph.}
\label{figure-four_loop_possibly_hidden_singular_graph_85774}
\end{figure}

At four-loop order, the algorithm identifies 1097 graphs potentially containing pinch Landau singularities. A key observation is that 1081 of these graphs can be obtained by adding one extra propagator to the graphs of figure~\ref{figure-possibly_hidden_singular_graphs}. Indeed, we explicitly checked, using \texttt{Maple}'s graph isomorphism command \cite{Maple}, whether for each of these 1097 graphs we could identify a subgraph isomorphic to one of the graphs in figure~\ref{figure-possibly_hidden_singular_graphs}, after deleting any one edge. This check failed only for 16 graphs. 
Therefore, the three-loop graphs potentially containing pinch Landau singularities appear as subgraphs of 1081 four-loop graphs. Three examples from the set of 1081 four-loop integrals are shown in figure~\ref{figure-four_loop_possibly_hidden_singular_graphs}, they contain $G_{\bullet\bullet}$, $G_{\bullet s}$, and $G_{ss}$ as subgraphs, respectively. Of the remaining 16 four-loop integrals, 4 integrals contain a four-point vertex, have twelve propagators, and are related to each other by crossing; an example is $\mathfrak{G}'$ shown in figure~\ref{figure-four_loop_possibly_hidden_singular_graph_85774}. The final 12 four-loop integrals each have thirteen propagators and contain one of these 4 four-loop integrals as a subgraph.

Upon analysing the integral $\mathfrak{G}'$, shown in figure~\ref{figure-four_loop_possibly_hidden_singular_graph_85774}, we find that when considering the coefficients of the Mandelstam invariants individually there is a pinch solution of the Landau equations. This implies that a hidden region may exist in the Regge limit described in section~\ref{section-high_energy_expansion_forward_scattering}, where a single Mandelstam invariant becomes small relative to the others. However, no simultaneous solution of the Landau equations is possible when two nonzero Mandelstam invariants are present in the $\mathcal{F}$ polynomial. This implies that a pinch solution does not exist for this on-shell integral in general kinematics, and that no hidden region is present in the on-shell expansion in wide-angle scattering, described in section~\ref{section-revisit_onshell_expansion_wideangle_scattering}.

\subsection{Identifying the pinch and dissecting the polytope}
\label{section-identifying_pinch_dissecting_polytope}

Before further discussing the interpretation of the pinch Landau singularities identified above, let us first consider how they impact the evaluation of integrals in parameter space.
Outside the same-sign regime (where all ${\cal F}$ terms have the same sign) Feynman integrals can develop simple poles on the real axis of the parameters which must be avoided using a contour compatible with the causal $i\varepsilon$ Feynman prescription. Specifically, the imaginary part of the $\mathcal{F}$ polynomial must be \emph{negative} in the vicinity of the pole.
As described in section~\ref{section-Landau_equations_parameter_space}, such an integration contour can be realised by giving a small imaginary part to the integration variables.
A commonly used prescription for the deformation is given by,
\begin{align}
\label{contour_deformation}
&\alpha_k \rightarrow \widetilde{\alpha}_k(\boldsymbol{\alpha)} = \alpha_k - i \tau_k(\boldsymbol{\alpha}),&
&\tau_k(\boldsymbol{\alpha}) = \lambda_k \alpha_k (1-\alpha_k) \frac{\partial \mathcal{F}(\boldsymbol{\alpha};\s)}{\partial \alpha_k}.&
\end{align}
The factors $\alpha_k(1-\alpha_k)$ in $\tau_k$ ensure that the contour deformation vanishes at the endpoints of integration ($\alpha_k =0$ and $\alpha_k = 1$). Upon inserting the new variables into $\mathcal{F}$ and expanding in a power series in $\tau$ we obtain,
\begin{align}
\label{eq:contour_deformation}
\begin{split}
\mathcal{F}(\widetilde{\boldsymbol{\alpha}};\s) &= \mathcal{F}(\boldsymbol{\alpha}) - i \sum_j \tau_j(\boldsymbol{\alpha}) \frac{\partial \mathcal{F}(\boldsymbol{\alpha};\s)}{\partial \alpha_j} + \mathcal{O}(\tau^2) \\
& = \mathcal{F}(\boldsymbol{\alpha}) - i \sum_j \lambda_j \alpha_j (1- \alpha_j) \left(\frac{\partial \mathcal{F}(\boldsymbol{\alpha};\s)}{\partial \alpha_j}\right)^2 + \mathcal{O}(\tau^2).
\end{split}
\end{align}
The reason for choosing the contour deformation of $\alpha_k$ to be proportional to the derivative of the ${\cal F}$ polynomial with respect to $\alpha_k$ in eq.~(\ref{contour_deformation}) now becomes clear: this prescription ensures that the first correction term in the expansion of $\mathcal{F}(\widetilde{\boldsymbol{\alpha}};\s)$ does not depend on the sign of $\partial \mathcal{F}/\partial \alpha_j$.
For sufficiently small $\lambda_k \ge 0$, such that terms of order $\tau^3$ can be neglected, this deformation therefore ensures that $\mathcal{F}(\widetilde{\boldsymbol{\alpha}};\s)$ has a negative imaginary part, as required by the Feynman prescription.
This contour deformation prescription\footnote{Or an equivalent prescription for $\mathcal{V}(\boldsymbol{\alpha};\s)=\mathcal{F}(\boldsymbol{\alpha};\s)/\mathcal{U}(\boldsymbol{\alpha})$.}~\cite{Soper:1999xk,Binoth:2005ff,Nagy:2006xy,Anastasiou:2007qb,Lazopoulos:2007ix,Beerli:2008zz,Borowka:2012yc,Borowka:2014aaa} is used in programs which numerically evaluate Feynman integrals, including FIESTA~\cite{SmnTtyk09FIESTA,SmnSmnTtyk11FIESTA2,Smn14FIESTA3,Smn16FIESTA4,Smn22FIESTA5}, \texttt{pySecDec}~\cite{pySecDec17} and \texttt{feyntrop}~\cite{BrskMchTld23}.
Examining the deformation, one may expect that the prescription would fail if there exists a point inside the integration domain for which $\mathcal{F} =0$ and $\alpha_j (1- \alpha_j)\partial \mathcal{F}/\partial \alpha_j =0$ for all $j$ simultaneously, which is closely related to the Landau condition in eq.~(\ref{eq:Landau_equation_parameter_representation}). The conclusion is that contour deformation is likely to be ineffective if a pinch solution to the Landau equations appears within the domain of integration.

Starting from three-loop order, for $2 \rightarrow 2$ massless scattering, examples do exist which satisfy the criteria $\mathcal{F} =0$ and $\alpha_j \partial \mathcal{F}/\partial \alpha_j =0$ for all $j$, for which the above prescription indeed fails. 
Note that this obstacle is a fundamental property of the integral and not an artifact of our specific choice for the contour deformation prescription.

One example is the Feynman integral corresponding to the crown diagram $G_{\bullet\bullet}$, shown in figure~\ref{possibly_hidden_singular_graph_dot_dot}. 
We may express the integral as,
\begin{align}
\label{eq:igbulletbullet}
\mathcal{I}_{G_{\bullet \bullet}}(\s) =& \int \frac{d^Dk_1}{i\pi^{D/2}}\frac{d^Dk_2}{i\pi^{D/2}}\frac{d^Dk_3}{i\pi^{D/2}} \frac{1}{D_0^{n_0}\ldots D_{7}^{n_{7}}}, \nonumber\\
=& \Gamma(2+3\epsilon) \int d\alpha_0\dots d\alpha_7 \,\,
\delta(1-\alpha_0\dots -\alpha_7) \,
\Big(
\mathcal{U}(\boldsymbol{\alpha})\Big)^{4\eps}\,
\Big(\mathcal{F}(\boldsymbol{\alpha};\s)\Big)^{-2-3\eps}. 
\end{align}
The inverse propagators are $D_j=q_j^2+i\varepsilon$ ($j=0,\dots,7$), with $q_{j}$ the $j$-th element of the set
\begin{eqnarray}
     \Big\{ k_1,\ p_1-k_1,\ k_2,\ p_2-k_2,\ k_3,\ p_3-k_3,\ k_1+k_2-k_3,\ p_4-k_1-k_2+k_3 \Big\}.
\end{eqnarray}
The Symanzik polynomials $\mathcal{U}$ and $\mathcal{F}$ read (under the parameterisation in figure~\ref{figure-three_loop_wide_angle_diagram23}):
\begin{align}
\label{eq:UF_Gbb}
\begin{split}
    &\mathcal{U}(\boldsymbol{\alpha}) = (\alpha_0+\alpha_1)(\alpha_2+\alpha_3)(\alpha_4+\alpha_5) + (\alpha_0+\alpha_1)(\alpha_2+\alpha_3)(\alpha_6+\alpha_7)\\
    &\qquad\qquad + (\alpha_0+\alpha_1)(\alpha_4+\alpha_5)(\alpha_6+\alpha_7) + (\alpha_2+\alpha_3)(\alpha_4+\alpha_5)(\alpha_6+\alpha_7), \\
    &\mathcal{F}(\boldsymbol{\alpha};\s) =  (-s_{12}) (\alpha_1\alpha_4 - \alpha_0\alpha_5)(\alpha_3\alpha_6 - \alpha_2\alpha_7) + (-s_{13})(\alpha_1\alpha_2 - \alpha_0\alpha_3)(\alpha_5\alpha_6 - \alpha_4\alpha_7)\,,
\end{split}
\end{align}
where $s_{12}\equiv (p_1+p_2)^2$, $s_{13}\equiv  (p_1-p_3)^2$, and $s_{14} \equiv (p_1-p_4)^2$, and
we have used the relation $s_{12} + s_{13} + s_{23} = 0$ to eliminate $s_{23}$, which leads to monomials of different sign multiplying the remaining invariants $s_{12}$ and $s_{13}$.
Note that, as a consequence of this, there is no analytic continuation prescription of the form $s_{ij} \rightarrow s_{ij} \pm i\varepsilon$ which would reproduce the overall $i\varepsilon$ in eq.~(\ref{UFterm_general_expression}) for all values of the Feynman parameters $\alpha_i \geqslant 0$.

Possible pinch singularities in $\mathcal{F}$ of eq.~(\ref{eq:UF_Gbb}), could be associated to the vanishing of subsets of the following polynomials:
\begin{align}
\label{v_iDef}
\begin{split}
    v_1=\alpha_1\alpha_4 - \alpha_0\alpha_5\,,\qquad v_2=\alpha_3\alpha_6 - \alpha_2\alpha_7\,,\\
    v_3=\alpha_1\alpha_2 - \alpha_0\alpha_3\,,\qquad v_4=\alpha_5\alpha_6 - \alpha_4\alpha_7\,.
\end{split}
\end{align}
Computing the derivatives of the $\mathcal{F}(\boldsymbol{\alpha};\s)$ polynomial with respect to each Feynman parameter we obtain,
\begin{subequations}
\begin{align}
  \frac{\partial \mathcal{F}(\boldsymbol{\alpha};\s)}{\partial \alpha_0} &= s_{12} x_5 (\alpha_3 \alpha_6 - \alpha_2 x_7) + s_{13} \alpha_3 (\alpha_5 \alpha_6 - \alpha_4 \alpha_7), \\
  & \hspace{3cm} \vdots \nonumber \\
  \addtocounter{equation}{6}
  \frac{\partial \mathcal{F}(\boldsymbol{\alpha};\s)}{\partial \alpha_7} &= s_{12} \alpha_2 (\alpha_1\alpha_4 - \alpha_0\alpha_5) + s_{13} \alpha_4 (\alpha_1\alpha_2 - \alpha_0\alpha_3).
\end{align}
\end{subequations}
We see that the Landau equations can be satisfied with all $\alpha_i > 0$ for generic $s_{12}, s_{13}$ if and only if $v_1 = v_2 = v_3 = v_4 =0$, corresponding to,
\begin{align}
\label{PinchSolution}
& \alpha_2 = \frac{\alpha_0\alpha_3}{\alpha_1}, &
& \alpha_4 = \frac{\alpha_0\alpha_5}{\alpha_1}, &
& \alpha_6 = \frac{\alpha_0\alpha_7}{\alpha_1}. &
\end{align}
This relations define the \emph{pinch surface}.
Note that the vanishing of any subset of three $v_i$ in eq.~(\ref{v_iDef}) implies the vanishing of the fourth.

Attempting to evaluate $\mathcal{I}_{G_{\bullet \bullet}}$ by numerically integrating over the parameters and using the contour deformation given in eq.~\eqref{eq:contour_deformation} to avoid singularities within the integration domain will fail.
This is because all $\partial{\mathcal{F}(\boldsymbol{\alpha};\s)}/\partial{\alpha_i}$ can vanish precisely on the hypersurface $\mathcal{F} = 0$, where deformation is needed. We conclude that there exists no contour deformation that can render eq.~(\ref{eq:igbulletbullet}) free of singularities in the domain of integration.

One way of avoiding this issue is to split, or \emph{dissect}, the integral such that the Landau singularity present inside the integration domain -- on the pinch surface given by eq.~(\ref{PinchSolution}) -- is mapped to a boundary in parameter space. In the new variables, the singularity will then appear as an endpoint singularity rather than a pinch singularity. To obtain a suitable decomposition of the integral, we can first try to apply the \texttt{PreResolve} procedure of \texttt{asy2}, described in section~\ref{section-motivating_examples}.
However, we observe that a linear change of variables (even if iterated) will not result in a polynomial with monomials of opposite sign eliminated. The singularities present go beyond the pairwise line singularities discussed in ref.~\cite{JtzSmnSmn12} and are therefore not currently treatable by \texttt{asy2}. Indeed, running \texttt{asy2} on this example, it correctly identifies that iterated linear changes of variables are insufficient to resolve the singularity and reports that preresolution has failed.

Instead, we first find it useful to linearise the cancellation, i.e. we perform a blowup that decreases the degree of the polynomial defining the variety of $\mathcal{F}$.
Under the following change of variables,
\begin{eqnarray}
\label{eq:change_of_variables_three_loop_hidden}
    \alpha_0 = y_0\cdot \alpha_1,\qquad \alpha_2 = y_2\cdot \alpha_3,\qquad \alpha_4 = y_4\cdot \alpha_5,\qquad \alpha_6 = y_6\cdot \alpha_7,
\end{eqnarray}
the Symanzik polynomials become,
\begin{subequations}
\begin{align}
\begin{split}
    &\mathcal{U}(\boldsymbol{\alpha}) = (y_0+1)(y_2+1)(y_4+1)\alpha_1\alpha_3\alpha_5 + (y_0+1)(y_2+1)(y_6+1)\alpha_1\alpha_3\alpha_7\\
    &\qquad\qquad + (y_0+1)(y_4+1)(y_6+1)\alpha_1\alpha_5\alpha_7 + (y_2+1)(y_4+1)(y_6+1)\alpha_3\alpha_5\alpha_7,
    \end{split}
    \\
    \begin{split}
    \mathcal{F}(\x;\s) = (-s_{12}) (y_4-y_0)(y_6-y_2) \alpha_1\alpha_3\alpha_5\alpha_7 + (-s_{13})(y_2-y_0)(y_6-y_4) \alpha_1\alpha_3\alpha_5\alpha_7.
    \end{split}
    \end{align}
\end{subequations}

We see that the polynomials defining the zero of $\mathcal{F}$ occurring due to cancellation become simpler, for example, up to overall factors $(\alpha_1\alpha_4 - \alpha_0\alpha_5) \rightarrow (y_4 - y_0) $.
Next, we dissect the integral by imposing a strict hierarchy between the even-numbered parameters.
For example, assuming $y_0 \geqslant y_2 \geqslant y_4 \geqslant y_6$ changes the boundaries of integration from $[0,\infty]$ in each variable to new boundaries depending on the variables.
We can then change variables according to,
\begin{subequations}
\label{eq:division_sector_three_loop_hidden}
\begin{align}
y_0 &= z_0 + z_2 + z_4 + z_6, \\
y_2 &= z_2 + z_4 + z_6, \\
y_4 &= z_4 + z_6, \\
y_6 &= z_6,
\end{align}
\end{subequations}
so the boundaries of integration for the new variables are again then mapped to $[0,\infty]$.
Considering all possible hierarchies will split the integral into $4!=24$ new integrals, each of which will have polynomials of definite sign multiplying the invariants $s_{12}$ and $s_{13}$.
Note also that the change of variables does not introduce any sign into the $\mathcal{U}$ polynomial.
The 24 new integrals have the following $\mathcal{F}$ polynomials,
\begin{subequations}
\label{eq:F_polynomials_after_decomposition}
\begin{align}
\mathcal{F}_{1}(\boldsymbol{\alpha};\s) &= \alpha_1 \alpha_3 \alpha_5 \alpha_7 \left[ -s_{12}(\alpha_0 + \alpha_2)(\alpha_2+\alpha_4) - s_{13}(\alpha_0 \alpha_4)\right], \\
\mathcal{F}_{2}(\boldsymbol{\alpha};\s) &= \alpha_1 \alpha_3 \alpha_5 \alpha_7 \left[ -s_{12}(\alpha_2)(\alpha_0+\alpha_2+\alpha_6) + s_{13}(\alpha_0 \alpha_6)\right], \\
\mathcal{F}_{3}(\boldsymbol{\alpha};\s) &= \alpha_1 \alpha_3 \alpha_5 \alpha_7 \left[ -s_{12}(\alpha_0 \alpha_2) - s_{13}(\alpha_0 + \alpha_4) (\alpha_2 + \alpha_4)\right], \\
\mathcal{F}_{4}(\boldsymbol{\alpha};\s) &= \alpha_1 \alpha_3 \alpha_5 \alpha_7 \left[ s_{12}(\alpha_0 \alpha_6) - s_{13} (\alpha_4)(\alpha_0 + \alpha_4 + \alpha_6)\right], \\
\mathcal{F}_{5}(\boldsymbol{\alpha};\s) &= \alpha_1 \alpha_3 \alpha_5 \alpha_7 \left[ s_{12}(\alpha_6)(\alpha_0+\alpha_2+\alpha_6) + s_{13}(\alpha_0 + \alpha_6)(\alpha_2+\alpha_6)\right], \\
\mathcal{F}_{6}(\boldsymbol{\alpha};\s) &= \alpha_1 \alpha_3 \alpha_5 \alpha_7 \left[ s_{12}(\alpha_0 + \alpha_6)(\alpha_4+\alpha_6) + s_{13}(\alpha_6)(\alpha_0 + \alpha_4+ \alpha_6)\right],
\end{align}
\end{subequations}
with the remaining 18 integrals equal to the above 6 up to a relabelling of the Feynman parameters. Each of the new integrals is free of pinch singularities within the integration domain.

\subsection{Numerically evaluating the dissected integral}
\label{section-numerically_evaluating_dissected_integral}

Before dissection, the direct application of the sector decomposition algorithm, as described in section~\ref{section-newton_polytopes_sector_decomposition}, fails to resolve the singularity associated to the pinch singularity and yields integrands unsuitable for numerical evaluation.
Turning now to the dissected integrals, for given values of $s_{12}$ and $s_{13}$ some of the integrals will have a same-sign regime and can be evaluated without contour deformation.
However, for any possible choice of values of the Mandelstam invariants, not all integrals will be simultaneously in a same-sign regime. Nevertheless, these integrals can now be evaluated with the contour deformation given in eq.~(\ref{eq:contour_deformation}).
Furthermore, it is also possible to find a prescription of the form $s_{ij} \rightarrow s_{ij} \pm i\varepsilon$ for each integral which allows them to be analytically continued.

Although the above procedure avoids the pinch singularity, the direct numerical evaluation of this three-loop 4-point integral is still challenging. Each of the sectors has a $1/\epsilon^5$ pole that cancels in the sum of sectors. Choosing the phase-space point $s_{12} = 1, s_{13}=-1/5$ we obtain,
\begin{subequations}
\label{eq:dissected_numerical_results}
\begin{align}
\mathcal{I}_1/C_\epsilon = &\,
\epsilon^{-5} \left[0.55555538\mathbf{27}\right]
&+&\,\epsilon^{-4} \left[-3.884290\mathbf{14} + 5.235983\mathbf{13}\, i\right] 
&+&\, \mathcal{O}(\epsilon^{-3}), \\
\mathcal{I}_2/C_\epsilon = &\,
\epsilon^{-5} \left[2.222232\mathbf{11}\right]
&+&\,\epsilon^{-4} \left[-7.92923\mathbf{11} + 20.94388\mathbf{18}\, i\right] 
&+&\, \mathcal{O}(\epsilon^{-3}), \\
\mathcal{I}_3/C_\epsilon = &\,
\epsilon^{-5} \left[-2.7777888\mathbf{83}\right]
&+&\,\epsilon^{-4} \left[18.519682\mathbf{69} - 15.708041\mathbf{67}\, i\right] 
&+&\, \mathcal{O}(\epsilon^{-3}), \\
\mathcal{I}_4/C_\epsilon = &\,
\epsilon^{-5} \left[2.2222211\mathbf{19}\right]
&+&\,\epsilon^{-4} \left[-13.294002\mathbf{23}\right] 
&+&\, \mathcal{O}(\epsilon^{-3}), \\
\mathcal{I}_5/C_\epsilon = &\,
\epsilon^{-5} \left[-2.7777713\mathbf{46}\right]
&+&\,\epsilon^{-4} \left[12.74345\mathbf{17} - 23.56186\mathbf{15}\, i\right] 
&+&\, \mathcal{O}(\epsilon^{-3}), \\
\mathcal{I}_6/C_\epsilon = &\,
\epsilon^{-5} \left[0.55555546\mathbf{19}\right]
&+&\,\epsilon^{-4} \left[-4.0702347\mathbf{61}\right] 
&+&\, \mathcal{O}(\epsilon^{-3}),
\end{align}
\end{subequations}
where $C_\epsilon = \Gamma(2+3\epsilon)$ is the integral prefactor given in eq.~\eqref{eq:igbulletbullet}, and the bold digits are used to report the error on two stated digits preceding them.
The precision stated above is obtained using contour deformation as implemented in pySecDec to evaluate the dissected integrals.
Alternatively, it is possible to dramatically improve the numerical precision using the techniques described in ref.~\cite{TomLoopsLegsProc}. Summing over the sectors using the higher-precision evaluation\footnote{Directly summing over the results in eq.~\eqref{eq:dissected_numerical_results} we would obtain the value $\mathcal{I}_{G_{\bullet \bullet}}^\mathrm{numeric} = \epsilon^{-4}\left[8.34150\mathbf{87}-52.3601\mathbf{10}\, i\right] + \mathcal{O}(\epsilon^{-3})$.}, the full numeric result is given by,
\begin{align}
\mathcal{I}_{G_{\bullet \bullet}}^\mathrm{numeric} &= 4\ (\I_1 + \I_2 + \I_3 + \I_4 + \I_5 + \I_6) \nonumber \\
&= \epsilon^{-4} \left[8.3400403920\mathbf{28} -  52.35987755983\mathbf{47} I \right] + \mathcal{O}(\epsilon^{-3}),
\end{align}
this agrees, within the numerical integration error, with the analytic result obtained in ref.~\cite{Bargiela:2021wuy}, with $x=-s_{13}/s_{12}$,
\begin{align}
\mathcal{I}_{G_{\bullet \bullet}}^\mathrm{analytic}(x)  
&= -\frac{8}{3} \left[\frac{ i\pi}{x(1-x)} + \frac{ \log x}{ 1-x} + \frac{ \log(1-x)}{x}\right]\frac{1}{\epsilon^4} + \mathcal{O}(\epsilon^{-3})\nonumber \\
&= \epsilon^{-4} \left[8.3400403922 - 52.3598775598 I \right] + \mathcal{O}(\epsilon^{-3}).
\end{align}
Dissecting the original Newton polytope of $G_{\bullet \bullet}$, as described in section~\ref{section-identifying_pinch_dissecting_polytope}, has mapped the pinch singularity, which was originally within the domain of integration, to the boundary of integration.
With the singularity now at the boundary, we have established that it is possible to resolve it using the method of sector decomposition, enabling the numerical evaluation of~$G_{\bullet \bullet}$.
In the following sections, we will investigate how the presence of the pinch singularity for the on-shell integral impacts the regions relevant for the MoR when the integral is expanded about a limit where the pinch is present.

\section{On-shell expansion for wide-angle scattering}
\label{section-revisit_onshell_expansion_wideangle_scattering}

In this section we study the implication of pinch Landau singularities in parameters space from the perspective of the MoR, focusing on the on-shell expansion of wide-angle scattering, the expansion discussed in refs.~\cite{GrdHzgJnsMaSchlk22,Ma23}. Generally, one starts with a graph with massless propagators contributing to an off-shell Green's function with $M$ external legs, out of which $K_0$ are strictly on-shell, while $K_\lambda$ are expanded about the on-shell limit,
\begin{eqnarray}
\label{eq:wideangle_onshell_kinematics}
\left\lbrace 
\begin{array}{ll}
p_i^2=0 
& \qquad i=1,\dots,K_0
\\
p_i^2\sim \lambda Q^2 &\qquad i=K_0+1,\dots,K,
\\
q_i^2\sim Q^2 
&\qquad i=K+1,\dots,M
\\
p_{i_1}\cdot p_{i_2}\sim Q^2\ \ 
&\qquad \forall \,\, i_1\neq i_2\,,
\end{array}
\right.
\end{eqnarray}
where the total number of massless jets, once the limit $\lambda\to 0$ is taken, is given by $K\equiv K_{0}+K_{\lambda}$.
Note that the wide-angle condition is incorporated in $p_{i_1}\cdot p_{i_2}\sim Q^2$, implying that while the jets $p_i$ are nearly lightlike, the angle between their three-momenta $\boldsymbol{p}_{i_1}$ and $\boldsymbol{p}_{i_2}$ is $\mathcal{O}(1)$ for any $i_{1}$ and~$i_{2}$. 

\begin{figure}[t]
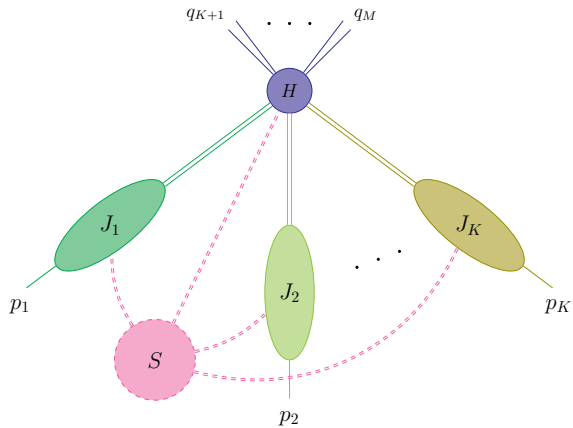

\centering
\include{figs/wide_angle_scattering_generic_region}
\caption{The general configuration of facet regions in the on-shell expansion of wide-angle scattering, where the graph $G$ is the union of the hard subgraph $H$, the jet subgraphs $J_1,\dots, J_K$, and the soft subgraph $S$. All the off-shell external momenta $q_{K+1},\dots,q_{M}$ attach to $H$, and each on-shell external momentum $p_i$ attaches to a corresponding $J_i$. Note that for facet regions, each of the subgraphs $H,J_1,\dots,J_K$ must be connected, while each connected component of $S$ must be connected to two or more jets.}
\label{figure-wide_angle_scattering_generic_region}
\end{figure}

Given any graph with the kinematics of~(\ref{eq:wideangle_onshell_kinematics}), the complete list of \emph{facet} regions can be described by the hard-collinear-soft picture in figure~\ref{figure-wide_angle_scattering_generic_region}, as proposed in ref.~\cite{GrdHzgJnsMaSchlk22} and proved in ref.~\cite{Ma23}. According to this picture, any region in the on-shell expansion of wide-angle scattering graphs consists of three types of modes:
\begin{align}
\label{WideAngleMods}
\begin{cases}
\text{hard:} & k_H^\mu\sim Q(1,1,1);\\
\text{collinear to }{p_i}: & k_{C_i}^\mu\sim Q(1,\lambda,\lambda^{1/2});\\
\text{soft:} & k_S^\mu\sim Q(\lambda,\lambda,\lambda).
\end{cases}
\end{align}
In the scaling above, we have used the $p_i$ lightcone coordinate, for $i=1,\dots,K$. Namely, $k^\mu = (k\cdot \overline{\beta}_i,\ k\cdot \beta_i,\ k\cdot \beta_{i\perp})$, where $\beta_i^\mu$ is a null vector in the direction of $p_i^\mu$, defined as $\beta_i^\mu=\frac{1}{\sqrt{2}}(1,\boldsymbol{p}_i/|\boldsymbol{p}_i|)$. For each $\beta_i^\mu$, we also define $\overline{\beta}_i^\mu \equiv \frac{1}{\sqrt{2}} (1, -\boldsymbol{p}_i/|\boldsymbol{p}_i|)$, so that $\beta_i \cdot \overline{\beta}_i =1$.
The hard ($H$), the $i$-th jet ($J_i$), and the soft ($S$) subgraphs in figure \ref{figure-wide_angle_scattering_generic_region} consist respectively of propagators carrying momenta $k_H^\mu$, $k_{C_i}^\mu$, and $k_S^\mu$, which scale according to (\ref{WideAngleMods}). All the off-shell external momenta $q_i$ attach to $H$, and each on-shell external momentum $p_i$ ($i\in\{1,\dots,K\}$) attaches to the corresponding $J_i$. Note that $H$, $J_1$, ..., $J_{K}$ are all connected, while $S$ can consist of multiple components, each of which is connected to two or more jets.\footnote{Further restrictions on the soft subgraph arise in cases where some external momenta are strictly on shell. See ref.~\cite{Ma23} for more detail.} Let us emphasise that the connectedness of the hard subgraph $H$ is a key property of facet regions. As we will see below, this property no longer holds for the hidden regions, i.e. regions that are associated with a pinch singularity in the limit $\lambda\to 0$, which therefore do not correspond to facets of the Newton polytope.

To study this expansion we focus in this section exclusively on the simplest case of relevance, namely $2\to 2$ massless scattering. In sections~\ref{section-facet_regions_hidden_regions} to~\ref{section-power_counting_momentum_space} we consider the general on-shell expansion, where $p_i^2\sim \lambda Q^2$ for all $i=1\ldots 4$, with~$p_{i_1}\cdot p_{i_2}\sim Q^2$. In sections~\ref{section-obtain_complete_set_regions_polytope_dissection} and \ref{section-numerical_evidence_discussion}, we further specialise to the case where only starts off shell, $p_1^2\sim \lambda Q^2$,  with the remaining three jets $i=2,3,4$ are strictly on shell ($p_2^2=p_3^2=p_4^2=0$) at the outset. In either of these scenarios the limit about which we are expanding, where all external legs are on shell, exposes the pinch singularity we studied in section~\ref{section-identifying_pinch_dissecting_polytope}. As we shall see, this is sufficient to introduce a hidden region into these asymptotic expansions.

\subsection{Facet regions and hidden regions in \texorpdfstring{$2\to 2$}{TEXT} scattering}
\label{section-facet_regions_hidden_regions}

As can be checked by the algorithm proposed in section~\ref{section-finding_diagrams_prone_to_hidden_landau_singularities}, there are no pinch Landau singularities in parameter space in on-shell $2\to 2$ massless scattering for any one- or two-loop graphs.
Based on the conjectured correspondence between Landau singularities and regions~\cite{GrdHzgJnsMaSchlk22} we are therefore led to exclude the possibility of hidden regions in such graphs, and trust the asymptotic expansion obtained based on facet regions alone.

The situation is rather different at three loops, where, based on the Landau equations we concluded in the previous section that certain graphs (figure~\ref{figure-possibly_hidden_singular_graphs}) may feature pinch singularities. In section~\ref{section-identifying_pinch_dissecting_polytope} we explicitly identified this pinch singularity in the simplest graph of this type, the three-loop 8-propagator crown graph $G_{\bullet \bullet}$.
Then, by dissecting the polytope at the singularity we were able to evaluate the on-shell integral using sector decomposition.
\begin{figure}[t]
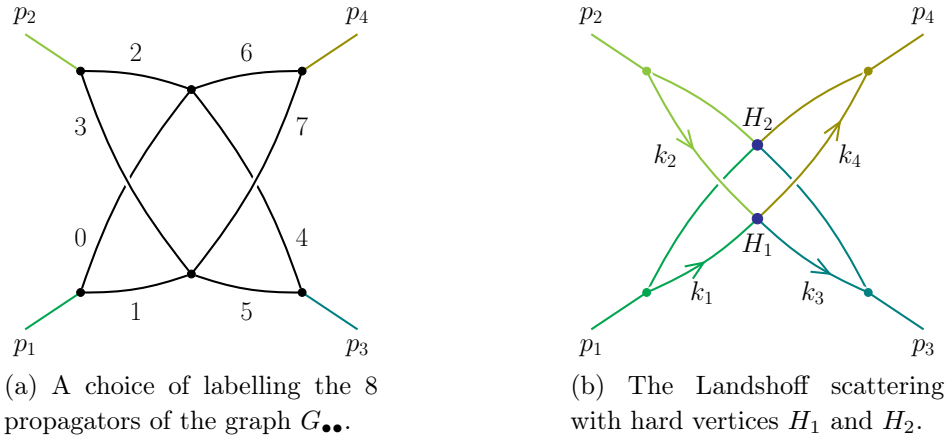

\centering
\hspace{-1em}
\begin{subfigure}[b]{0.32\textwidth}
\include{figs/three_loop_wide_angle_diagram23}
\vspace{-3em}
\caption{A choice of labelling the 8 propagators of the graph $G_{\bullet\bullet}$.}
\label{figure-three_loop_wide_angle_diagram23}
\end{subfigure}
\hspace{6em}
\begin{subfigure}[b]{0.32\textwidth}
\include{figs/three_loop_wide_angle_Landshoff_scattering}
\vspace{-3em}
\caption{The Landshoff scattering with hard vertices $H_1$ and~$H_2$.}
\label{figure-three_loop_wide_angle_Landshoff_scattering}
\end{subfigure}
\caption{The graph $G_{\bullet\bullet}$ with the propagators labelled $e_0,e_1,\dots,e_7$, and its associated (Landshoff scattering) loop-momentum configuration where the hard subgraph has two distinct components, $H_1$ and $H_2$. Note that the collinear line momenta attached to $H_1$, which are labelled as $k_1,\dots,k_4$, depend on each other via momentum conservation, because this graph has three loops in total.}
\label{figure-diagram23_and_Landshoff_scattering}
\end{figure}

Considering now the on-shell \emph{expansion} of $G_{\bullet \bullet}$ where one or more of the external lines is initially off shell, $p_i^2={\cal O}(\lambda Q^2)$, and expanding in powers of $\lambda$, we expect a hidden region to emerge, which is induced by the aforementioned pinch singularity. Let us then consider this putative hidden region in some detail. 

By labelling the propagators of $G_{\bullet \bullet}$ as in figure~\ref{figure-three_loop_wide_angle_diagram23}, the corresponding Symanzik polynomials are:
\begin{subequations}
\label{eq:8_propagator_wideangle_Symanzik_polynomials}
\begin{align}
\begin{split}
\label{eq:8_propagator_wideangle_Symanzik_polynomialsU}
    &\mathcal{U}(\x) = (x_0+x_1)(x_2+x_3)(x_4+x_5) + (x_0+x_1)(x_2+x_3)(x_6+x_7)\\
    &\qquad\qquad + (x_0+x_1)(x_4+x_5)(x_6+x_7) + (x_2+x_3)(x_4+x_5)(x_6+x_7),
    \end{split}
    \\
    \begin{split}
     \label{eq:8_propagator_wideangle_Symanzik_polynomialsF}
    &\mathcal{F}(\x;\s) = (-p_1^2) \left\{ x_0x_1\big[ (x_2+x_3)(x_4+x_5) + (x_2+x_3)(x_6+x_7) + (x_4+x_5)(x_6+x_7) \big] \right.\\
    &\phantom{\mathcal{F}(\x;\s) = (-p_1^2) \{}\left. + (x_1x_2x_4 + x_0x_3x_5) (x_6 + x_7) \right\}\\
    &\qquad\qquad + (-p_2^2) \left\{ x_2x_3 \big[ (x_0+x_1)(x_4+x_5) + (x_4+x_5)(x_6+x_7) + (x_0+x_1)(x_6+x_7) \big] \right.\\
    &\phantom{\mathcal{F}(\x;\s) = (-p_1^2) \{}\left. + (x_1x_4x_7 + x_0x_5x_6) (x_2 + x_3) \right\}\\
    &\qquad\qquad + (-p_3^2) \left\{ x_4x_5 \big[ (x_0+x_1)(x_2+x_3) + (x_2+x_3)(x_6+x_7) + (x_0+x_1)(x_6+x_7) \big] \right.\\
    &\phantom{\mathcal{F}(\x;\s) = (-p_1^2) \{}\left. + (x_0x_3x_6 + x_1x_2x_7) (x_4 + x_5) \right\}\\
    &\qquad\qquad + (-p_4^2) \left\{ x_6x_7 \big[ (x_0+x_1)(x_4+x_5) + (x_4+x_5)(x_2+x_3) + (x_0+x_1)(x_2+x_3) \big] \right.\\
    &\phantom{\mathcal{F}(\x;\s) = (-p_1^2) \{}\left. + (x_2x_4x_7 + x_3x_5x_6) (x_0 + x_1) \right\}\\
    &\qquad\qquad + (-s_{12}) (x_0x_5 - x_1x_4)(x_2x_7 - x_3x_6) + (-s_{13})(x_0x_3 - x_1x_2)(x_4x_7 - x_5x_6),
     \end{split}
     \end{align}
    \end{subequations}
where we have used the relation between the Mandelstam variables to eliminate $s_{14}$ from $\mathcal{F}(\x;\s)$ in favour of $-s_{12}-s_{13} + \sum_{i=1}^4 p_i^2$.
This readily introduces explicit minus signs into the last line of eq.~(\ref{eq:8_propagator_wideangle_Symanzik_polynomialsF}), leading to potential cancellation as discussed in section~\ref{section-finding_diagrams_prone_to_hidden_landau_singularities}. The facet regions of $G_{\bullet\bullet}$, as one can check, are all compatible with the configuration of hard, jet, and soft modes shown in figure~\ref{figure-wide_angle_scattering_generic_region}. More information on these regions will be provided in section~\ref{section-obtain_complete_set_regions_polytope_dissection} below (see table~\ref{tab:regions_onshell_original} there).

Our hypothesis is that in addition to the facet regions, $G_{\bullet \bullet}$ features a hidden region whose properties we now explore.
Based on our analysis in Feynman parameter space, the pinch occurs at (\ref{PinchSolution}), \emph{within} the domain of integration, rather than on its boundary. Specifically, using the $\delta(1-\sum_{k=0}^7\alpha_k)$ constraint and the symmetry of the graph, it is easy to see that the solution is at $\alpha_k=\frac18$ for all $k$. Converting this to Lee-Pomeransky parameters according to (\ref{eq:Feynman_LP_relation}) we can deduce that the only consistent scaling vectors are ones where all parameters $x_k$ (for $k=0$ to $7$) scale in the same way. In momentum space this corresponds (via eq.~(\ref{x_scaling_rule})) to all propagators having the same virtuality scaling~\cite{GrdHzgJnsMaSchlk22}. 
Given the nearly-lightlike external  kinematics, it is natural to expect that all propagators would be jet like, that is, every line momentum is collinear to one external momentum, with virtuality $\mathcal{O}(\lambda Q^2)$, such that in figure~\ref{figure-diagram23_and_Landshoff_scattering} propagators 0 and 1 are both collinear to $p_1$, propagators 2 and 3 are collinear to $p_2$, 4~and 5 are collinear to $p_3$, and 6 and 7 are collinear to $p_4$. Given the collinear scaling law of all propagators, along with the usual correspondence~(\ref{x_scaling_rule}), we deduce that each Lee-Pomeransky parameter $x_k$ must be $\mathcal{O}(\lambda^{-1})$, yielding the corresponding region vector:
\begin{eqnarray}
\label{eq:Gdotdot_Landshoff_scattering_region_vector}
    \v_\text{Landshoff} = (-1,-1,-1,-1,-1,-1,-1,-1;1)\,,
\end{eqnarray}
where we named the hidden region vector $\v_\text{Landshoff}$ in order to capture its physical interpretation as one corresponding to Landshoff scattering, where there are multiple -- in this case two -- separate hard connected components $H_1$ and $H_2$, as shown in figure~\ref{figure-three_loop_wide_angle_Landshoff_scattering}. 

We stress that the momentum-space interpretation of the hidden region as corresponding to a scattering process with two \emph{disjoint} hard subgraphs clarifies why it cannot be associated with a facet of the Newton polytope: all facet region in the on-shell expansion in wide-angle scattering involve a single connected hard subgraph~\cite{GrdHzgJnsMaSchlk22,Ma23}, consistent with figure~\ref{figure-wide_angle_scattering_generic_region}. Further analysis of the momentum configuration in the Landshoff scattering region will be presented in section~\ref{section-power_counting_momentum_space}.

Given the scaling rule (\ref{eq:Gdotdot_Landshoff_scattering_region_vector}) of the parameters, each individual term in the last line of eq.~(\ref{eq:8_propagator_wideangle_Symanzik_polynomialsF}), namely 
\begin{eqnarray}
\label{eq:super_leading_three_loop_Landshoff}
    (-s_{12}) (x_0x_5 - x_1x_4)(x_2x_7 - x_3x_6) + (-s_{13})(x_0x_3 - x_1x_2)(x_4x_7 - x_5x_6),
\end{eqnarray}
is $\mathcal{O}(\lambda^{-4})$, while all the other terms  in~$\mathcal{P}$ (including all of the terms in ${\cal U}$) are $\mathcal{O}(\lambda^{-3})$, and hence can na\"ively be discarded. 
Below we shall call the individual terms in eq.~(\ref{eq:super_leading_three_loop_Landshoff}) \emph{superleading terms}. If we simply use them to substitute the Lee-Pomeransky polynomial $\mathcal{P}(\x;\s)$, the integral would be scaleless. As a result, a na\"ive conclusion would be that the vector $\v_\text{Landshoff}$ does not correspond to any region. 
This conclusion is however false, as it ignores the possibility of mutual cancellation between these terms, which would imply that their individual $\mathcal{O}(\lambda^{-4})$ scaling does not represent the actual scaling of the combinations in eq.~(\ref{eq:super_leading_three_loop_Landshoff}). Indeed, if these terms conspire to cancel, i.e. they admit\footnote{Note that the fourth condition follows from the first three.} 
\begin{equation}
\label{cancellation def}
    \frac{x_0x_5}{x_1x_4} -1\ \sim\ \frac{x_2x_7}{x_3x_6} - 1\ \sim\ \frac{x_0x_3}{x_1x_2} - 1\ \sim\  \frac{x_4x_7}{x_5x_6} - 1\ \sim\ {\cal O}(\sqrt{\lambda})
\end{equation}
then the two combinations in eq.~(\ref{eq:super_leading_three_loop_Landshoff}) behave as 
\begin{eqnarray}
\label{eq:wideangle_superleading_terms_cancellation_scaling}
    (x_0x_5 - x_1x_4)(x_2x_7 - x_3x_6)\sim (x_0x_3 - x_1x_2)(x_4x_7 - x_5x_6)\sim \lambda^{-3},
\end{eqnarray}
in which case the last line of eq.~(\ref{eq:8_propagator_wideangle_Symanzik_polynomialsF}) would be of the \emph{same} order of magnitude as the remaining terms in  $\mathcal{P}$. This would lead to a scaleful integral, and we hence discovered a \emph{hidden region} with the scaling vector of (\ref{eq:Gdotdot_Landshoff_scattering_region_vector}) in which all the terms in the ${\cal P}$ polynomial survive!

To verify this and exclude the existence of additional hidden regions, which might not correspond to the scaling of eq.~(\ref{eq:Gdotdot_Landshoff_scattering_region_vector}), in section~\ref{section-obtain_complete_set_regions_polytope_dissection} we will dissect the Newton polytope according to section~\ref{section-identifying_pinch_dissecting_polytope} and examine the lower facets of the new polytope in each sector. Before doing so, in sections~\ref{section-power_counting_parameter_space} and~\ref{section-power_counting_momentum_space} we shall perform two independent power-counting analyses in parameter space and in momentum space respectively, demonstrating that the leading asymptotic behaviour of the graph $G_{\bullet\bullet}$ is exclusively from this hidden region, namely,
\begin{equation}
\label{Gbb_anticipating_scaling}
\I(s,t,\lambda p_i^2) \sim \lambda^{\mu[G_{\bullet\bullet},\text{Landshoff}]},\qquad \text{with}\qquad \mu[G_{\bullet\bullet},\text{Landshoff}] =-\frac12 -3\epsilon,
\end{equation}
and that this hidden region therefore dominates the asymptotic expansion of this integral.

\subsection{Power counting in parameter space}
\label{section-power_counting_parameter_space}

Let us first carry out the power counting analysis in Lee-Pomeransky representation. To begin with, according to eq.~(\ref{eq:lee_pomeransky_definition}), the integral takes the form:
\begin{equation}
\label{eq:lee_pomeransky_definition_8propdiag}
\I_{G_{\bullet\bullet}} (s,t,\lambda p_i^2) = \frac{\Gamma(D/2)}{\Gamma\left(2D-8\right)}  \int_0^\infty \,\prod_{e=0}^{7} \mathrm{d} x_e \,\cdot \Big( \mathcal{P}(\x; s,t,\lambda p_i^2) \Big)^{-D/2}.
\end{equation}
where, in the Landshoff region, according to eqs.~(\ref{eq:super_leading_three_loop_Landshoff}) and (\ref{eq:wideangle_superleading_terms_cancellation_scaling}), the integrand scales as
\begin{eqnarray}
\left(\mathcal{P}^{(\text{Landshoff} )}(\x; s,t,\lambda p_i^2)\right)^{D/2} \sim \lambda^{(-3)\cdot (-D/2)} =\lambda^{ 6-3\epsilon}.
\label{eq:Landshoff_scattering_parameter_integrand_scaling}
\end{eqnarray}

To obtain the contribution to the degree of divergence from the integration measure, we shall change the variables $\{x_0$,$x_2$,$x_4$,$x_6\}$ into $\{y_0$,$y_2$,$y_4$,$y_6\}$ respectively, such that
\begin{eqnarray}
    x_0 = y_0\cdot x_1,\qquad x_2 = y_2\cdot x_3,\qquad x_4 = y_4\cdot x_5,\qquad x_6 = y_6\cdot x_7.
\label{eq:Landshoff_scattering_change_variables}
\end{eqnarray}
It then follows that $y_i$ is $\mathcal{O}(1)$ for every $i=0,2,4,6$, and from eq.~(\ref{eq:wideangle_superleading_terms_cancellation_scaling}), they further satisfy
\begin{subequations}
\begin{eqnarray}
\label{eq:y_difference_product_scaling_a}
  &  (y_4-y_0)(y_6-y_2)x_1x_3x_5x_7\sim\lambda^{-3},
    \\&
\label{eq:y_difference_product_scaling_b}
    (y_2-y_0)(y_6-y_4)x_1x_3x_5x_7\sim \lambda^{-3}.
\end{eqnarray}
\end{subequations}
As $x_1x_3x_5x_7\sim \lambda^{-4}$, it is clear that $(y_2-y_0)(y_6-y_4)\sim \lambda$ and $(y_4-y_0)(y_6-y_2)\sim \lambda$. Furthermore, it can be deduced that for any $i,j\in\{0,2,4,6\}$, the dominant contribution arises from $y_i-y_j\sim \lambda^{1/2}$. To see this, let us set 
\[
y_2-y_0\sim \lambda^{1/2+a}
\qquad \text{and}\qquad 
y_6-y_4\sim \lambda^{1/2-a}
\]
for some $a\in \mathbb{R}$, consistently with eq.~(\ref{eq:y_difference_product_scaling_b}). If $a>0$, then $\lambda^{1/2-a}\gg \lambda^{1/2+a}$, and either $y_4$ or $y_6$ can be varied freely in a range of $\mathcal{O}(\lambda^{1/2-a})$. So at a fixed $y_0$, either of the following two scenarios holds:
\begin{align*}
&y_4-y_0\sim \lambda^{1/2-a} \qquad \qquad \text{or} \\
& y_6-y_2 = \underbrace{(y_2-y_0)}_{{\cal O}({\lambda^{1/2+a}})} - \underbrace{(y_6-y_0) }_{{\cal O}({\lambda^{1/2-a}})}\sim 
\lambda^{1/2-a}\,.
\end{align*}
However, in each of these cases one can use eq.~(\ref{eq:y_difference_product_scaling_a}) 
to deduce the scaling of the remaining differences, namely
\begin{align*}
&\text{if}\quad y_4-y_0\sim \lambda^{1/2-a} 
&\text{then}\qquad\qquad
&y_6-y_2 \sim\lambda^{1/2+a}
\\
&
&
\text{and\,then}\quad
&y_6-y_0 = (y_6-y_2) + (y_2-y_0) \sim\lambda^{1/2+a}\,,
\\
&\text{if}\quad 
y_6-y_2 \sim\lambda^{1/2-a}
& \text{then}\qquad\qquad
& y_4-y_0 \sim 
\lambda^{1/2+a}\,.
\end{align*}
In either case it follows that the integration measure for all the $y$ variables can then be rewritten as:
\begin{align}
\begin{split}
    dy_0dy_2dy_4dy_6 &= dy_0 \,d(y_2-y_0) \, d(y_4-y_0) \, d(y_6-y_0)\nonumber\\
    &\sim \lambda^{1/2+a} \cdot \lambda^{1/2-a} \cdot \lambda^{1/2+a} \\
    & = \lambda^{3/2+a}
    \qquad \qquad \text{for}\,\, a>0.
    \end{split}
\end{align}
Using the same reasoning, one obtains $dy_0dy_2dy_4dy_6\sim \lambda^{3/2-a}$ for $a<0$. Therefore, the measure $dy_0dy_2dy_4dy_6$ attains its maximum $\lambda^{3/2}$ at $a=0$, namely, $y_i-y_j\sim \lambda^{1/2}$ for any $i,j\in\{0,2,4,6\}$. 
Therefore, the leading behaviour of the integral subject to the constraints (\ref{eq:y_difference_product_scaling_a}) and (\ref{eq:y_difference_product_scaling_b}) is governed by $a=0$. This implies that there is a hidden region with the scaling vector (\ref{eq:Gdotdot_Landshoff_scattering_region_vector}) in the original variables, with $a=0$, i.e. with all $y_i-y_j\sim \lambda^{1/2}$. Note, however, that this analysis does not exclude the existence of other, power-suppressed, scaleful hidden regions, which share the same scaling in the original variables, but with $a \neq 0$.
In fact, in section~\ref{section-numerically_evaluating_dissected_integral} we shall explicitly encounter such regions.
For the $a=0$ case, which dominates the asymptotic expansion, the whole integration measure is then:
\begin{eqnarray}
    \prod_{i=0}^7 dx_i = \left(\prod_{i=1,3,5,7}x_idx_i\right) \cdot dy_0dy_2dy_4dy_6 \sim \lambda^{-8}\cdot\lambda^{3/2} = \lambda^{-13/2}.
\label{eq:Landshoff_scattering_parameter_integration_measure_scaling}
\end{eqnarray}
The degree of divergence for this hidden region, according to 
eqs.~(\ref{eq:lee_pomeransky_definition_8propdiag}), (\ref{eq:Landshoff_scattering_parameter_integrand_scaling}) and (\ref{eq:Landshoff_scattering_parameter_integration_measure_scaling}), is then
\begin{eqnarray}
\label{eq:power_counting_Landshoff_parameter_space}
    \mu[G_{\bullet\bullet},\text{Landshoff}] = -\frac{13}{2} + (-3)\cdot\left(-\frac{D}{2}\right) = -\frac{1}{2}-3\epsilon\,,
\end{eqnarray}
as anticipated in eq.~(\ref{Gbb_anticipating_scaling}) above. 

\subsection{Power counting in momentum space}
\label{section-power_counting_momentum_space}

To check the result in eq.~(\ref{eq:power_counting_Landshoff_parameter_space}) and gain further physical insight into its origin, we shall analyse the asymptotic behaviour of this region directly in momentum space, following the work of Botts and Sterman~\cite{BottsStm89}. We consider the center-of-mass frame for the $1+2\to3+4$ process, with 
\begin{equation}
\label{pConserv}
    p_1+p_2=p_3+p_4\,.
\end{equation}
We denote the scattering angle as $\theta$, and define $v_i^\mu$ to be the unit lightlike vector representing the direction of the $p_i^\mu$ jet. Namely,
\begin{subequations}
\label{eq:scattering_kinematics_parameterisation}
\begin{align}
&p_i^2 = \lambda Q^2,\qquad p_i\cdot v_i\sim \lambda Q, \qquad p_i\cdot \overline{v}_i\sim Q,\qquad p_i\cdot v_{i\perp} \sim \sqrt{\lambda} Q\,; \\
&v_i^2 = 0,\qquad v_1\cdot v_2 = v_3\cdot v_4 = 1,\,\qquad v_1\cdot v_3 = v_2\cdot v_4 = \frac{1}{2}(1-\textup{cos}\ \theta)\,,
\end{align}
\end{subequations}
where for later convenience, for each $i\in \{1,\dots,4\}$, we have introduced $\overline{v}_i^\mu$ as the unit lightlike vector that is back-to-back with $v_i^\mu$. Furthermore, we define $u_i^\mu$ to be the unit vector in the scattering plane which is normal to both $v_i^\mu$ and $\overline{v}_i^\mu$, and $n_i^\mu$ to be unit vectors in the remaining $D-3$ dimensional space normal to the scattering plane. It then follows that
\begin{subequations}
\begin{align}
    & \overline{v}_1^\mu = v_2^\mu,\qquad \overline{v}_2^\mu = v_1^\mu,\qquad \overline{v}_3^\mu = v_4^\mu,\qquad\overline{v}_4^\mu = v_3^\mu.\\
    & u_1^\mu = u_2^\mu,\qquad u_3^\mu = u_4^\mu,\qquad u_i^2 = n_i^2 =1,\\
    & v_i\cdot u_i= \overline{v}_i\cdot u_i=0,\quad  v_i\cdot n_j= \overline{v}_i\cdot n_j=u_i\cdot n_j=0.
\end{align}
\end{subequations}
Based on these notations, we define the three independent loop momenta $k_1$, $k_2$, and $k_3$ to be the line momenta between the external $p_i$ vertices for $i=1,2,3$, respectively, and the four-point vertex $H_1$ shown in figure~\ref{figure-three_loop_wide_angle_Landshoff_scattering}. It is convenient to define in a similar way a fourth loop momentum (for the line connecting $p_4$ and $H_1$), such that momentum conservation at $H_1$ implies
\begin{equation}
\label{kConserv}
k_1+k_2=k_3+k_4\,,
\end{equation}
where eqs.~(\ref{pConserv}) and (\ref{kConserv}) readily imply momentum conservation at the vertex $H_2$. We parameterise the loop momenta as follows:
\begin{eqnarray}
\label{eq:parameterisation_jet_like_k_i}
    k_i^\mu = Q\left(
    \xi_i v_i^\mu + \lambda \kappa_i\overline{v}_i^\mu 
    + \sqrt{\lambda}\tau_i u_i^\mu 
    + \sqrt{\lambda}\nu_i n^\mu_i \right),\qquad i=1,2,3,4.
\end{eqnarray}
As $k_i^\mu$ is collinear to $p_i^\mu$, we have the scaling parameter $\lambda\to 0$ while $\xi_i,\kappa_i,\tau_i,\nu_i\sim 1$. In particular, the large component varies over the whole range $0<\xi_i<1$.

The integral corresponding to figure~\ref{figure-diagram23_and_Landshoff_scattering} takes the form
\begin{eqnarray}
\displaystyle{
\I_{G_{\bullet\bullet}}(s,t,\lambda p_i^2) = \int \frac{d^Dk_1}{i\pi^{D/2}} \frac{d^Dk_2}{i\pi^{D/2}} \frac{d^Dk_3}{i\pi^{D/2}} \underbrace{  \prod_{i=1}^4 \frac{1}{(k_i^2+i\varepsilon)((p_i-k_i)^2+i\varepsilon)}}_{\sim \lambda^{-8}},}
\label{eq:Landshoff_scattering_momentum_integrand_scaling}
\end{eqnarray}
where used the collinear scaling of the propagators in eq.~(\ref{eq:parameterisation_jet_like_k_i})
to determine the
scaling of integrand.

The integration measure over the three independent loop momenta can then be written using the parameterisation of (\ref{eq:parameterisation_jet_like_k_i}) as:
\begin{eqnarray}
\label{measure_first_expression}
    \int d^Dk_1 d^Dk_2 d^Dk_3 = Q^{3D} \int \prod_{i=1}^3 d\xi_i d\kappa_i d\tau_i d\nu_i \,\nu_i^{D-4} d\Omega_i^{(D-3)}\,,
\end{eqnarray}
where we introduced polar coordinates for the $D-3$ dimensional normal space parameterised by the projections $\nu_i$ parallel to $n_i$ and the differential solid angle $d\Omega_i^{(D-3)}$. In (\ref{measure_first_expression}) we have not included the $\lambda$ scaling. Doing so requires some care, as we now explain.

In traditional power counting for a given pinch surface, a key step is to identify the \emph{normal coordinates} and the \emph{internal coordinates}~\cite{LbyStm78,Stm78I,Stm95book}. The former are by definition the ones that tend to zero with the scaling parameter $\lambda$. As can be seen in the parameterisation of the jet loop momenta (\ref{eq:parameterisation_jet_like_k_i}), among all components, the only internal coordinate is the one proportional to $v_i^\mu$, i.e. the large momentum component governed by $\xi_i$; all the others are suppressed in the limit $\lambda\to 0$, and are therefore normal coordinates. 
However, in the Landshoff scattering picture we study here, as soon as the jet-scaling condition is imposed, momentum conservation (\ref{kConserv}) implies that $\xi_1$, $\xi_2$, and $\xi_3$ cannot anymore all be regarded as internal coordinates simultaneously, since their magnitudes are constrained. Indeed, inserting the parameterisation (\ref{eq:parameterisation_jet_like_k_i}) into (\ref{kConserv}), it is direct to check that $\xi_i$ for $i\neq1$ can be expressed as
\begin{subequations}
\label{eq:xi234_dependences_on_xi1_tau_kappa}
\begin{align}
 &\xi_2 =\xi_1 -\frac{1}{2}\sqrt{\lambda} \cos ^2(\theta ) \left( \tan \left(\frac{\theta }{2}\right) \Delta\tau- \cot \left(\frac{\theta }{2}\right) \Sigma\tau\right)+\lambda (\kappa_2-\kappa_1),
 \\
 &\xi_3 =\xi_1+ \frac{1}{2} \sqrt{\lambda} \tan \left(\frac{\theta }{2}\right) \Delta\tau+\lambda (\kappa_2-\kappa_4),
 \\&
 \xi_4 = \xi_1 -\frac{1}{2} \sqrt{\lambda} \cot \left(\frac{\theta }{2}\right) \Sigma\tau+\lambda (\kappa_2-\kappa_3),
\end{align}
\end{subequations}
where we defined
\[
 \Delta\tau\equiv \tau_1+\tau_2-\tau_3-\tau_4\,, \qquad\quad \Sigma\tau = \tau_1+\tau_2+\tau_3+\tau_4.
 \]
A first observation from eq.~(\ref{eq:xi234_dependences_on_xi1_tau_kappa}) is that all the $\xi_i$ are equal up to $\mathcal{O}(\sqrt{\lambda})$ terms. Note that this statement is valid only for generic $\theta$; for forward or backward scattering ($\theta\to0$ or $\pi$), it is invalidated. We will revisit this aspect in section~\ref{section-example_8_propagator_graph}, while here we proceed assuming that $\theta$ is indeed generic. 

The relations (\ref{eq:xi234_dependences_on_xi1_tau_kappa}) also allow us to perform the change of integration variables $\{\xi_2,\xi_3\}\to \{\kappa_4,\tau_4\}$, with the Jacobian, 
\begin{equation}
\label{Jacobian_theta}
\mathrm{det}\left ( \frac{\partial(\xi_2,\xi_3)}{\partial(\kappa_4,\tau_4)} \right ) = \lambda^{3/2} \cos(\theta) \cot(\theta)\,.
\end{equation}
Once this is done the only internal coordinate is $\xi_1$, while all the other integration variables are normal coordinates. The scaling of the integration measure in eq.~(\ref{measure_first_expression}) can now be obtained:
\begin{eqnarray}
    &&\int \prod_{i=1}^3 d\xi_i\ d\kappa_i d\tau_i d\nu_i 
    \,\nu_i^{D-4}  = \int_0^1 d\xi_1  \underbrace{\left(\int \prod_{i=1}^3 \ (\lambda d\kappa_i)(\lambda^{\frac12} d\tau_i) (\lambda^{\frac12-\eps} d\nu_i\, \nu_i^{-2\eps}) \right)}_{\sim \lambda^{6-3\epsilon}}\nonumber\\
    &&\hspace{7cm}\cdot \int d\kappa_4 d\tau_4\,\, \underbrace{\mathrm{det}\left ( \frac{\partial(\xi_2,\xi_3)}{\partial(\kappa_4,\tau_4)} \right )}_{\sim \lambda^{3/2}}\,.
\label{eq:Landshoff_scattering_momentum_integration_measure_scaling}
\end{eqnarray}
Eq.~(\ref{eq:Landshoff_scattering_momentum_integration_measure_scaling}) fully captures the constraints due to momentum conservation, subject to the assumption that $k_i$ are jet-like momenta admitting the scaling rule of eq.~(\ref{eq:parameterisation_jet_like_k_i}).

Therefore, the corresponding degree of divergence follows from eqs.~(\ref{eq:Landshoff_scattering_momentum_integrand_scaling}) and (\ref{eq:Landshoff_scattering_momentum_integration_measure_scaling}):
\begin{eqnarray}
\label{eq:power_counting_Landshoff_momentum_space}
    \mu[G_{\bullet\bullet},\text{Landshoff}] = 6-3\epsilon +\frac{3}{2} -8 = -\frac{1}{2}-3\epsilon,
\end{eqnarray}
which coincides with the result in parameter space, eq.~(\ref{eq:power_counting_Landshoff_parameter_space}). Note that this result regarding the asymptotic behaviour of figure~\ref{figure-three_loop_wide_angle_diagram23} is also consistent with the literature, e.g.~\cite{Hld64,BottsStm89}.

\subsection{Obtaining the complete set of regions via polytope dissection}
\label{section-obtain_complete_set_regions_polytope_dissection}

In sections~\ref{section-facet_regions_hidden_regions}-\ref{section-power_counting_momentum_space}, we argued and provided evidence, in both parameter space and momentum space, that $G_{\bullet \bullet}$ contains a Landshoff region, which does not appear as a facet of the Newton polytope, i.e. it is a hidden region. 
With the polytope dissection discussed in section~\ref{section-identifying_pinch_dissecting_polytope} in hand, it is now possible to systematically identify that such a region does exist and to check for other potentially hidden regions using the new polytopes.

\begin{table}[h]
    \small
    \centering
    \begin{tabular}{l|l}
        \bf $\v_\mathrm{R}$ ($x_0, x_1, \ldots, x_7 ;\lambda$)      & \bf degree of divergence \\\hline
        $(-2,-1,-2,-1,-2,-1,-2,-1;1)$ & $-6\epsilon$\\
        $(-1,-2,-1,-2,-1,-2,-1,-2;1)$ & $-6\epsilon$\\
        $(-1,-1,-1,0,-1,0,-1,0;1)$    & $1-3\epsilon$\\
        $(-1,-1,0,-1,0,-1,0,-1;1)$    & $1-3\epsilon$\\
        $(-1,-1,0,0,0,0,0,0;1)$       & $-\epsilon$\\
        $(0,0,0,0,0,0,0,0;1)$         & $0$\\
    \end{tabular}
    \caption{Regions obtained by directly applying the MoR to the on-shell expansion of graph $G_{\bullet \bullet}$ with $p_1^2 \sim \lambda Q^2$ and $p_i^2=0\ (i=2,3,4)$.}
    \label{tab:regions_onshell_original}
\end{table} 

Firstly, let us directly apply the MoR to $G_{\bullet \bullet}$. We consider the case with a single leg, $p_1$, slightly off-shell, $p_1^2 \sim \lambda Q^2$, and all other legs strictly on-shell $p_i^2 = 0$ for $i=2,3,4$. The MoR identifies the six regions shown in table~\ref{tab:regions_onshell_original}.
We observe that the Landshoff region, having a uniform scaling vector (\ref{eq:Gdotdot_Landshoff_scattering_region_vector}), is not a facet of the original integrand and is therefore absent.\footnote{As explained in section~\ref{section-facet_regions_hidden_regions}, the Landshoff region \emph{cannot} be characterised by a facet of the original polytope, as it must admit the relations (\ref{cancellation def}), in addition to the scaling of $x_i$.}
According to the table, the leading power behaviour of the facet regions of the integral in the small $\lambda$ limit is $\mathcal{O}(\lambda^0)$.

\begin{table}[h]
    \small
    \centering
    \begin{tabular}{l|l|l}
        \multirow{2}{5.5cm}{\bf $\v_\mathrm{R}$ ($y_0, x_1, y_2, x_3, y_4, x_5, y_6, x_7 ;\lambda$)} &  \multirow{2}{5cm}{\bf $\v_\mathrm{R}$ ($x_0, x_1, \ldots, x_7; \lambda$)} & \bf degree of \\
         & & \bf divergence \\\hline
        $(1/2,-1,1/2,-1,1/2,-1,0,-1;1)$       & $(-2,-2,-2,-2,-2,-2,-2,-2;2)$ & $-1/2-3\epsilon$\\
        $(0,-1,1,-1,1,-1,0,-1;1)$    & $(-1,-1,-1,-1,-1,-1,-1,-1;1)$ & $-3\epsilon$\\
        $(1,-1,1,-1,0,-1,0,-1;1)$         & $(-1,-1,-1,-1,-1,-1,-1,-1;1)$ & $-3\epsilon$\\
        $(-1,-1,-1,-1,-1,-1,-1,-1;1)$ & $(-2,-1,-2,-1,-2,-1,-2,-1;1)$ & $-6\epsilon$\\
        $(1,-2,1,-2,1,-2,1,-2;1)$         & $(-1,-2,-1,-2,-1,-2,-1,-2;1)$ & $-6\epsilon$\\
        $(0,-1,0,0,0,0,0,0;1)$ & $(-1,-1,0,0,0,0,0,0;1)$ & $-\epsilon$\\
        $(0,0,0,0,0,0,0,0;1)$    & $(0,0,0,0,0,0,0,0;1)$ & $0$\\
    \end{tabular}
    \caption{On-shell expansion of $G_{\bullet \bullet}$, with $p_1^2 \sim \lambda Q^2$ and $p_i^2=0\ (i=2,3,4)$, first integral dissection $\mathcal{I}_1$.}
    \label{tab:regions_onshell_split1}
\end{table}

Next, we will dissect the polytope using the procedure described in section~\ref{section-identifying_pinch_dissecting_polytope} and compare the regions we obtain from the new set of integrals.
Examining eq.~\eqref{eq:8_propagator_wideangle_Symanzik_polynomialsF}, we note that after applying the change of variables given in eq.~\eqref{eq:change_of_variables_three_loop_hidden} and dissecting the integral according to eq.~\eqref{eq:division_sector_three_loop_hidden}, we obtain a set of 24 new integrals, $\mathcal{I}_1, \ldots, \mathcal{I}_{24}$, one for each of the orderings of $\{y_0,y_2,y_4,y_6\}$.
These integrals have $\mathcal{F}$ polynomials similar to eq.~\eqref{eq:F_polynomials_after_decomposition}, but with additional terms proportional to $(-p_1^2)$ multiplied by non-negative polynomials of the parameters.
Each of the new integrals has a same-sign regime for some choice of the value of $p_1^2, s$ and $t$ (although not the same choice for all integrals). We therefore expect that each integral separately can be analytically continued from a same-sign regime to the region of interest, and the application of the MoR should now truly identify all scaleful regions: there are no hidden regions in the new integrals.
In conclusion, we expect that all the regions of the original integral, including hidden regions, would be obtained as facet regions of the new polytopes.

Applying the MoR to each of the 24 new integrals, we obtain 24 sets of regions. In table~\ref{tab:regions_onshell_split1} we display the regions for the first integral dissection, $\mathcal{I}_1$, in terms of both the new parameters, $y_0, x_1, y_2, x_3, y_4, x_5, y_6, x_7$, in which it has been computed, and after translating back to the original parameters, $x_0,\ldots,x_7$.  We find that the Landshoff region, which did not appear as a facet of the Newton polytope of the original integrand, now appears as a facet of the dissected integrand polytopes, and is detected geometrically using the straightforward application of the MoR algorithm as described in section~\ref{section-Newton_polytope_approach}.

We note in passing that there are some other interesting changes in the set of region vectors upon replacing the original integral by the 24 new ones. In particular, the region $(-1,-1,-1,0,-1,0,-1,0;1)$ in the original polytope (see table~\ref{tab:regions_onshell_original}), is absent in $\mathcal{I}_1$ but is present in other integral dissections, e.g. $\mathcal{I}_{10}$. The region $(-1,-1,0,-1,0,-1,0,-1;1)$ in the original polytope, is entirely absent in the integral dissections. Its absence should not be a problem for recovering the correct asymptotic expansion: importantly, non-analytic behaviour of the form $\lambda^{-3\epsilon}$ is of course still present. 

The Landshoff region is present in all 24 integral dissections, ${\cal I}_1, \ldots, {\cal I}_{24}$ and has a degree of divergence of $\mu = -1/2-3\epsilon$, making it the leading region in the small $\lambda$ limit for $G_{\bullet \bullet}$.
We observe that there is a many-to-one map when going from the new variables to the old variables and, therefore, there can be several regions in the new variables that have the same scaling in the old variables.
Comparing to the analysis presented in section~\ref{section-power_counting_parameter_space}, we note that indeed the first region vector reported in table~\ref{tab:regions_onshell_split1} is the Landshoff region, compatible with the one predicted in~eq.~(\ref{eq:Gdotdot_Landshoff_scattering_region_vector}) (the factor of $2$ in their normalisation is immaterial), and this region drives the behaviour of the integral at small~$\lambda$.
Examining the scaling of the new parameters for this region vector, they are each precisely as predicted in section~\ref{section-power_counting_parameter_space}, and the degree of divergence corresponds to the case $a=0$ in the analysis presented there.
The second and third regions reported in the table also correspond to region vectors with uniform scaling of the propagators in the original variables. We find that they correspond to the case $a=\pm \frac{1}{2}$ in section~\ref{section-power_counting_parameter_space}.

In summary, we have shown that the Landshoff region is the unique hidden region in the on-shell expansion of $G_{\bullet \bullet}$ when a single leg is taken slightly off-shell.
Repeating the above analysis for the case where all legs are taken slightly off-shell, $p_i^2 \sim \lambda Q^2$ for $i=1,\ldots,4$, we obtain a total of 73 facet regions by applying the MoR to the original integral, and as expected, we observe that the Landshoff region is missed.
After dissecting the polytope and obtaining regions for each of the 24 new integrals, we find that only 71 regions are present. The Landshoff region is present in each of the 24 new integral dissections. It is the leading-power region in the small-$\lambda$ limit and has a degree of divergence $\mu = -1/2-3\epsilon$, as expected.

\subsection{Numerical evidence and discussion}
\label{section-numerical_evidence_discussion}

The power-counting technique, which we have applied to the Landshoff scattering region, can also be implemented to other facet regions of $G_{\bullet\bullet}$. One can check that the degree of divergence $\mu\geqslant 0$ for these regions, which implies that the infrared singularities are at most logarithmic. In other words, the Landshoff region encodes the entire power divergence of $G_{\bullet\bullet}$. Therefore, we have a perfect example where the hidden region not only exists, but also plays the most important role in this asymptotic expansion; without it even the leading asymptotic behaviour would not be captured.

\begin{figure}[t]
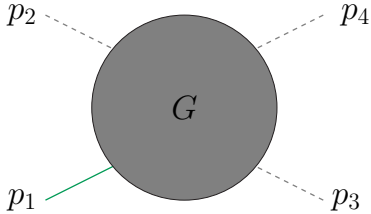

\centering
\include{figs/generic_one_offshell}
\vspace{-1em}
\caption{The generic picture for expanding one external momentum ($p_1$) in the on-shell expansion, all others ($p_2$, $p_3$, and $p_4$) exactly on-shell. }
\label{figure-generic_one_offshell}
\end{figure}
Below we show some numerical evidence for the statements above by considering the kinematics that the external momenta $p_2$, $p_3$, and $p_4$ are put exactly on shell, while \hbox{$p_1^2 \neq 0$} (see figure~\ref{figure-generic_one_offshell}). By running pySecDec, one can evaluate both the entire original integral $\mathcal{I}(G_{\bullet\bullet})$, and the leading term in the power expansion of the Landshoff scattering region, for a given~$p_1^2$. By approaching the limit $p_1^2 \to 0$, their asymptotic behaviours can be compared.

\begin{figure}[h!]
\centering
\begin{subfigure}[b]{0.495\textwidth}
\includegraphics[width=\textwidth]{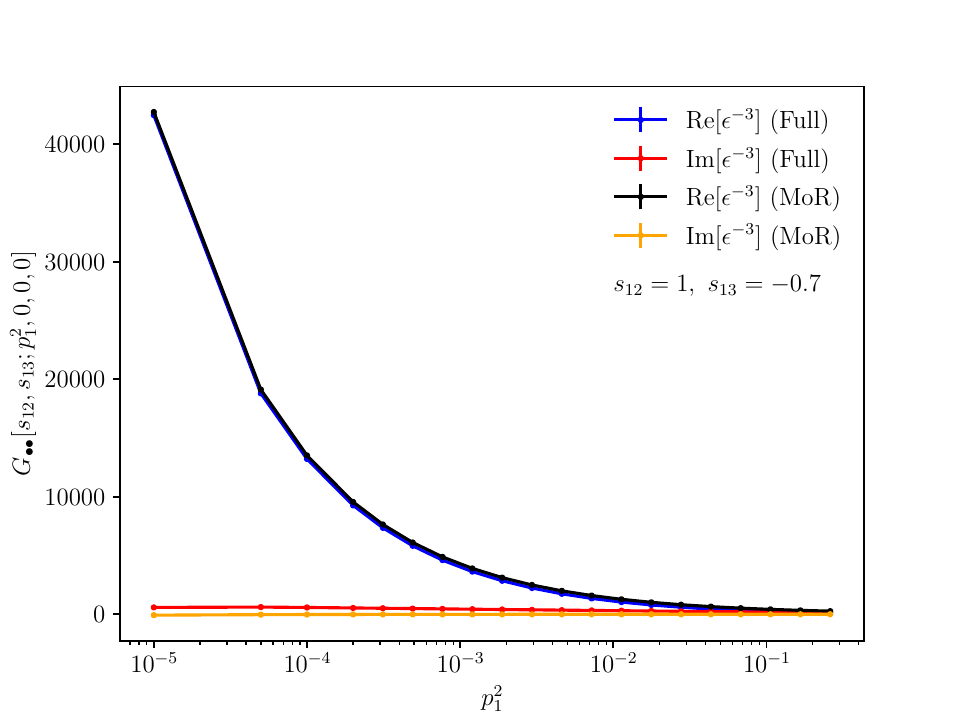}
\end{subfigure}
\begin{subfigure}[b]{0.495\textwidth}
\includegraphics[width=\textwidth]{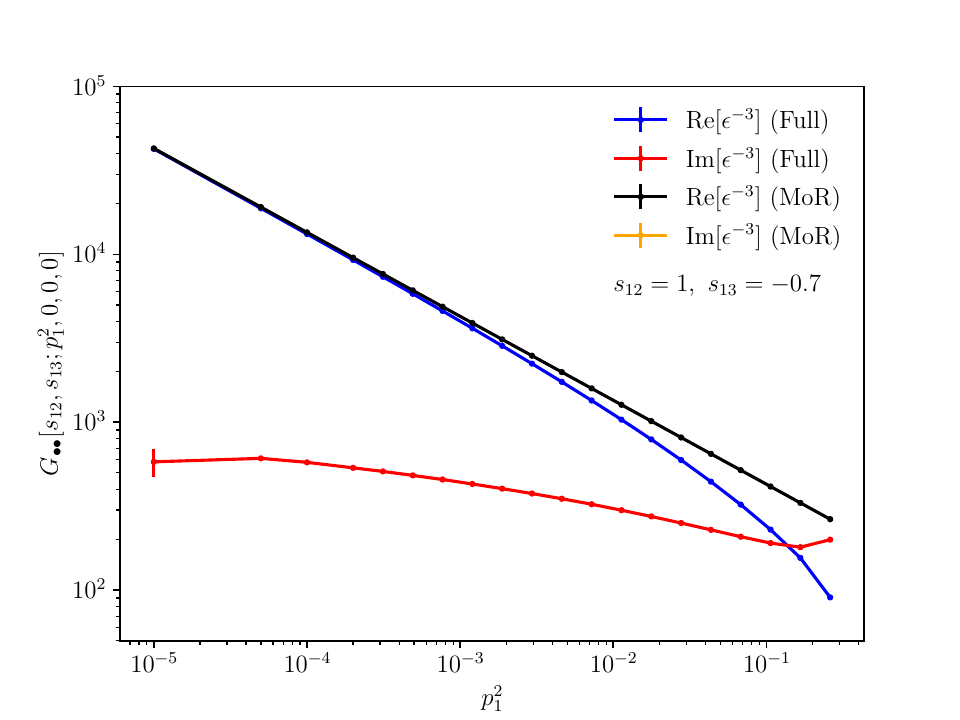}
\end{subfigure}
\caption{Numerical evaluation of the ${\cal O}(\epsilon^{-3})$ contribution to ${\cal I}(G_{\bullet \bullet})$ with one off-shell leg (leg~1), plotted as a function of $p_1^2$, which is gradually taken small, down to $p_1^2/s_{12}=10^{-5}$. The value of the integral ${\cal I}(G_{\bullet \bullet})$ at ${\cal O}(\epsilon^{-3})$ is shown on a linear vertical axis on the left panel and on a logarithmic vertical axis on the right panel. The blue and red points correspond to the full integral, while the black and yellow ones to the leading term in the asymptotic expansion of the integral according to the MoR, i.e. the one scaling as $\lambda^{-1/2}$.}
\label{figure-numeric_evidences}
\end{figure}

We take $G = G_{\bullet\bullet}$ in figure~\ref{figure-generic_one_offshell}, with results shown in figure~\ref{figure-numeric_evidences}. The integral $\mathcal{I}(G_{\bullet\bullet})$, as a function of $p_1^2$, is depicted by the blue (real part) and red (imaginary part) curves, respectively. In contrast, the contribution from the Landshoff scattering region is depicted by the black (real part) and yellow (imaginary part) curves, respectively. The following observations are straightforward from the plots. First, the contribution from the Landshoff scattering region exhibits a power-like behaviour, which is a good approximation to the integral $\mathcal{I}(G_{\bullet\bullet})$ at small~$p_1^2$. Second, the imaginary part of the Landshoff scattering region is negligible compared to the real part, which implies that the leading contribution to the asymptotic expansion is purely real when normalised as in eq.~\eqref{eq:igbulletbullet}. Third, the imaginary part of $\mathcal{I}(G_{\bullet\bullet})$, as well as the difference between the real parts of $\mathcal{I}(G_{\bullet\bullet})$ and the Landshoff scattering region, originate from the contributions of other regions, which are subleading in the asymptotic expansion; they become important only as $p_1^2$ increases.

\begin{figure}[t]
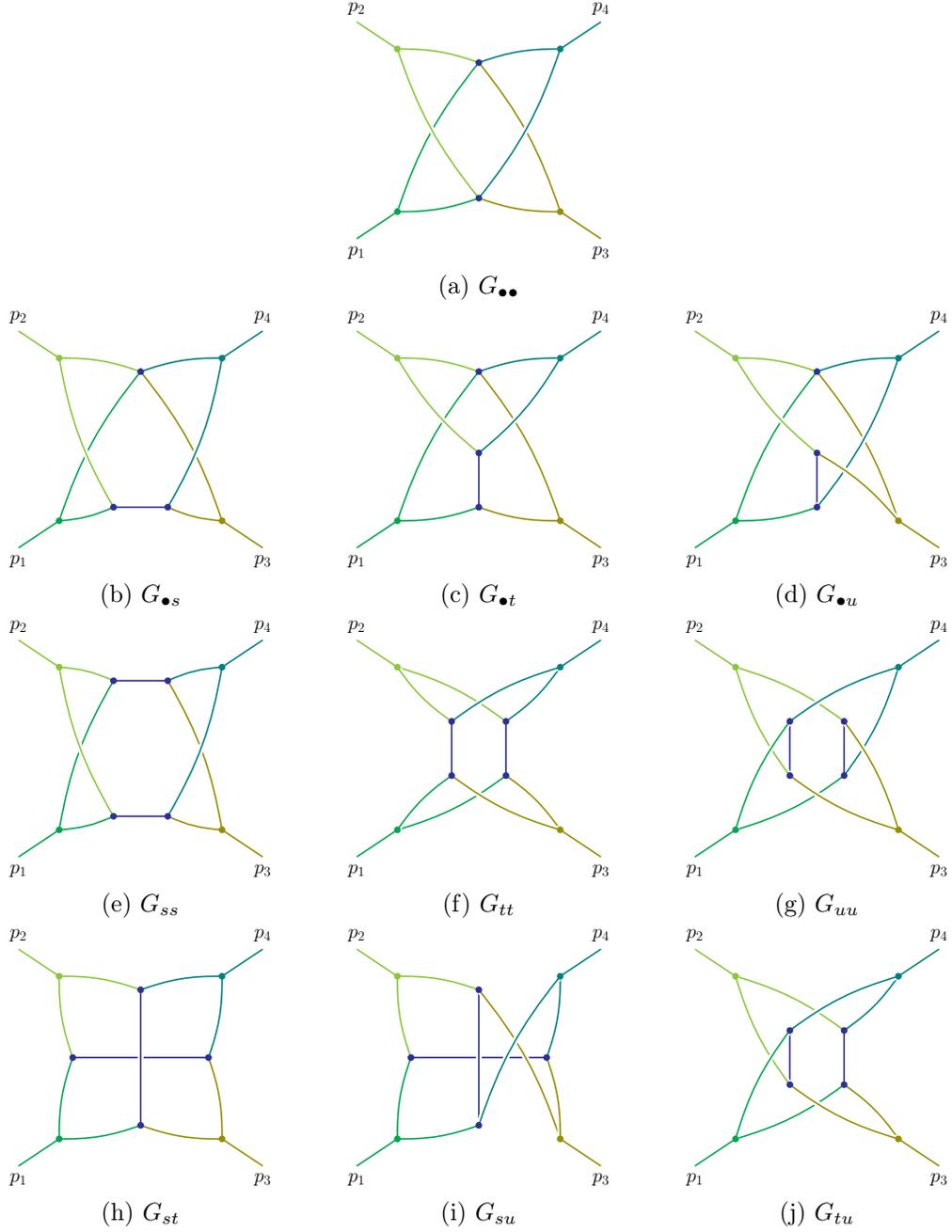

\centering
\begin{subfigure}[b]{0.25\textwidth}
\centering
\include{figs/possibly_hidden_singular_graph_dot_dot_wideangle}
\vspace{-3em}\caption{$G_{\bullet \bullet}$}
\label{possibly_hidden_singular_graph_dot_dot_wideangle}
\end{subfigure}
\\
\begin{subfigure}[b]{0.25\textwidth}
\centering
\include{figs/possibly_hidden_singular_graph_dot_s_wideangle}
\vspace{-3em}\caption{$G_{\bullet s}$}
\label{possibly_hidden_singular_graph_dot_s_wideangle}
\end{subfigure}
\qquad
\begin{subfigure}[b]{0.25\textwidth}
\centering
\include{figs/possibly_hidden_singular_graph_dot_t_wideangle}
\vspace{-3em}\caption{$G_{\bullet t}$}
\label{possibly_hidden_singular_graph_dot_t_wideangle}
\end{subfigure}
\qquad
\begin{subfigure}[b]{0.25\textwidth}
\centering
\include{figs/possibly_hidden_singular_graph_dot_u_wideangle}
\vspace{-3em}\caption{$G_{\bullet u}$}
\label{possibly_hidden_singular_graph_dot_u_wideangle}
\end{subfigure}
\\
\begin{subfigure}[b]{0.25\textwidth}
\centering
\include{figs/possibly_hidden_singular_graph_s_s_wideangle}
\vspace{-3em}\caption{$G_{s s}$}
\label{possibly_hidden_singular_graph_s_s_wideangle}
\end{subfigure}
\qquad
\begin{subfigure}[b]{0.25\textwidth}
\centering
\include{figs/possibly_hidden_singular_graph_t_t_wideangle}
\vspace{-3em}\caption{$G_{t t}$}
\label{possibly_hidden_singular_graph_t_t_wideangle}
\end{subfigure}
\qquad
\begin{subfigure}[b]{0.25\textwidth}
\centering
\include{figs/possibly_hidden_singular_graph_u_u_wideangle}
\vspace{-3em}\caption{$G_{u u}$}
\label{possibly_hidden_singular_graph_u_u_wideangle}
\end{subfigure}
\\
\begin{subfigure}[b]{0.25\textwidth}
\centering
\include{figs/possibly_hidden_singular_graph_s_t_wideangle}
\vspace{-3em}\caption{$G_{s t}$}
\label{possibly_hidden_singular_graph_s_t_wideangle}
\end{subfigure}
\qquad
\begin{subfigure}[b]{0.25\textwidth}
\centering
\include{figs/possibly_hidden_singular_graph_s_u_wideangle}
\vspace{-3em}\caption{$G_{s u}$}
\label{possibly_hidden_singular_graph_s_u_wideangle}
\end{subfigure}
\qquad
\begin{subfigure}[b]{0.25\textwidth}
\centering
\include{figs/possibly_hidden_singular_graph_t_u_wideangle}
\vspace{-3em}\caption{$G_{t u}$}
\label{possibly_hidden_singular_graph_t_u_wideangle}
\end{subfigure}
\caption{The Landshoff scattering pictures corresponding to the graphs in figure~\ref{figure-possibly_hidden_singular_graphs}.}
\label{figure-possibly_hidden_singular_graphs_wideangle}
\end{figure}
Having understood ${\cal I}(G_{\bullet \bullet})$ in momentum space as Landshoff scattering, namely collinear split of each of the four jet momenta subject to momentum conservation at the two hard vertices, it is readily clear that the same (Landshoff) momentum configuration defines a region in each of the remaining graphs. We describe these regions in figure~\ref{figure-possibly_hidden_singular_graphs_wideangle} by colouring the edges of the graphs, such that the four jets are coloured in \textbf{\color{Green}green}, \textbf{\color{LimeGreen}lime}, \textbf{\color{olive}olive}, and \textbf{\color{teal}teal} respectively, while the hard subgraph is coloured in \textbf{\color{Blue}blue}. It is evident that each of these coloured graphs feature two disconnected hard subgraphs (clearly, upon contraction of these hard subgraphs one recovers~$G_{\bullet \bullet}$). 

We now argue that each Landshoff scattering region in figure~\ref{figure-possibly_hidden_singular_graphs_wideangle} is the unique hidden region of the corresponding graph of figure~\ref{figure-possibly_hidden_singular_graphs}. Thus, figure~\ref{figure-possibly_hidden_singular_graphs_wideangle} exhausts in fact the entire set of hidden regions for massless $2\to 2$ scattering at the three-loop level. To see this, we first observe that for any such region, the superleading terms, each of which is $\mathcal{O}(\lambda^{-4})$, are always identical to those of $G_{\bullet\bullet}$, namely, eq.~(\ref{eq:super_leading_three_loop_Landshoff}). It then allows us to employ the same change of variables shown in eq.~(\ref{eq:change_of_variables_three_loop_hidden}) and divide the integration range into the same sectors as eq.~(\ref{eq:division_sector_three_loop_hidden}). The new 24 integrals correspond to $\mathcal{F}$ polynomials that are identical to those in eq.~(\ref{eq:F_polynomials_after_decomposition}) up to $\mathcal{O}(\lambda^{-3})$  terms (and terms that are further suppressed) involving the Feynman parameters of the propagators which do not feature in $G_{\bullet \bullet}$.
While these additional terms in $\mathcal{F}$ modify the integral, they do not affect the location of the pinch, and hence the dissection of the polytope. Finally, it is straightforward to check that all the corresponding new integrals are free from pinch singularities within the integration domain.

As a result of dissecting the polytope, it is a natural possibility that additional facet regions can appear that correspond neither to a hidden region associated with cancellation between monomials in ${\cal F}$, nor to facets of the original polytope (that is, after mapping the vectors obtained from facets of the new polytope back into the original parameters).
For example, the regions of $G_{\bullet s}$, in the original parameters, are listed in table~\ref{table-wideangle_facet_hidden_vectors_Gdots}, where we note that besides the Landshoff scattering region vector $(-1,-1,-1,-1,-1,-1,-1,-1,0;1)$, many additional vectors also appear from the facets of the new polytopes. It can be verified that these vectors are \emph{incompatible} with a cancellation solution for general kinematics (i.e. arbitrary values of $s_{12}$ and $s_{13})$: under any of these scaling vectors, the individual leading monomials in the $\mathcal{F}$ polynomial are of the same order as the leading monomials in $\mathcal{U}$, so any putative cancellation solution among these leading terms can only hold for specific values of Mandelstam variables. Such cancellation structures cannot contribute to hidden regions of~$G_{\bullet s}$ and we regard them as spurious.
\begin{table}[t]
\begin{center}
\begin{tabular}{ |c||c|c|c| } 
\hline
 & \multirow{2}{8em}{\quad$\boldsymbol{v} (x_0,\dots,x_8;\lambda)$} & \multirow{2}{6.2em}{region or not?} \\
 & & \\
\hline
\multirow{4}{4em}{$\begin{matrix} \text{original}\\ \text{polytope} \end{matrix}$} & $(-2,-1,-2,-1,-2,-1,-2,-1,-2;1)$ & $\greencheckmark[ForestGreen]$ \\ 
 & \qquad\quad \multirow{2}{4em}{$\vdots$} & \qquad\quad\ \ \multirow{2}{4em}{$\vdots$} \\
 &  & \\
 & $(0,0,0,0,0,0,0,0,0;1)$ & $\greencheckmark[ForestGreen]$ \\
\hline
\multirow{5}{4em}{$\begin{matrix} \text{dissected}\\ \text{polytopes} \end{matrix}$} & $(-1,-1,-1,-1,-1,-1,-1,-1,0;1)$ & $\greencheckmark[ForestGreen]$ \\
 & $(-2,-1,-2,-1,-1,-1,0,0,-2;1)$ & $\crossmark[Red]$ \\
 & \qquad\quad \multirow{2}{4em}{$\vdots$} & \qquad\quad\ \ \multirow{2}{4em}{$\vdots$} \\
 &  & \\
 & $(0,1,0,0,-1,0,-1,-1,-1;1)$ & $\crossmark[Red]$ \\
\hline
\end{tabular}
\end{center}
\vspace{-1em}\caption{The vectors that are possibly relevant for the regions in the on-shell expansion of $G_{\bullet s}$. Each vector is written in the format $(x_0,\dots,x_8)$ and is expressed after mapping back to the original variables. The first group of vectors correspond to the lower facets of the original polytope, all of which are regions of $G_{\bullet s}$. In contrast, the second group of vectors are hidden inside the original polytope, which arise only as lower facets of the dissected polytope. Among all these regions, the only one corresponding to a hidden region (Landshoff scattering) is $(-1,-1,-1,-1,-1,-1,-1,-1,0;1)$. All the other vectors in the second group are considered spurious because they are inconsistent with any cancellations for general kinematics.}
\label{table-wideangle_facet_hidden_vectors_Gdots}
\end{table}

\begin{figure}[t]
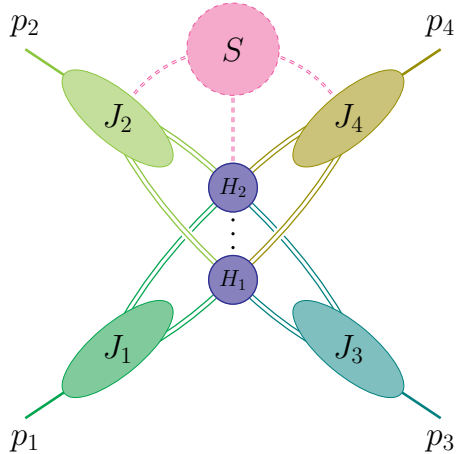

\centering
\include{figs/wide_angle_scattering_hidden_region}
\vspace{-1cm}\caption{The general configuration of hidden regions in the on-shell expansion of $2\to 2$ wide-angle scattering, as we propose at the end of section~\ref{section-revisit_onshell_expansion_wideangle_scattering}. The graph $G$ is the union of the hard subgraph $H$, the jet subgraphs $J_1,\dots, J_4$, and the soft subgraph $S$. In contrast to the facet regions in figure~\ref{figure-wide_angle_scattering_generic_region}, the hard subgraph has multiple connected components $H_1,H_2,\dots$, and all the four jets are attached to each of these components.}
\label{figure-wide_angle_scattering_hidden_region}
\end{figure}

Based on the analysis above, we conjecture that in the context of the on-shell expansion for $2\to 2$ wide-angle scattering, a generic hidden region must be described by figure~\ref{figure-wide_angle_scattering_hidden_region}. In this picture, there are multiple hard components, each of which is attached to the four jets. Additionally, there can be a soft subgraph attached to the hard and jet subgraphs. In conclusion, figures~\ref{figure-wide_angle_scattering_generic_region} and \ref{figure-wide_angle_scattering_hidden_region}, which characterise the facet regions and hidden regions, respectively, conjecturally constitute the entire list of regions in the on-shell expansion for $2 \to 2$ wide-angle scattering.

\section{Regge-limit expansion for \texorpdfstring{$2\to 2$}{TEXT} scattering}
\label{section-high_energy_expansion_forward_scattering}

In this section, we study the Regge-limit expansion for $2\to 2$ scattering, as in eq.~(\ref{pConserv}), where every external momentum is strictly massless and on shell, with $p_3$ (nearly) collinear to $p_1$ and $p_4$ nearly collinear to $p_2$, namely,
\begin{eqnarray}
\label{Regge_limit}
    p_i^2=0\ \ (i=1,2,3,4),\qquad  \frac{-t}{s} \sim  \frac{-t}{-u} \sim \lambda,
\end{eqnarray}
where $s\equiv s_{12}\equiv (p_1+p_2)^2>0$, $t\equiv s_{13}\equiv  (p_1-p_3)^2<0$, and $u\equiv s_{14} \equiv (p_1-p_4)^2<0$. In contrast to the wide-angle scattering kinematics, eq.~(\ref{eq:wideangle_onshell_kinematics}), the Regge-limit expansion famously features Glauber modes that contribute to the region structure in certain graphs. 

Glauber mode may appear in facet regions or in hidden regions. Our focus here is on the latter. Based on the analysis in section~\ref{section-identifying_pinch_dissecting_polytope}, the first graphs potentially featuring hidden regions in the Regge limit occur at three loops, and are shown in figure~\ref{figure-possibly_hidden_singular_graphs}.
Here we will interpret these in both parameter and momentum space. Notably, we will show that all these hidden regions feature a Glauber-mode loop momentum and can be related to the Landshoff scattering regions in wide-angle scattering. We then derive the asymptotic behaviour of these regions using power counting.

We point out that the asymptotic behaviour of Feynman integrals for scattering in the Regge limit has been investigated by many authors over the years, with early work in refs.~\cite{AmtFbnStgln62asymptotic,AmtFbnStgln62theory,GrbPmrch62,LeeSyr62,Pkhn63I,Pkhn63II,IslLdsfTlr63,Hld63,Mdst63,Tktpls63,Tktpls63nonplanar,Mnk64,Hprcht65,Lam69,Ldsf74,Cls77book,Pkhn80book}. Our results for the asymptotic behaviour is consistent with their findings.
A key difference is that our investigation utilises the geometric MoR and the approach to hidden regions developed in the present paper, which also offers a transparent connection to the on-shell expansion wide-angle scattering. In addition, the availability of analytic computations of massless four-point three-loop integrals~\cite{Henn:2020lye,Bargiela:2021wuy} allows us to test and verify the asymptotic expansion techniques for the first time.

\subsection{Hidden regions in the Regge-limit expansion}
\label{section-hidden_regions_Glauber_mode}

As we have illustrated in section~\ref{section-finding_diagrams_prone_to_hidden_landau_singularities}, up to the three-loop order, all the graphs prone to having hidden regions in the Regge-limit expansion are shown in figure~\ref{figure-possibly_hidden_singular_graphs}. Below we focus particularly on two representative graphs: $G_{\bullet\bullet}$ (figure~\ref{possibly_hidden_singular_graph_dot_dot}) and $G_{tt}$ (figure~\ref{possibly_hidden_singular_graph_t_t}).

\subsubsection{Example 1: \texorpdfstring{The crown graph $G_{\bullet\bullet}$}{TEXT}}
\label{section-example_8_propagator_graph}

\begin{figure}[t]
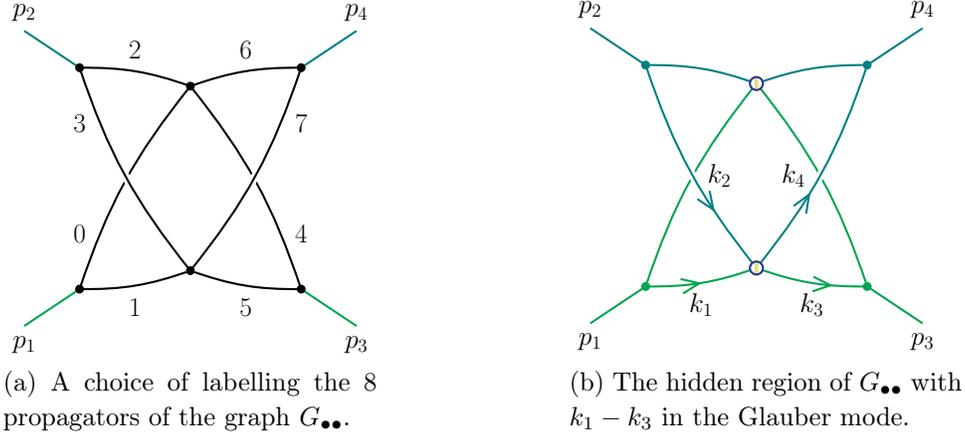

\centering
\hspace{-1em}
\begin{subfigure}[b]{0.32\textwidth}
\include{figs/Gdotdot_forward_scattering_parameterisation}
\vspace{-3em}
\caption{A choice of labelling the 8 propagators of the graph $G_{\bullet\bullet}$.}
\label{figure-Gdotdot_forward_scattering_parameterisation}
\end{subfigure}
\hspace{6em}
\begin{subfigure}[b]{0.32\textwidth}
\include{figs/Gdotdot_forward_scattering_hidden}
\vspace{-3em}
\captionsetup{width=1.05\linewidth}
\caption{The hidden region of $G_{\bullet\bullet}$ with $k_1-k_3$ in the Glauber mode.}
\label{figure-Gdotdot_forward_scattering_hidden}
\end{subfigure}
\caption{The three-loop nonplanar graph $G_{\bullet\bullet}$ and its hidden region. There are two jets in this region, one attached by $p_1$ and $p_3$ while the other by $p_2$ and $p_4$. The momentum that is exchanged between the jets, $k_1-k_3$, is in the Glauber mode.}
\label{figure-Gdotdot_forward_scattering}
\end{figure}

The Symanzik polynomials for figure~\ref{figure-Gdotdot_forward_scattering_parameterisation}, after replacing $s_{23}$ by $-s_{12}-s_{13}$, is 
\begin{eqnarray}
\label{eq:8_propagator_forward_Symanzik_polynomials}
    &&\mathcal{U}(\x) = (x_0+x_1)(x_2+x_3)(x_4+x_5) + (x_0+x_1)(x_2+x_3)(x_6+x_7) \nonumber\\
    &&\qquad\qquad + (x_0+x_1)(x_4+x_5)(x_6+x_7) + (x_2+x_3)(x_4+x_5)(x_6+x_7),\nonumber\\
    &&\mathcal{F}(\x;\s) = (-s_{12}) (x_0x_5 - x_1x_4)(x_2x_7 - x_3x_6) + (-s_{13})(x_0x_3 - x_1x_2)(x_4x_7 - x_5x_6).\nonumber\\
\end{eqnarray}
The list of facet regions with their degrees of divergence is shown in table~\ref{tab:regions_forward_original} in appendix~\ref{appendix-examples_regions_gbb_regge}. Note that in each of these regions, the loop momenta and propagator momenta are either in the collinear-1 mode (collinear to $p_1$ and $p_3$), the collinear-2 mode (collinear to $p_2$ and $p_4$), or the hard mode.

Let us then study the hidden regions, which must arise from cancellation among the superleading terms. Inspired by the Landshoff scattering regions, which are hidden regions for wide-angle scattering in the same $G_{\bullet\bullet}$ graph, we consider the case where all the Lee-Pomeransky parameters $x_i$ are of $\mathcal{O}(\lambda^{-1})$, as in eq.~(\ref{eq:Gdotdot_Landshoff_scattering_region_vector}). 
Since in the Regge limit (\ref{Regge_limit}) we consider $s=s_{12}$ as ${\cal O} (\lambda^0)$, while $t=s_{13}$ as ${\cal O} (\lambda^1)$, the four monomials involving $s_{12}$ in $\mathcal{F}$ in eq.~(\ref{eq:8_propagator_forward_Symanzik_polynomials}) are individually of $\mathcal{O}(\lambda^{-4})$, while those involving $s_{13}$ there, as well as the terms in ${\cal U}$ are of $\mathcal{O}(\lambda^{-3})$. A scaleful integral is then obtained subject to the following condition:
\begin{eqnarray}
\label{eq:forward_superleading_terms_cancellation_scaling}
    (x_0x_5 - x_1x_4)(x_2x_7 - x_3x_6)\sim \lambda^{-3}.
\end{eqnarray}
With the constraint (\ref{eq:forward_superleading_terms_cancellation_scaling}), all the terms in the Lee-Pomeransky polynomial ${\cal P= U+F}$ in eq.~(\ref{eq:8_propagator_forward_Symanzik_polynomials}) admit uniform $\lambda^{-3}$ scaling behaviour. Note that in contrast to the Landshoff scattering case, where two separate constraints (\ref{eq:wideangle_superleading_terms_cancellation_scaling}) arise, the Regge-limit expansion involves a single constraint, eq.~(\ref{eq:forward_superleading_terms_cancellation_scaling}).    

In the corresponding configuration in momentum space, each line momentum is collinear to the external momenta it is adjacent to, as shown in figure~\ref{figure-Gdotdot_forward_scattering_hidden}. We have a similar yet distinct situation compared to the Landshoff scattering case, figure~\ref{figure-diagram23_and_Landshoff_scattering}, where there are four jets, each associated with a single external momentum $p_i$. 
In the Regge limit there are just two jets, one associated with the external momenta $p_1$ and $p_3$ and the other with $p_2$ and~$p_4$. Inside each jet, all the propagators are (nearly) collinear. Below we shall show that between these  two jets, some exchanged loop momenta admit the \emph{Glauber mode} scaling law $k\sim Q(\lambda,\lambda,\lambda^{1/2})$. In figure~\ref{figure-diagram23_and_Landshoff_scattering} this is represented by the symbol \GlauberTextHard\, which stands for a hard vertex with a Glauber-mode momentum exchange in the $t$ channel.
An interesting feature of this region is that while one loop momentum is a Glauber mode, all the propagator momenta are of either the collinear-1 and the collinear-2 modes.

Let us now conduct a power-counting analysis of this region, first in parameter space and then in momentum space. The parameter-space analysis follows closely  section~\ref{eq:Gdotdot_Landshoff_scattering_region_vector}. Upon dissecting the Newton polytope, one can show that the configuration described above is a region, and furthermore, the unique hidden region of $G_{\bullet\bullet}$ in the Regge limit. Let us now compute the degree of divergence of this hidden region. As the superleading terms, eq.~(\ref{eq:forward_superleading_terms_cancellation_scaling}), are also superleading in the Landshoff scattering case, we can perform the same change of variables as eq.~(\ref{eq:Landshoff_scattering_change_variables}), namely,
\begin{eqnarray}
    x_0 = y_0\cdot x_1,\qquad x_2 = y_2\cdot x_3,\qquad x_4 = y_4\cdot x_5,\qquad x_6 = y_6\cdot x_7.
\label{eq:Landshoff_scattering_change_variables_repeat}
\end{eqnarray}
We then obtain
\begin{eqnarray}
    (x_1x_4-x_0x_5)(x_3x_6-x_2x_7) = (y_4-y_0)(y_6-y_2)x_1x_3x_5x_7\sim \lambda^{-3}.
\end{eqnarray}
This further implies $y_4-y_0\sim \lambda^{1/2+a}$ and $y_6-y_2\sim \lambda^{1/2-a}$ for some $a\in (-\frac{1}{2},\frac{1}{2})$. Independently of $a$, the integration measure is:
\begin{eqnarray}
    \prod_{i=0}^7 dx_i &&= \bigg(\prod_{i=1,3,5,7}x_idx_i\bigg) \cdot dy_0dy_2dy_4dy_6\nonumber\\
    &&= \bigg(\prod_{i=1,3,5,7}x_idx_i\bigg) \cdot dy_0dy_2d(y_4-y_0)d(y_6-y_2) \sim \lambda^{-8}\cdot\lambda^{1} = \lambda^{-7}.
\label{eq:Gdotdot_forward_parameter_integration_measure_scaling}
\end{eqnarray}
Meanwhile, the scaling of the integrand follows eq.~(\ref{eq:Landshoff_scattering_parameter_integrand_scaling}) identically, which is $\lambda^{6-3\epsilon}$. The corresponding degree of divergence is then straightforward:
\begin{eqnarray}
\label{eq:Gdotdot_power_counting_Glauber_parameter_space}
    \mu[G_{\bullet\bullet},\text{Glauber}] = -7 + 6 - 3\epsilon = -1-3\epsilon.
\end{eqnarray}

Let us now rederive this result in momentum space. Similarly to the case of the Landshoff scattering, we parameterise the loop momenta using eq.~(\ref{eq:parameterisation_jet_like_k_i}), where $\lambda$ is a small scaling parameter defined such that the virtuality of each line momentum within a jet is $\mathcal{O}(\lambda Q^2)$. Taking this over to the Regge limit, the expansion parameter must also be proportional to the ratio of kinematic invariants
\begin{equation}
    \lambda\sim \frac{-t}{s}= \frac{-(p_1- p_3)^2}{(p_1 + p_2)^2} \sim \frac{p_1\cdot p_3}{p_1\cdot p_2} \sim \frac{\theta^2}{4} \qquad  \text{for} \quad \theta\to 0\,
\end{equation}
where the relation with $\theta$ follows from eq.~(\ref{eq:scattering_kinematics_parameterisation}), expanded to leading order in~$\theta$.
For convenience, let us then set 
\begin{equation}
\tan\left(\frac{\theta}{2}\right)=\sqrt{\lambda}\,,
\end{equation}
in eq.~(\ref{eq:xi234_dependences_on_xi1_tau_kappa}). We then have:
\begin{subequations}
\label{xi_i_forward}
\begin{align}
 &\xi_2 =\xi_1 - \frac{1}{2}\left(\frac{1-\lambda}{1+\lambda}\right)^2 \left(\lambda \Delta\tau - \Sigma\tau \right)+\lambda (\kappa_2-\kappa_1),
 \\
 &\xi_3 =\xi_1+ \frac{1}{2} \lambda \Delta\tau+\lambda (\kappa_2-\kappa_4),
 \\&
 \xi_4 = \xi_1 -\frac{1}{2} \Sigma\tau+\lambda (\kappa_2-\kappa_3).
\end{align}
\end{subequations}
From this one observes a key feature of the Regge limit kinematics, namely that $\xi_3=\xi_1 + \mathcal{O}(\lambda)$ and $\xi_4=\xi_2 + \mathcal{O}(\lambda)$. This is consistent with our qualitative picture of the region, namely that the propagators collinear to the $p_1$ (and $p_3$) jet have the same large momentum component, up to terms of ${\cal O}(\lambda)$, and similarly, those collinear to the $p_2$ (and $p_4$) jet share the same large momentum component (again, up to terms of ${\cal O}(\lambda)$). Since also $v_3=v_1+{\cal O}(\lambda)$ and $\overline v_3=\overline{v}_1+{\cal O}(\lambda)$, upon using eq.~(\ref{eq:parameterisation_jet_like_k_i}), one obtains
\begin{align}
\label{eq:forward_scattering_Glauber_loop_momentum}
    k_4^\mu-k_2^\mu &= k_1^\mu-k_3^\mu \nonumber \\
    &= Q\left((\xi_1-\xi_3) v_1^\mu + \lambda (\kappa_1-\kappa_3)\overline{v}_1^\mu + \sqrt{\lambda}(\tau_1-\tau_3) u_1^\mu + \sqrt{\lambda}(\nu_1 n_1^\mu -\nu_3 n_3^\mu)  \right)  + {\cal O}(\lambda)\nonumber\\
    &\sim Q\left( \lambda,\lambda,\sqrt{\lambda} \right),
\end{align}
where the last line follows from eq.~(\ref{xi_i_forward}), namely the loop momentum $k_1^\mu-k_3^\mu$ has both lightcone components of order ${\cal O}(\lambda)$, while the transverse components remain larger, of order ${\cal O}(\sqrt{\lambda})$. Hence this is a Glauber-mode momentum, as anticipated.

We note that none of the propagators in this graph carry a momentum with Glauber scaling, and yet one of the loops is a Glauber mode. The complete picture is that in the Regge-limit expansion two of the three independent loop momenta, say $k_1^{\mu}$ and $k_2^{\mu}$ are jet modes, while the third, $k_1^\mu-k_3^\mu$, is a Glauber mode.

Let us then use the above conclusions to determine the asymptotic behaviour of the graph associated with this region. Using the parameterisation of eq.~(\ref{eq:parameterisation_jet_like_k_i}) the integration measure of the two jet (collinear) loop momenta and the Glauber loop momentum, respectively, are given by:
\begin{align}
\label{eq:forward_scattering_momentum_integration_measure_scaling}
\begin{cases}
\text{for}\,\, k_i,\, i=1,2\, \text{(collinear)} & d\xi_i\ d\kappa_i d\tau_i d\nu_i \,\nu_i^{D-4} \\
& \qquad = (d\xi_i) (\lambda d\kappa_i)(\lambda^{\frac12} d\tau_i) (\lambda^{\frac12-\epsilon} d\nu_i \,\nu_i^{-2\eps}) \sim \lambda^{2-\epsilon}, \\
\text{for} \,\,  k_1-k_3\, \text{(Glauber)} & d(\xi_1-\xi_3)\ d(\kappa_1-\kappa_3)  d(\tau_1-\tau_3) d\nu_{13} \,\nu_{13}^{D-4} \\
& \qquad = (\lambda d\xi_{13}) (\lambda d\kappa_3)(\lambda^{\frac12} d\tau_3) (\lambda^{\frac12-\epsilon} d\nu_{13} \,\nu_{13}^{-2\eps}) \sim \lambda^{3-\epsilon},
\end{cases}
\end{align}
where $\xi_{13}\equiv \xi_1-\xi_3$ and  $\nu_{13}^2\equiv{-(\nu_1n_1-\nu_3 n_3)^2}$. Note that the sole difference between the collinear and Glauber scaling of the measure is due to the absence of a large momentum component in the Glauber case. 
Turning now to the integrand, since each propagator contributes $\mathcal{O}(\lambda^{-1})$, the integrand scale as $\mathcal{O}(\lambda^{-8})$, exactly as in eq.~(\ref{eq:Landshoff_scattering_momentum_integrand_scaling}). Therefore, using (\ref{eq:forward_scattering_momentum_integration_measure_scaling}) we obtain
\begin{eqnarray}
\label{eq:power_counting_Glauber_momentum_space_8props_forward}
    \mu[G_{\bullet\bullet},\text{Glauber}] = 2\cdot (2-\epsilon) +  ( 3-\epsilon) -8  = -1-3\epsilon,
\end{eqnarray}
which coincides with the result in parameter space, eq.~(\ref{eq:Gdotdot_power_counting_Glauber_parameter_space}).

Through the momentum space analysis we see that the Glauber region in the Regge limit is indeed related to -- and can be inferred from -- the Landshoff scattering region in general kinematics, by tuning the scattering angle $\theta$. Using the parameterisation in eq.~(\ref{eq:parameterisation_jet_like_k_i}), we have seen that the large components of the momenta $k_i$, which are $\xi_i = k_i\cdot \overline{v}_i$ for $i=1,2,3,4$, are equal up to $\mathcal{O}(\sqrt{\lambda}Q)$ terms for generic $\theta$. Meanwhile, the difference between $k_1$ and $k_3$, and hence between $k_2$ and $k_4$, are hard modes. 
In contrast, as $\theta\to 0$, $\xi_2$ and $\xi_4$ differ from $\xi_1$ and $\xi_3$ by the same large ($\mathcal{O}(\lambda^0)$) momenta, while $k_1-k_3$ (and hence $k_2-k_4$) admit  Glauber scaling. In summary, in this graph the Landshoff and Glauber regions can be regarded of the same origin.

\subsubsection{Example 2: \texorpdfstring{Mandelstam's graph $G_{tt}$}{TEXT}}
\label{section-example_10_propagator_graph}

With the same procedure, we shall now study the hidden regions of the graph $G_{tt}$. We refer to this graph as Mandelstam's graph, since it is the simplest of a class of graphs pointed out by Mandelstam as ones whose Regge asymptotic is driven by Regge cuts~\cite{Mdst63}.  Although this is not the focus of our investigation, this context gives another impetus to consider this particular example.

\begin{figure}[t]
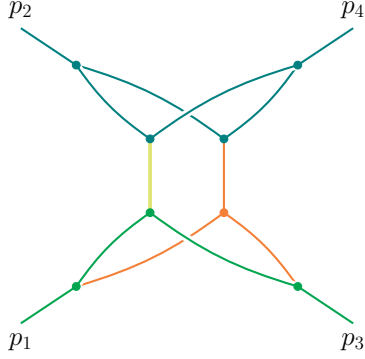

\centering
\include{figs/Gtt_facet_region_representative}
\vspace{-1cm}\caption{One of the facet regions of $G_{tt}$, which corresponds to the region vector $(-2,-3,-1,-1,-2,-3,-1,-1,-1,-3;1)$. In addition to the {\color{teal} \bf collinear} (to $p_2$) mode, this region also features a {\color{GreenYellow} \bf Glauber} mode, a {\color{Orange} \bf (collinear-1)$\cdot$soft} mode, and a {\color{Green} \bf (collinear-1)$^2$} mode, as coloured in the figure. The scaling law of all modes is given in eq.~(\ref{ReggeMods}). }
\label{figure-Gtt_facet_region_representative}
\end{figure}

\begin{figure}[t]
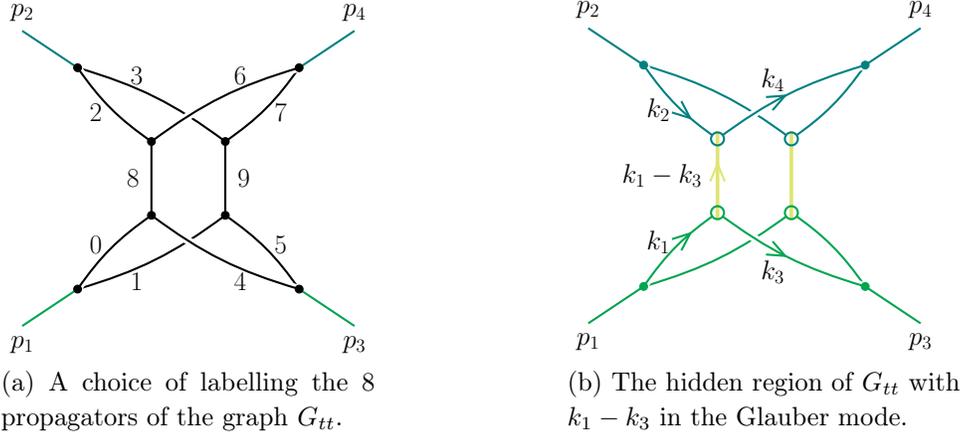

\centering
\hspace{-1em}
\begin{subfigure}[b]{0.32\textwidth}
\include{figs/Gtt_forward_scattering_parameterisation}
\vspace{-3em}
\caption{A choice of labelling the 8 propagators of the graph $G_{tt}$.}
\label{figure-Gtt_forward_scattering_parameterisation}
\end{subfigure}
\hspace{6em}
\begin{subfigure}[b]{0.32\textwidth}
\include{figs/Gtt_forward_scattering_hidden}
\vspace{-3em}
\captionsetup{width=1.05\linewidth}
\caption{The hidden region of $G_{tt}$ with $k_1-k_3$ in the Glauber mode.}
\label{figure-Gtt_forward_scattering_hidden}
\end{subfigure}
\caption{The three-loop nonplanar graph $G_{tt}$ (Mandelstam's graph) and its (unique) hidden region. There are two jets in this region, one associated with the external momenta~$p_1$ and $p_3$, while the other with $p_2$ and $p_4$. In the hidden region, among the ten propagators, eight are part of the jets (four in each) and the remaining two mediate in the $t$ channel between the jets, carrying Glauber momenta.}
\label{figure-Gtt_forward_scattering}
\end{figure}

Using the parameterisation in figure~\ref{figure-Gtt_forward_scattering_parameterisation}, $G_{tt}$ has the following Symanzik polynomials:
\begin{eqnarray}
\label{eq:10_propagator_forward_Symanzik_polynomials}
    &&\mathcal{U}(\x) = (x_0+x_1)(x_2+x_3)(x_4+x_5) + (x_0+x_1)(x_2+x_3)(x_6+x_7) \nonumber\\
    &&\qquad\quad + (x_0+x_1)(x_4+x_5)(x_6+x_7) + (x_2+x_3)(x_4+x_5)(x_6+x_7) \nonumber\\
    &&\qquad\quad + (x_0+x_1+x_4+x_5)(x_2+x_3+x_6+x_7)(x_8+x_9),\nonumber\\
    &&\mathcal{F}(\x;\s) = (-s_{12}) (x_0x_5 - x_1x_4)(x_2x_7 - x_3x_6) + (-s_{13}) \Big[(x_0x_3 - x_1x_2)(x_4x_7 - x_5x_6)\nonumber\\
    &&\qquad\quad +(x_0+x_1+x_4+x_5) (x_3x_7x_8+x_2x_6x_9) + (x_2+x_3+x_6+x_7) (x_1x_5x_8+x_0x_4x_9) \Big]. \nonumber\\
\end{eqnarray}
Here we have replaced $s_{23}$ by $-s_{12}-s_{13}$. The list of facet regions with their degrees of divergence, is given in table~\ref{tab:regions_forward_original_Gtt} of appendix~\ref{appendix-examples_regions_gtt_regge}.
From a detailed study of these region vectors~\cite{GrdHzgJnsMaprepare}, one can observe two following striking features. First, propagator momenta in the Glauber mode is already present in the facet regions. Second, new on-shell modes scaling as \emph{soft times collinear} and as \emph{collinear squared} show up. The complete set of modes for this graph in the Regge limit is:
\begin{align}
\label{ReggeMods}
\begin{cases}
\text{hard}: & k_H^\mu\sim Q(1,1,1);
\\
\text{collinear-1}: & k_{C_1}^\mu\sim Q(1,\lambda,\lambda^{1/2});\\
\text{collinear-2}: & k_{C_2}^\mu\sim Q(\lambda,1,\lambda^{1/2});\\
\text{Glauber}: & k_G^\mu\sim Q(\lambda,\lambda,\sqrt{\lambda});\\
\text{soft}: & k_S^\mu\sim Q(\lambda,\lambda,\lambda);\\
\text{collinear-1} \cdot \text{soft}:  \hspace*{30pt} 
& 
k_{C_1S}^\mu\sim Q(\lambda,\lambda^2,\lambda^{3/2});
\\
\text{collinear-2} \cdot \text{soft}: 
& 
k_{C_2S}^\mu\sim Q(\lambda^2,\lambda,\lambda^{3/2});
\\
\text{(collinear-1)}^2:  & 
k_{C_1C_1}^\mu\sim Q(1,\lambda^2,\lambda);
\\
\text{(collinear-2)}^2:   & 
k_{C_2C_2}^\mu\sim Q(\lambda^2,1,\lambda),
\end{cases}
\end{align}
where the corresponding scaling $u_{R,e}$ of the Lee-Pomeransky parameter~$x_e$ is readily given by  the inverse of the virtuality of the propagator according to the usual rule (\ref{x_scaling_rule}). As an example, consider the second facet region vector in table~\ref{tab:regions_forward_original_Gtt} of appendix~\ref{appendix-examples_regions_gtt_regge}, given by $(-2,-3,-1,-1,-2,-3,-1,-1,-1,-3;1)$. The corresponding graph representing this region vector is shown in figure~\ref{figure-Gtt_facet_region_representative}, which indeed features four different modes of those listed in~(\ref{ReggeMods}).
We point out that even more modes are relevant for the facet regions at higher loop orders. We will present an analysis of the full set of modes and regions in detail in future work~\cite{GrdHzgJnsMaprepare}. Our primary focus here is on the hidden region.

Motivated by the hidden region of $G_{\bullet\bullet}$, let us consider the case where all the $x_i$ are~$\mathcal{O}(\lambda^{-1})$,
\begin{eqnarray}
\label{eq:Gtt_Landshoff_scattering_region_vector}
    \v_\text{Glauber} = (-1,-1,-1,-1,-1,-1,-1,-1,-1,-1;1)\,.
\end{eqnarray}
We shall see that this is the only hidden region vector of $G_{tt}$ in the Regge limit. We named it $\v_\text{Glauber}$ as it is the only region of this graph featuring two Glauber propagators.

We first observe that the terms proportional to $s_{12}$ in the ${\cal F}$ polynomial, eq.~(\ref{eq:10_propagator_forward_Symanzik_polynomials}), are \emph{identical} to those in $G_{\bullet\bullet}$, eq.~(\ref{eq:8_propagator_forward_Symanzik_polynomials}). Hence, the region vector (\ref{eq:Gtt_Landshoff_scattering_region_vector}) would render the individual monomials involving $s_{12}$ superleading. As argued in the case of $G_{\bullet\bullet}$, to obtain a scaleful integral for this region vector, the combination
of terms proportional to $s_{12}$ must conspire to cancel at leading power in $\lambda$, i.e. it must admit the constraint in eq.~(\ref{eq:forward_superleading_terms_cancellation_scaling}).  Then \emph{all} the $\mathcal{U}$ and $\mathcal{F}$ terms contribute at ${\cal O}(\lambda^{-3})$ to the Lee-Pomeransky polynomial. 

Based on the knowledge of the hidden region of $G_{\bullet\bullet}$, it is also straightforward to deduce that the momentum configuration of the hidden region of $G_{tt}$ in the Regge limit is the one shown in figure~\ref{figure-Gtt_forward_scattering_hidden}. Each of the two jets $J_i$ ($i=1,2$) consists of four propagators  which are collinear to $p_i$, respectively. 
The two remaining propagators, $e_8$ and $e_9$ in figure~\ref{figure-Gtt_forward_scattering_parameterisation},  carry Glauber momenta, as shown in eq.~(\ref{eq:forward_scattering_Glauber_loop_momentum}). Using the usual correspondence~(\ref{x_scaling_rule}), this momentum-space picture is consistent with the region vector of eq.~(\ref{eq:Gtt_Landshoff_scattering_region_vector}), as both collinear and Glauber modes scale as $\lambda^{-1}$. 

Note that in figure~\ref{figure-Gtt_forward_scattering_hidden}, we have used the symbol \GlauberTextJetA\ (\GlauberTextJetB) to denote any jet-1 (jet-2) vertex with a Glauber momentum exchange in the $t$ channel. Similar to the hidden region of $G_{\bullet\bullet}$ (see figure~\ref{figure-Gdotdot_forward_scattering_hidden}), the two jets $J_1$ and $J_2$ share a loop through which they exchange Glauber momenta in the $t$ channel. The key difference is that here we have two Glauber propagators ($e_8$ and $e_9$), while no such propagator is present in the case of $G_{\bullet\bullet}$ in figure~\ref{figure-Gdotdot_forward_scattering_hidden}.

By dissecting the Newton polytope as in section~\ref{section-identifying_pinch_dissecting_polytope}, one can check that figure~\ref{figure-Gtt_forward_scattering_hidden} is the unique hidden region in $G_{tt}$.\footnote{In more detail, the vector $(-1,-1,-1,-1,-1,-1,-1,-1,-1,-1;1)$ arises from the lower facets of the new polytopes after dissection. Note that there are some additional new vectors from the lower facets (for example, the vector $(-2,-1,-2,-1,-2,-1,-2,-1,-2,-2;1)$ from table~\ref{tab:regions_forward_split3_4437}), but they are incompatible with a cancellation solution for arbitrary values of $s_{12}$, as they render the individual leading monomials in $\mathcal{F}$ of the same order as the leading monomials in $\mathcal{U}$.} We now compute its degree of divergence. In Lee-Pomeransky representation, we can apply the same change of variables in eq.~(\ref{eq:Landshoff_scattering_change_variables_repeat}). Then, similar to eq.~(\ref{eq:Gdotdot_forward_parameter_integration_measure_scaling}), the scaling of the integration measure is
\begin{eqnarray}
    \prod_{i=0}^9 dx_i &&= \bigg(\prod_{i=1,3,5,7,8,9}x_idx_i\bigg) \cdot dy_0dy_2dy_4dy_6\nonumber\\
    &&= \bigg(\prod_{i=1,3,5,7,8,9}x_idx_i\bigg) \cdot dy_0dy_2d(y_4-y_0)d(y_6-y_2) \sim \lambda^{-10}\cdot\lambda^{1} = \lambda^{-9}.
\label{eq:Gtt_forward_parameter_integration_measure_scaling}
\end{eqnarray}
Meanwhile, the scaling of the integrand follows eq.~(\ref{eq:Landshoff_scattering_parameter_integrand_scaling}) identically, and is $\sim \lambda^{6-3\epsilon}$. The corresponding degree of divergence is then
\begin{eqnarray}
\label{eq:Gtt_power_counting_Glauber_parameter_space}
    \mu[G_{tt},\text{Glauber}] = -9 + 6 - 3\epsilon = -3-3\epsilon.
\end{eqnarray}
In contrast to the case of $G_{\bullet\bullet}$, where the hidden region is the unique source of the leading contribution, here the hidden region of $G_{tt}$ is not anymore leading. Instead, as can be seen from table~\ref{tab:regions_forward_original_Gtt}, the leading regions are the four facet regions with collinear$\cdot$soft modes (see figure~\ref{figure-Gtt_facet_region_representative} as one example), with degree of divergence $-4-7\epsilon$, which is enhanced compared to $\mu[G_{tt},\text{Glauber}]$.

Let us rederive the result above in momentum space. As we have explained above and displayed in figure~\ref{figure-Gtt_forward_scattering_hidden}, eight of the propagators are collinear (four to each of the two jets), while the remaining two carry Glauber momentum. The integrand then behaves as
\begin{eqnarray}
    \frac{1}{((k_1-k_3)^2+i\varepsilon) ((k_2-k_4)^2+i\varepsilon)}\prod_{i=1}^4 \frac{1}{(k_i^2+i\varepsilon)((p_i-k_i)^2+i\varepsilon)}\sim \lambda^{-10}.
\label{eq:forward_scattering_10props_momentum_integrand_scaling}
\end{eqnarray}
Turning now to the integration measure, the computation is exactly the same as in $G_{\bullet\bullet}$ in eq.~(\ref{eq:forward_scattering_momentum_integration_measure_scaling}), i.e.~two collinear loops each counting as $\mathcal{O}(\lambda^{2-\epsilon})$, and one Glauber loop contributing $\mathcal{O}(\lambda^{3-\epsilon})$. Therefore, 
\begin{eqnarray}
\label{eq:power_counting_Glauber_momentum_space_10props_forward}
    \mu[G_{tt},\text{Glauber}] = 2(2-\epsilon)  + (3-\epsilon) -10  = -3-3\epsilon,
\end{eqnarray}
reproducing eq.~(\ref{eq:Gtt_power_counting_Glauber_parameter_space}). 

In the next section, we will use analytic results for the integrals $G_{\bullet\bullet}$ and $G_{tt}$ to identify the leading contributions in the Regge limit. We will be able to explicitly verify the conclusions we have reached above using the MoR regarding the leading and next-to-leading terms in the power expansion of these integrals, be it due to facet regions or hidden ones. 

\subsection{Asymptotic expansion of analytically known three-loop integrals}
\label{section-asymptotic_expansion_analytically_known_three_loop_integrals}

The availability of analytic results for some\footnote{Analytic results for four-point topologies at three loops are currently available for strictly on-shell integrals~\cite{Henn:2020lye,Bargiela:2021wuy}, planar integrals with one off-shell leg~\cite{DiVita:2014pza,Canko:2021xmn}, and for some nonplanar topologies with one off-shell~\cite{Syrrakos:2023mor,Henn:2023vbd}.
Complete results for the nonplanar integrals with off-shell external legs, considered in the previous section, are not yet available. In particular, the integrals $G_{tt}$ and $G_{\bullet \bullet}$ belong to integral family $1(d)$ of ref.~\cite{Henn:2020lye}, but with an off-shell leg; results for this topology are not provided in ref.~\cite{Henn:2023vbd}.
The integral $G_{st}$ belongs to integral family $1(h)$ of ref.~\cite{Henn:2020lye}, but with an off-shell leg. The $G_{tt}$ and $G_{st}$ integrals are discussed in detail in ref.~\cite{Syrrakos:2023mor}, where they are called NPL2\_8121 and NPL2\_16297, respectively, the $G_{\bullet \bullet}$ is a subsector of $G_{tt}$, it is shown in their figure 5.
In contrast, the on-shell integrals we are concerned with in the Regge-limit expansion are available, and will be used here.} of the integrals considered here provides us with a great opportunity to verify the results we obtained by asymptotic expansions. 

Master integrals for three-loop massless four-point topologies have been computed in refs.~\cite{Henn:2020lye,Bargiela:2021wuy}. In general, they are defined by 
\begin{equation}
    J_{\{n_i\}}=
   \int  \prod_{j=1}^3 \left(\frac{ e^{\gamma_E\epsilon}}{i\pi^{D/2} } \,d^{D}k_j  \right) \frac{1}{D_1^{n_1}\ldots D_{15}^{n_{15}}}\,,
\end{equation}
with the inverse massless propagators $D_j(q_j)=q_j^2+i\varepsilon$ with $q_j$ being the $j$-th element of the set
\begin{align}
\begin{split}
\Big\{k_1, k_2, k_3, k_1 - p_1, k_2 - p_1, k_3 - p_1, k_1 - p_1 - p_2, k_3 - p_1 - p_2, k_1 - k_2, k_2 - k_3, \\
k_1 - k_2 - p_3, k_2 - k_3 + p_1 + p_2 + p_3, k_2 + p_3, k_1 - k_3, k_2 - p_1 - p_2\Big\}.
\end{split}
\end{align}
In the notation of ref.~\cite{Bargiela:2021wuy}, the graph $G_{\bullet\bullet}$ is 
\[
J_{G_{\bullet\bullet}}(s_{12},s_{13};\epsilon)=\tt{{\rm INT}[``{\rm NPL2}", 8, 4009, 8, 0, {1, 0, 0, 1, 0, 1, 0, 1, 1, 1, 1, 1, 0, 0, 0}]},
\]  
while the graph $G_{tt}$ is 
\[
J_{G_{tt}}(s_{12},s_{13};\epsilon)=\tt{{\rm INT}[``{\rm NPL2}", 10, 8121, 10, 0, {1, 0, 0, 1, 1, 1, 0, 1, 1, 1, 1, 1, 1, 0, 0}]}.
\]
Below we shall examine the analytic structure of the expansions of $G_{\bullet\bullet}$ and $G_{tt}$ and compare it with our MoR analysis. We shall work exclusively in the physical region $s_{12}>0$ and $s_{13}<0$, which is compatible with the kinematic region defined in eq.~(\ref{Regge_limit}) above, in which the Regge-limit expansion is considered.

\subsubsection{Expansion of \texorpdfstring{$J_{G_{\bullet\bullet}}$}{TEXT}}
\label{section-expansion_Gdotdot_results}

We express $J_{G_{\bullet\bullet}}(s_{12},s_{13};\epsilon)$ in terms of a dimensionless integral $I_{G_{\bullet\bullet}}(x)$ 
with $x=-s_{13}/s_{12}$, namely,
\begin{equation}
J_{G_{\bullet\bullet}}(s_{12},s_{13};\epsilon)= s_{12}^{-2 - 3\epsilon} \, I_{G_{\bullet\bullet}}(x;\epsilon) = s_{12}^{-2 - 3\epsilon} \sum_{n=-4}^{\infty} I_{G_{\bullet\bullet}}^{(n)}(x)\, \epsilon^{n},
\end{equation}
where $I_{G_{\bullet\bullet}}^{(n)}(x)$ are defined as the Laurent coefficients in the $\epsilon\to 0$ expansion of $I_{G_{\bullet\bullet}}(x;\epsilon)$, which starts at ${\cal O}(\epsilon^{-4})$. 
Note that $I_{G_{\bullet \bullet}}=e^{3 \gamma_E \epsilon} \mathcal{I}_{G_{\bullet \bullet}}$, where $\mathcal{I}_{G_{\bullet \bullet}}$ was discussed in section~\ref{section-identifying_pinch_dissecting_polytope}.
The leading-order result reads
\begin{equation}
\label{IXXOminus4}
I_{G_{\bullet\bullet}}(x;\epsilon) = -\frac{8}{3} \left[\frac{ i\pi}{x(1-x)} + \frac{ \log x}{ 1-x} + \frac{ \log(1-x)}{x}\right]\frac{1}{\epsilon^4} +\cdots\,,
 \end{equation}
where we observe the symmetry $x\to 1-x$ which is part of the permutation symmetry of this integral.

Next, performing an expansion of the Laurent coefficients $I_{G_{\bullet\bullet}}^{(n)}(x)$ of the computed integral for $x\to 0$ we obtain, at each order $n$, an expansion in powers of $x$ accompanied by polynomials in logarithms $L_x=\log(x)$. For example, the leading order in $\epsilon$ in (\ref{IXXOminus4}) expands into 
\begin{equation}
    I_{G_{\bullet\bullet}}^{(-4)}(x)= - \frac{8 i {\pi}}{3 x} + \frac{8}{3} - \frac{8 L_x}{3} - \frac{8 i {\pi}}{3} + \left(\frac{4}{3} - \frac{8 L_x}{3} - \frac{8 i {\pi}}{3}\right) x + \left(\frac{8}{9} - \frac{8 L_x}{3} -    \frac{8 i {\pi}}{3}\right) x^2 +\cdots \,,
\end{equation}
which, of course, in this case, only contains $L_x$ to powers zero or one. Note that for any $n$, $I_{G_{\bullet\bullet}}^{(n)}(x)$ starts with $x^{-1}$ and we denote this the Leading Power (LP). Similarly, subleading corrections in the power expansion, $\sim x^{k}$, would be denoted by ${\rm N}^{k+1}{\rm LP}$. In general, we have
\begin{align}
\label{IbbAnalyticExpansion}
\begin{split}
I_{G_{\bullet\bullet}}(x;\epsilon)  &= \sum_{n=-4}^{\infty} I_{G_{\bullet\bullet}}^{(n)}(x)\, \epsilon^{n} = \sum_{n=-4}^{\infty} \sum_{k=-1}^{\infty} I_{G_{\bullet\bullet}}^{(n,k)}(L_x)\, x^{k}\, \epsilon^{n} \\
& = x^{-1} {\rm LP} \left\{I_{G_{\bullet\bullet}}\right\}(L_x;\epsilon) + x^{0} {\rm NLP} \left\{I_{G_{\bullet\bullet}}\right\}(L_x;\epsilon) + x^{1} {\rm NNLP} \left\{I_{G_{\bullet\bullet}}\right\}(L_x;\epsilon) +\cdots\,, 
\end{split}
\end{align}
where the functions ${\rm N}^{k+1}{\rm LP} \left\{I_{G_{\bullet\bullet}}\right\}(L_x;\epsilon)$ resum all powers of $\epsilon$ for fixed $k$.

The resummation of the $\epsilon$ expansion consists of a sum of distinct exponentials 
\[
\exp{\left\{-p_R \epsilon L_x\right\}}=x^{-p_R\epsilon}\,,
\]
where the exponents are linear in $\epsilon$, and the coefficients $p_R$, are non-negative\footnote{Generally $p_R$ can be integers of either sign, but in the case of on-shell four-point scattering it is expected that they are non-negative. See ref.~\cite{Henn:2020lye}.} integers which characterise the different regions, each multiplied by a function (a Laurent expansion) of~$\epsilon$. The special case $p_R=0$ is associated with the hard region, which is analytic, while $p_R\neq0$ corresponds to infrared regions, whose expansion involves logarithms of $x$.
In some cases (where dimensional regularisation cannot alone be used to compute individual regions) explicit ``bare logarithms'' which cannot be resummed into said exponentials remain in the function multiplying them. We shall see examples of this below. It should be stressed that a unique resummation based solely on the $\epsilon$ expansion coefficients would in principle require all-order information. However, in practice, given that the leading powers are only controlled by a small number of regions, plausible resummed expressions can be obtained even with a small number of expansion coefficients.   

Returning to the above example, using the known Laurent coefficients in $\epsilon$ through ${\cal O}(\epsilon^0)$, we find that the leading power (the coefficient of $x^{-1}$) in eq.~(\ref{IbbAnalyticExpansion}) can be resummed as follows:
\begin{align}
\label{IXX_LP}
\begin{split}
{\rm LP} \left\{I_{G_{\bullet\bullet}}\right\}(L_x;\epsilon) =& i \pi  e^{-3 \epsilon L_x} \Bigg(-\frac{8}{3 \epsilon^4}+\frac{16}{\epsilon^3}+\frac{2 \left(\pi ^2-144\right)}{3 \epsilon^2}-\frac{4 \left(-58 \zeta (3)+3 \pi ^2-432\right)}{3 \epsilon} \\
&+\frac{1}{60} \left(-27840 \zeta (3)+71 \pi ^4+1440 \pi ^2-207360\right) +\cdots \Bigg)\,,
\end{split}
\end{align}
where the factor $e^{-3 \epsilon L_x}$, which resums all logarithms, indicates that the origin of these terms is in some infrared region.
Note that the result is purely imaginary in physical kinematics where $x>0$.
In turn, the NLP (the coefficient of $x^0$) resums into
\begin{align}
\label{IXX_NLP}
\begin{split}
{\rm NLP} \left\{I_{G_{\bullet\bullet}}\right\}(L_x;\epsilon) =& -\frac{16}{9 \epsilon^5}-\frac{8  (i\pi -2 )}{3 \epsilon^4} +\frac{8 \left(-18+9 i \pi +5 \pi ^2\right)}{9\epsilon^3}\\
& +\frac{2  \left(232  \zeta (3)+216 -108 i\pi -60  \pi ^2 +21 i\pi ^3\right)}{9 \epsilon^2} +{\cal O}(\epsilon^{-1}) \\
& +e^{-3 \epsilon L_x}\,\Bigg[ \,L_x\, \Bigg(\frac{8 }{3 \epsilon^4}-\frac{2 \pi ^2 }{3 \epsilon^2}-\frac{232  \zeta (3)}{3 \epsilon} +{\cal O}(\epsilon^0) \Bigg) +\Bigg(\frac{16}{9 \epsilon^5}-\frac{8}{3 \epsilon^4} \\
& -\frac{8 \left(5 \pi ^2-9 i \pi \right)}{9 \epsilon^3}+\frac{2 \left(-340 \zeta (3)-216 i \pi +3 \pi ^2\right)}{9 \epsilon^2} +{\cal O}(\epsilon^{-1}) \Bigg)\Bigg]\,,
\end{split}
\end{align}
where again we used the $\epsilon$ expansion coefficient through finite terms. We note that the ${\cal O}(\epsilon^{-1})$ term in eq.~(\ref{IXX_NLP}) cannot be fixed based on the information available.
The first two lines in (\ref{IXX_NLP}) represent the hard region where no $\exp{\left\{-p_R \epsilon L_x\right\}}$ occurs (i.e. $p_R=0$), while the third and fourth lines represent one (or more) infrared regions characterised by $e^{-3 \epsilon L_x}$, just as in (\ref{IXX_LP}) at leading power.

The key conclusion we draw from the analytic results in eqs.~(\ref{IbbAnalyticExpansion})-(\ref{IXX_NLP}),
is that the 
asymptotic behaviour of $G_{\bullet\bullet}$ in the Regge limit is given by
\begin{equation}
    I_{G_{\bullet\bullet}}(x;\epsilon)\sim x^{-1-3\epsilon}\quad\qquad \text{for}\quad x\to 0\,,
\end{equation}
in agreement with the MoR predictions and power counting analysis, which we performed in section~\ref{section-hidden_regions_Glauber_mode} in both parameter space and momentum space, leading to eqs.~(\ref{eq:Gdotdot_power_counting_Glauber_parameter_space}) and~(\ref{eq:power_counting_Glauber_momentum_space_8props_forward}), respectively. Recall that tables~\ref{tab:regions_forward_original} and~\ref{tab:regions_forward_split1} demonstrate that the hidden (Glauber) region alone is responsible for this leading asymptotic behaviour of $G_{\bullet\bullet}$ in the Regge limit, while all facet regions are subleading, and that the hidden region is exposed by dissecting the polytope as proposed. The analytic result thus provides a final confirmation that the MoR has been successfully used to determine the asymptotic behaviour of the integral.    

\subsubsection{Expansion of \texorpdfstring{$J_{G_{tt}}$}{TEXT}}
\label{section-expansion_Gtt_results}

We proceed to analyse the Regge limit of the analytic results~\cite{Henn:2020lye,Bargiela:2021wuy} for the second diagram, $G_{tt}$, which we studied using the MoR in section~\ref{section-example_10_propagator_graph} above. We express $J_{G_{tt}}(s_{12},s_{13};\epsilon)$ in terms of a dimensionless integral $I_{G_{tt}}(x)$ with $x=-s_{13}/s_{12}$:
\begin{equation}
    J_{G_{tt}}(s_{12},s_{13};\epsilon) =s_{12}^{-4-3\epsilon} I_{G_{tt}}(x) = s_{12}^{-4-3\epsilon} \sum_{n=-6}^{\infty} I_{G_{tt}}^{(n)}(x)\, \epsilon^{n}\,.
\end{equation}
The first couple of orders in the $\epsilon$ expansion read:
\begin{equation}
\label{Mand_eps_leading_terms}
    I_{G_{tt}}(x)= \frac{-16\epsilon^{-6}}{9 (1 - x) x^2} - \frac{2\epsilon^{-5}}{(1-x)x^2} \left( \frac{i\pi}{3} + \frac{1 - x + 8 x^2}{5x^2} + \frac{2 \log(x)}{3} - \frac{5 \log(1-x)}{3} \right) +\cdots\,.
\end{equation}
Resumming the $\epsilon$ expansion for the first three terms in the power expansion, using the information available from the $\epsilon$ expansion through finite terms, we find:
\begin{align}
\label{Mand_power_expansion}
\begin{split}
I_{G_{tt}}(x;\epsilon)  &=\sum_{n=-6}^{\infty} I_{G_{tt}}^{(n)}(x)\, \epsilon^{n} = \sum_{n=-4}^{\infty} \sum_{k=-4}^{\infty} I_{G_{tt}}^{(n,k)} (L_x)\, x^{k}\, \epsilon^{n} \\
&  = x^{-4} {\rm LP} \left\{I_{G_{tt}}\right\}(L_x;\epsilon) +x^{-3} {\rm NLP} \left\{I_{G_{tt}}\right\}(L_x;\epsilon) + x^{-2} {\rm NNLP} \left\{I_{G_{tt}}\right\}(L_x;\epsilon) +\cdots
\end{split}
\end{align}
where the leading power term contains a single scaling:
\begin{align}
\begin{split}
{\rm LP} \left\{I_{G_{tt}}\right\}(L_x;\epsilon) =\,&e^{-7\epsilon L_x}\bigg( -\frac{2}{5 \epsilon^5}+\frac{6 (1-4 i \pi )}{5 \epsilon^4}+\frac{-108+432 i \pi +779 \pi ^2}{30 \epsilon^3} \\
&+\frac{116 \zeta (3)+108-432 i \pi -779 \pi ^2+812 i \pi ^3}{10 \epsilon^2}+\cdots \bigg)\,,
\end{split}
\end{align}
while at NLP we encounter three different scalings,
\begin{align}
\label{MandlNLP}
\begin{split}
{\rm NLP} \left\{I_{G_{tt}}\right\}(L_x;\epsilon)&= e^{-7\epsilon L_x}\left(\frac{4}{5} \frac{1}{\epsilon^{4}} -\frac{12(1-4\pi i)}{5\epsilon^{3}} +\cdots\right) +e^{-4\epsilon L_x} \left(\frac{4\pi i}{\epsilon^{3}} +\cdots\right)\\
&+e^{-3\epsilon L_x} \left(\frac{-20\pi i}{\epsilon^{4}} + \frac{8(-9 i\pi  + 135 \pi^2 + 8 i \pi^3)}{9 \epsilon^3} +\cdots\right)\,,
\end{split}
\end{align}
where the third, $e^{-3\epsilon L_x}$, does not appear as a facet of the original polytope, but does appear after dissecting the polytope, and is consistent with the predicted behaviour of the hidden (Glauber) region according to eqs.~(\ref{eq:Gtt_power_counting_Glauber_parameter_space}) and~(\ref{eq:power_counting_Glauber_momentum_space_10props_forward}).
 
At NNLP the analytic result displays four different scalings,
\begin{align}
\begin{split}
{\rm NNLP} \left\{I_{G_{tt}}\right\}(L_x;\epsilon) =\,&e^{-7\epsilon L_x} \left(-\frac{2}{45}\frac{1}{\epsilon^{6}} + \frac{1}{\epsilon^{5}}\left(-\frac{17}{45} - \frac{8 i\pi}{15}\right)+\cdots\right) \\
& +e^{-4\epsilon L_x} \left(\frac{16}{9}\frac{1}{\epsilon^{6}}  + 8\frac{1}{\epsilon^{5}}  \left(11 + 15 i\pi\right)+\cdots \right) \\
& +e^{-3\epsilon L_x}\left( \frac{638}{9}\frac{1}{\epsilon^{6}} +\frac{1}{3} \frac{1}{\epsilon^{5}}\left(903 + 1108 i\pi + 112 L_x\right)+\cdots\right) \\
& +e^{-2\epsilon L_x}\left(-\frac{372}{5}\frac{1}{\epsilon^{6}}  + \frac{2}{5} \frac{1}{\epsilon^{5}}(-784 - 957 i\pi + 80 L_x)+\cdots\right),
\end{split}
\end{align}
where the explicit non-exponentiated $L_x$ term in the series multiplying both $e^{-3\epsilon L_x}$ and $e^{-2\epsilon L_x}$ terms is indicative of the necessity of analytic regulators in computing the contributions of the relevant regions separately.

The main conclusion is that, once again, we find complete agreement 
with the MoR predictions based on  the power counting analysis and the dissection of the polytope reported 
in section~\ref{section-example_10_propagator_graph} and
in tables \ref{tab:regions_forward_original_Gtt} and~\ref{tab:regions_forward_split3_4437}. 

We note that the hidden region in both $G_{\bullet\bullet}$ (the sole contribution to the leading power, eq.~(\ref{IXX_LP})) and $G_{tt}$ (the $e^{-3\epsilon L_x}$ term at NLP in eq.~(\ref{MandlNLP})) involve at least one power of $i\pi$. This is consistent with the expectation: recall that loops producing a factor of $i\pi$ is a hallmark of a Glauber mode.

\subsection{Hidden regions in other three-loop graphs}
\label{section-further_discussion}

Let us now consider the full set of three-loop graphs in figure~\ref{figure-possibly_hidden_singular_graphs}, which have been identified as potentially featuring a hidden region. It turns out that the analysis of such regions in the Regge limit closely follows what we have done in section~\ref{section-hidden_regions_Glauber_mode} with $G_{\bullet\bullet}$ and $G_{tt}$, and we therefore briefly summarise the results.

\begin{figure}[t]
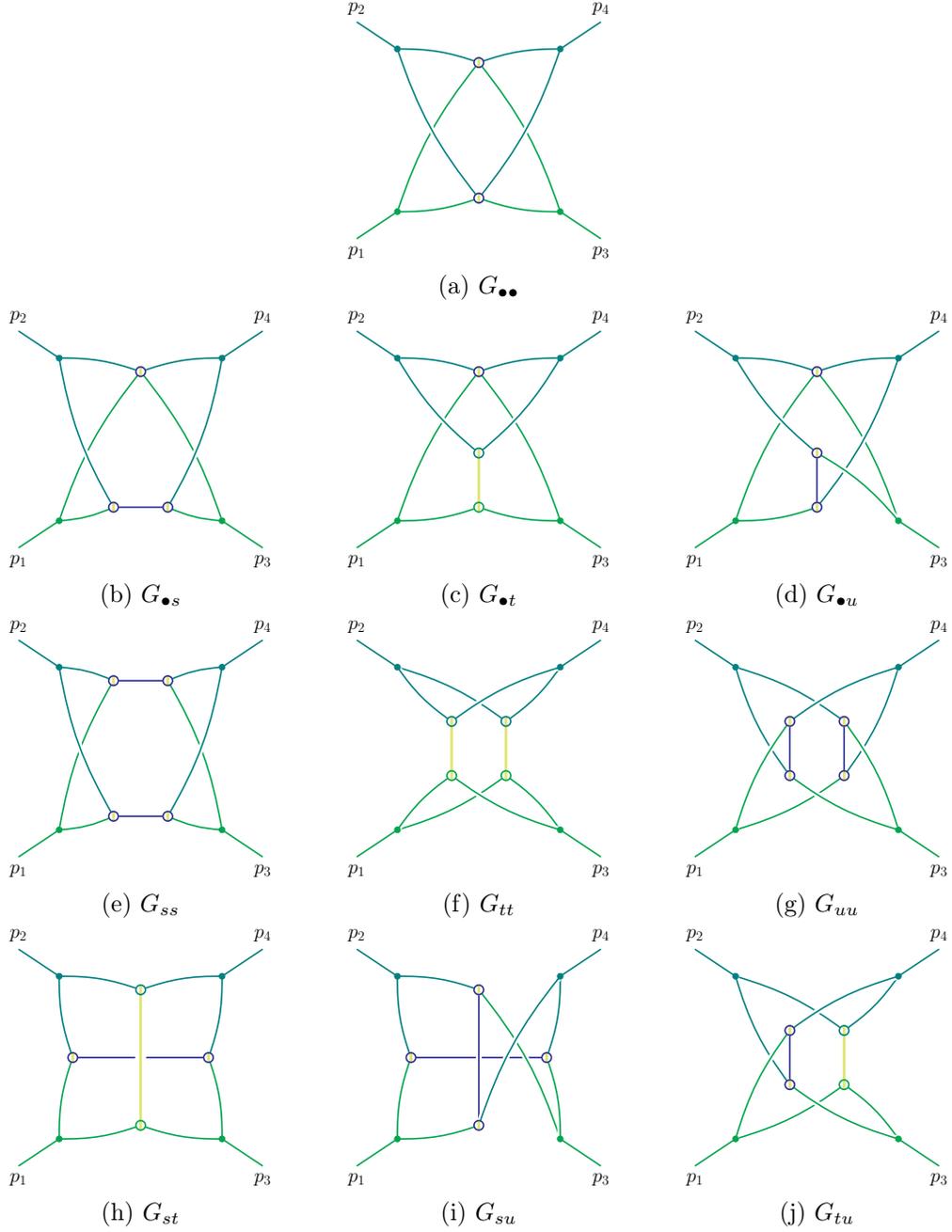

\centering
\begin{subfigure}[b]{0.25\textwidth}
\centering
\include{figs/possibly_hidden_singular_graph_dot_dot_forward}
\vspace{-3em}\caption{$G_{\bullet \bullet}$}
\label{possibly_hidden_singular_graph_dot_dot_forward}
\end{subfigure}
\\
\begin{subfigure}[b]{0.25\textwidth}
\centering
\include{figs/possibly_hidden_singular_graph_dot_s_forward}
\vspace{-3em}\caption{$G_{\bullet s}$}
\label{possibly_hidden_singular_graph_dot_s_forward}
\end{subfigure}
\qquad
\begin{subfigure}[b]{0.25\textwidth}
\centering
\include{figs/possibly_hidden_singular_graph_dot_t_forward}
\vspace{-3em}\caption{$G_{\bullet t}$}
\label{possibly_hidden_singular_graph_dot_t_forward}
\end{subfigure}
\qquad
\begin{subfigure}[b]{0.25\textwidth}
\centering
\include{figs/possibly_hidden_singular_graph_dot_u_forward}
\vspace{-3em}\caption{$G_{\bullet u}$}
\label{possibly_hidden_singular_graph_dot_u_forward}
\end{subfigure}
\\
\begin{subfigure}[b]{0.25\textwidth}
\centering
\include{figs/possibly_hidden_singular_graph_s_s_forward}
\vspace{-3em}\caption{$G_{s s}$}
\label{possibly_hidden_singular_graph_s_s_forward}
\end{subfigure}
\qquad
\begin{subfigure}[b]{0.25\textwidth}
\centering
\include{figs/possibly_hidden_singular_graph_t_t_forward}
\vspace{-3em}\caption{$G_{t t}$}
\label{possibly_hidden_singular_graph_t_t_forward}
\end{subfigure}
\qquad
\begin{subfigure}[b]{0.25\textwidth}
\centering
\include{figs/possibly_hidden_singular_graph_u_u_forward}
\vspace{-3em}\caption{$G_{u u}$}
\label{possibly_hidden_singular_graph_u_u_forward}
\end{subfigure}
\\
\begin{subfigure}[b]{0.25\textwidth}
\centering
\include{figs/possibly_hidden_singular_graph_s_t_forward}
\vspace{-3em}\caption{$G_{s t}$}
\label{possibly_hidden_singular_graph_s_t_forward}
\end{subfigure}
\qquad
\begin{subfigure}[b]{0.25\textwidth}
\centering
\include{figs/possibly_hidden_singular_graph_s_u_forward}
\vspace{-3em}\caption{$G_{s u}$}
\label{possibly_hidden_singular_graph_s_u_forward}
\end{subfigure}
\qquad
\begin{subfigure}[b]{0.25\textwidth}
\centering
\include{figs/possibly_hidden_singular_graph_t_u_forward}
\vspace{-3em}\caption{$G_{t u}$}
\label{possibly_hidden_singular_graph_t_u_forward}
\end{subfigure}
\caption{The hidden regions corresponding to the graphs in figure~\ref{figure-possibly_hidden_singular_graphs}, each featuring the exchange of a Glauber loop momentum between the upper and lower jets.}
\label{figure-possibly_hidden_singular_graphs_forward}
\end{figure}

We find that each of the graphs in figure~\ref{figure-possibly_hidden_singular_graphs} has a single hidden region. Recall that in the wide-angle scattering context these hidden regions were all related to Landshoff scattering, as described by figure~\ref{figure-possibly_hidden_singular_graphs_wideangle}. Similarly, in the Regge limit these hidden regions are all characterised by single loop involving Glauber scaling, and they are shown in figure~\ref{figure-possibly_hidden_singular_graphs_forward}. 
As can be read from the figure, each of these regions features a lower jet $J_{1}$ (consisting of the incoming $p_1$, the outgoing $p_3$ and a few collinear propagators) and an upper jet $J_{2}$ (consisting of the incoming $p_2$, the outgoing $p_4$ and a few propagators) and possibly a hard subgraph. A key feature is that Glauber momentum is exchanged between $J_{1}$ and $J_{2}$ at the vertices \GlauberTextHard\ and/or \GlauberTextJetA\ (\GlauberTextJetB). For the graphs $G_{\bullet t}$, $G_{tt}$, $G_{st}$, and $G_{tu}$, there are also propagators carrying Glauber momenta.

To identify the hidden region(s) for each graph, we have dissected the original Newton polytope, as described by section~\ref{section-identifying_pinch_dissecting_polytope}, and examined the lower facets of the new polytopes in each sector. For example, in addition to the 10 facet region vectors of $G_{\bullet s}$ (see table~\ref{table-forward_facet_hidden_vectors_Gdots}), there is a unique new vector arising from the facets of the new polytopes, which is $(-1,-1,-1,-1,-1,-1,-1,-1,0;1)$. This is exactly the aforementioned hidden region, which describes the Glauber exchange between $J_{1}$ and $J_{2}$. 
\begin{table}[htb]
\begin{center}
\begin{tabular}{ |c||c|c|c| } 
\hline
 & \multirow{2}{8em}{\quad$\boldsymbol{v}\ (x_0,\dots,x_8,\lambda)$} & \multirow{2}{6.2em}{region or not?} \\
 & & \\
\hline
\multirow{10}{4em}{$\begin{matrix} \text{original}\\ \text{polytope} \end{matrix}$} & $(-2,-1,-2,-1,-2,-1,-2,-1,-2;1)$ & $\greencheckmark[ForestGreen]$ \\ 
 & $(-2,-1,-2,-1,-1,-1,-1,0,-2;1)$ & $\greencheckmark[ForestGreen]$ \\
 & $(-2,-1,-2,-1,-1,0,-1,-1,-2;1)$ & $\greencheckmark[ForestGreen]$ \\
 & $(-1,-2,-1,-2,-1,-2,-1,-2,0;1)$ & $\greencheckmark[ForestGreen]$ \\
 & $(-1,-1,-1,0,-2,-1,-2,-1,-2;1)$ & $\greencheckmark[ForestGreen]$ \\
 & $(-1,-1,-1,0,-1,0,-1,0,-1;1)$ & $\greencheckmark[ForestGreen]$ \\
 & $(-1,-1,-1,0,0,0,0,0,-1;1)$ & $\greencheckmark[ForestGreen]$ \\
 & $(-1,-1,0,-1,0,-1,0,-1,0;1)$ & $\greencheckmark[ForestGreen]$ \\
 & $(-1,-1,0,0,0,0,0,0,0;1)$ & $\greencheckmark[ForestGreen]$ \\
 & $(0,0,0,0,0,0,0,0,0;1)$ & $\greencheckmark[ForestGreen]$ \\
\hline
\multirow{2}{4em}{$\begin{matrix} \text{dissected}\\ \text{polytopes} \end{matrix}$} & \multirow{2}{16em}{$(-1,-1,-1,-1,-1,-1,-1,-1,0;1)$} & \multirow{2}{1em}{$\greencheckmark[ForestGreen]$} \\
 & & \\
\hline
\end{tabular}
\end{center}
\vspace{-1em}\caption{The vectors that are possibly relevant for the regions in the Regge-limit expansion of $G_{\bullet s}$. Each vector is written in the format $(x_0,\dots,x_8)$. The first group consists of 10 vectors, each corresponding to the lower facets of the original polytope. In contrast, the second group consists of only one vector $(-1,-1,-1,-1,-1,-1,-1,-1,0;1)$, which is exposed as a region via lower facets of the dissected polytope. The associated hidden region features a Glauber exchange between the upper and lower jets, as depicted in figure~\ref{possibly_hidden_singular_graph_dot_s_forward}.
The right column confirms that these vectors do correspond to regions: in contrast to table~\ref{table-wideangle_facet_hidden_vectors_Gdots} where spurious regions arise in the dissected polytope (and subsequently discarded) no such vectors appear here.}
\label{table-forward_facet_hidden_vectors_Gdots}
\end{table}

We emphasise that the Glauber mode does not feature only in hidden regions. As we have discussed above, the vector $(-2,-3,-1,-1,-2,-3,-1,-1,-1,-3;1)$, which corresponds to a facet region of $G_{tt}$ in the Regge-limit expansion, has one line momentum in the Glauber mode (see figure~\ref{figure-Gtt_facet_region_representative}). The rather rich mode and region structure of the Regge limit at three loops (see (\ref{ReggeMods})) and beyond will be investigated in more detail in a forthcoming publication~\cite{GrdHzgJnsMaprepare}.

\section{Conclusions and outlook}
\label{section-conclusion_outlook}

Recognising the value of parametric representations of Feynman integrals and their geometrical interpretation in treating singularities as endpoint divergences, in this paper we focused on the exceptions: singularities which manifest themselves as pinches in parameter space, and obstruct the application of existing strategies based on the Newton polytope. Earlier work in this direction~\cite{JtzSmnSmn12,AnthnrySkrRmn19} in the context of the MoR focused on linear cancellation between pairs of Mandelstam invariants, while our analysis identified more general cancellation patterns, including higher-degree polynomials involving multiple Feynman parameters.
While the former, linear, cancellation occurs in special kinematic limits, such as the threshold expansion and the forward limit, giving rise, respectively, to potential and Glauber modes --- both off-shell modes --- we have shown that the latter also impact integrals in wide-angle scattering, in which case no off-shell modes arise, namely all loop momenta and propagators are still either hard, collinear or soft. The main distinctive feature of these so-called \emph{Landshoff regions} is the presence of multiple hard connected subgraphs, which cannot occur in regions corresponding to facets of the Newton polytope in the on-shell expansion~\cite{GrdHzgJnsMaSchlk22,Ma23}. In the Regge limit, these new cancellation patterns give rise to Glauber modes.

The first problem we addressed in this paper is the identification of graphs in which a pinch in parameter space can occur, which could hinder the Newton-polytope based numerical evaluation or the determination of a complete set of regions in the MoR.
Our key tool has been the Landau equations, but since these are notoriously difficult to solve in general, we merely aimed at identifying a necessary condition for such a solution to exist within the domain of integration. The condition we identified is a requirement for cancellation to take place between monomials of opposite signs in the ${\cal F}$ polynomial (see section~\ref{section-finding_diagrams_prone_to_hidden_landau_singularities}). 
Based on this, we set up an algorithm to distinguish between graphs in which the pattern of signs of monomials associated with a given Mandelstam invariant in the ${\cal F}$ polynomial facilitates a pinch solution, from graphs where such a solution is excluded at the outset. With this we have been able to show that pinch singularities in massless $2\to 2$ scattering appear first at three loops, and are restricted at this order to a very special class of graphs collected in figure~\ref{figure-possibly_hidden_singular_graphs}. 
These graphs ultimately originate in a single topology, $G_{\bullet\bullet}$. Running the same algorithm at four loops gives rise to a larger set of 1097 graphs. However, almost all of them (1081 in total) contain the aforementioned three-loop topologies as subgraphs, while the remaining ones (see figure~\ref{figure-four_loop_possibly_hidden_singular_graph_85774} for example) do not actually admit pinches in wide-angle scattering. Some still have such singularities in the Regge limit. 

The second problem we addressed was the numerical computation of integrals which feature a pinch in parameter space within the domain of integration. While straightforward application of existing sector decomposition tools to integrals such as those in figure~\ref{figure-possibly_hidden_singular_graphs} is doomed to fail due to said pinch singularity, we devised a procedure by which they can be evaluated. 
The key steps are first, identifying the singularity as one generated by cancellation between monomials of opposite signs, then linearising the polynomial around the singularity, and finally, separating the integration domain at the singularity, so as to map it to the boundary of the integration domain. This amounts to dissecting the Newton polytope and performing the integration in each sector independently, with the key advantage being that the previous cancellation responsible for the singularity has been eliminated. In section~\ref{section-numerically_evaluating_dissected_integral} this technique was successfully applied to the $G_{\bullet\bullet}$ integral.

The third problem we addressed in this paper is the determination of a complete set of regions needed for asymptotic expansions of integrals around a limit in which a pinch in parameter space occurs. The key example has been the on-shell expansion of $G_{\bullet\bullet}$,  starting with a configuration in which (at least) one of the legs is off shell. We have shown that in such an on-shell expansion, the presence of a pinch singularity at the limit, induces a hidden region, namely a region which cannot be detected as a facet of the Newton polytope. We have shown that the method above, based on dissecting the polytope and considering each sector separately, exposes the hidden regions and allows us to obtain a reliable asymptotic expansion. 
By performing a careful power counting analysis in both parameter space and momentum space, we determined the asymptotic behaviour of the hidden region of $G_{\bullet\bullet}$. We have found that this region uniquely governs the leading asymptotic behaviour of the integral, and we provided robust numerical evidence (figure~\ref{figure-numeric_evidences}) that this region alone reproduces the full integral near the on-shell limit.

We then analysed the remaining graphs in figure~\ref{figure-possibly_hidden_singular_graphs} in the context of the on-shell expansion in wide-angle scattering.
The geometric approach and the momentum space one nicely complement each other, leading to a clear physical picture of Landshoff scattering in this class of graphs.
We demonstrated that each of these graphs features a single hidden region, shown in figure~\ref{figure-possibly_hidden_singular_graphs_wideangle}, which is related to the hidden region of $G_{\bullet\bullet}$ by simply contracting the additional hard propagators. 

We further applied the newly developed set of tools to the Regge limit in $2\to 2$ scattering.
In section~\ref{section-hidden_regions_Glauber_mode} we analysed in detail the hidden region of two integrals in the Regge limit, $G_{\bullet\bullet}$ and $G_{tt}$. We found that the asymptotic behaviour of $G_{\bullet\bullet}$ is dominated by the hidden region, while all facet regions are suppressed by at least one power of $x\equiv -t/s$. In turn, for $G_{tt}$ the hidden region only contributes at next-to-leading power. 
As in the wide-angle case, we derived the asymptotic behaviour of the hidden region in both parameter space and momentum space. Our conclusions agree with the results of earlier work on the asymptotic behaviour of high-energy processes; see~\cite{Cls77book,Pkhn80book}.
One of the important new insights we gained in our analysis of the hidden region, was the relation between the Regge limit and wide-angle scattering. In particular, also in the Regge limit each of the 10 three-loop graphs of figure~\ref{figure-possibly_hidden_singular_graphs} features a single hidden region. In the Regge limit case, the latter always involves a single Glauber loop momentum and two jet loop momenta, as shown in figure~\ref{figure-possibly_hidden_singular_graphs_forward}.
Finally, in section~\ref{section-asymptotic_expansion_analytically_known_three_loop_integrals} we gained further insight regarding the asymptotic behaviour of $G_{\bullet\bullet}$ and $G_{tt}$ and the contribution it receives from the hidden and the facet regions, using analytic results for these integrals~\cite{Henn:2020lye,Bargiela:2021wuy}, which have been obtained through ${\cal O}(\epsilon^0)$. By expanding these results in powers of $x$, and fitting the log-power expansion to an ansatz which encodes the non-analytic behaviour associated with each infrared region, $\exp{\left\{-p_R \epsilon L_x\right\}}=x^{-p_R\epsilon}$, we were able to resum the logarithms $L_x=\log(x)$ at each power, and confirm our conclusions regarding the order at which each region enters the power expansion. 

\bigbreak
Let us now turn to a broader picture and provide some comments regarding the interpretation of our results and some outlook. Our work provides an analysis of pinch singularities in parameter space for massless $2\to 2$ scattering processes, considering both wide-angle and the Regge limit scenarios. We focused primarily on three-loop integrals, where such singularities first occur. An intriguing question arises regarding the behaviour of these pinch singularities in higher-loop integrals and ones with additional external momenta. One may speculate whether for wide-angle scattering \emph{all} pinch singularities conform to the Landshoff scattering picture of figure~\ref{figure-wide_angle_scattering_hidden_region}. At three loops we saw explicitly that all graphs having such a pinch singularity readily conform to this structure, having precisely two independent hard scattering subgraphs (figure~\ref{figure-possibly_hidden_singular_graphs_wideangle}). 
Preliminary evidence that this picture extends beyond three loops comes from those four-loop graphs prone to having a pinch in parameter space (see section~\ref{section-finding_diagrams_prone_to_hidden_landau_singularities}): each of these can be obtained by adding an extra loop to one of the three-loop graphs of figure~\ref{figure-possibly_hidden_singular_graphs}, which is consistent with the pinch singularity corresponding to a four-loop Landshoff scattering picture. Further detailed analyses are required to validate this conjecture.

In the on-shell expansion in wide-angle scattering, only the three basic types of modes, soft, collinear and hard have so far been observed. In the case of facet regions we formulated in ref.~\cite{GrdHzgJnsMaSchlk22} a simple diagrammatic algorithm to identify all regions. It would clearly be of great interest to address the question whether the three basic modes are also sufficient to characterise all hidden regions at higher loop orders and in graphs with more external legs. Ultimately, we would like to extend said diagrammatic algorithm to describe hidden regions as well.

Our findings for the $2\to 2$ Regge limit in this paper may provide insights into the infrared structure in this limit. First, for the hidden regions identified in this work at three loops, we observed an interesting relation, in both momentum space and parameter space, between Landshoff scattering in the wide-angle case, and Glauber exchange in the Regge limit. More generally, one may expect that hidden regions would be more common in the Regge limit as compared to wide-angle scattering, as fewer Landau equations need to be simultaneously satisfied. Some evidence for this was observed at four loops upon analysing the graph $\mathfrak{G}'$ in figure~\ref{figure-four_loop_possibly_hidden_singular_graph_85774}.

Generally, the region structure in the Regge limit is rather complex. In particular, it involves a much richer set of modes as compared to the on-shell expansion. This is so for both the hidden region, which involves a Glauber mode, and for facet regions, which involve both Glauber modes and additional new modes, as is evident from the $G_{tt}$ example (see eq.~(\ref{ReggeMods})). 
These findings may be of direct relevance to the formulation of effective field theories such as Glauber SCET~\cite{RstStw16}, and for the analysis of the high-energy limit of QCD amplitudes.
We recall that Mandelstam's graph, $G_{tt}$, analysed here, is the simplest in a family of diagrams that was shown
(by Stanley Mandelstam)~\cite{Mdst63} to possess Regge cuts. More complete understanding of the analytic structure of this class of integrals may shed light on the Regge limit, a topic of recent renewed interest in perturbative QCD~\cite{Gao:2024qsg,Moult:2022lfy,Milloy:2022fns,Falcioni:2021dgr,Caola:2021izf,Fadin:2023aen,Fadin:2021csi}. It would also be of great interest to develop a purely graph-theoretical algorithm to identify regions in the Regge limit, extending the wide-angle on-shell expansion algorithm of ref.~\cite{GrdHzgJnsMaSchlk22}. The challenge is of course to capture not only the facet regions but also the hidden ones.

\acknowledgments

We would like to thank Charalampos Anastasiou, Thomas Becher, Holmfridur Hannesdottir, Andrew McLeod, Erik Panzer, Johannes Schlenk, and George Sterman for useful discussions. In particular, we thank Piotr Bargiela for his valuable help with the analytic three-loop four-point results. 

EG and FH are supported by the STFC Consolidated Grant ``Particle Physics at the Higgs Centre''. FH is also supported by the UKRI FLF grant ``Forest Formulas for the LHC'' (Mr/S03479x/1). SJ is supported by the Royal Society University Research Fellowship (URF/R1/201268) and by the UK Science and Technology Facilities Council under contract ST/T001011/1. YM is supported by the Swiss National Science Foundation through its project funding scheme, grant number 10001706.

\appendix
\section{Region vectors for some presented examples}
\label{appendix-examples_regions}

In this appendix, we summarise the results obtained using the pySecDec Expansion by Regions code, both using the original Newton polytope and the Newton polytopes obtained by dissecting the original integral. We shall see that the hidden regions, which are absent in the original polytope, are discovered by dissecting the polytope. When reporting the results of the dissected polytope, we shall include, in two columns, both the result in terms of the new variables, defined by eq.~(\ref{eq:change_of_variables_three_loop_hidden}), in which the geometric expansion-by-region routine has been operating, and the vectors obtained after scaling back to the original set of variables. 

We provide examples of both the on-shell expansion in wide-angle scattering 
(for simplicity, we used $p_1^2 \sim \lambda Q^2$ and $p_i^2=0\ (i=2,3,4)$, where the total number of regions is small compared to the case where all four external lines start off shell)
and the Regge-limit expansion.

Tables~\ref{tab:regions_forward_original}, \ref{tab:regions_onshell_original_3902}, \ref{tab:regions_forward_original_3902}, and \ref{tab:regions_forward_original_Gtt} show the vectors from the original Newton polytope, which are in the form of ($x_0, \ldots, x_N; \lambda$), with $x_e$ the Lee-Pomeransky parameter associated with edge $e$, and the last entry $\lambda$ representing the normalisation of the vector.

Tables~\ref{tab:regions_forward_split1}, \ref{tab:regions_onshell_split1_3902}, \ref{tab:regions_forward_split3_3902}, and \ref{tab:regions_forward_split3_4437} demonstrate how hidden region vectors can be obtained from the new polytopes. In each table we show a particular dissection (one of 24 obtained via a shift of the form of eq.~(\ref{eq:division_sector_three_loop_hidden})) and for each integral dissection, the region vector $\v_R$ is written in terms of the new parameters $\{y_e\}$ in the first column, and translated back to the original parameters $\{x_e\}$ in the second column. The parameters are related via eq.~(\ref{eq:Landshoff_scattering_change_variables}).

\subsection{Regge-limit expansion of \texorpdfstring{$G_{\bullet \bullet}$}{TEXT}}
\label{appendix-examples_regions_gbb_regge}

\begin{table}[H]
    \small
    \centering
    \begin{tabular}{l|l}
        \bf $\v_\mathrm{R}$ ($x_0, x_1, \ldots, x_7 ;\lambda$)      & \bf degree of divergence \\\hline
        $(-1,-1,-1,0,-1,-1,-1,0;1)$ & $-3\epsilon$\\
        $(-1,-1,0,-1,-1,-1,0,-1;1)$ & $-3\epsilon$\\
        $(-1,0,-1,-1,-1,0,-1,-1;1)$    & $-3\epsilon$\\
        $(0,-1,-1,-1,0,-1,-1,-1;1)$    & $-3\epsilon$\\
        $(0,0,0,0,0,0,0,0;1)$         & $0$\\
    \end{tabular}
    \caption{Regions obtained by directly applying the MoR to the Regge-limit expansion of graph $G_{\bullet \bullet}$.}
    \label{tab:regions_forward_original}
\end{table}

\begin{table}[H]
    \small
    \centering
    \begin{tabular}{l|l|l}
        \multirow{2}{5.5cm}{\bf $\v_\mathrm{R}$ ($y_0, x_1, y_2, x_3, y_4, x_5, y_6, x_7 ;\lambda$)} &  \multirow{2}{5cm}{\bf $\v_\mathrm{R}$ ($x_0, x_1, \ldots, x_7; \lambda$)} & \bf degree of \\
         & & \bf divergence \\\hline
        $(0,-1,0,-1,0,-1,1,-1;1)$    & $(-1,-1,-1,-1,-1,-1,-1,-1;1)$ & $-1-3\epsilon$\\
        $(1,-1,0,-1,0,-1,0,-1;1)$ & $(-1,-1,-1,-1,-1,-1,-1,-1;1)$ & $-1-3\epsilon$\\
        $(-1,0,0,-1,-1,0,0,-1;1)$       & $(-1,0,-1,-1,-1,0,-1,-1;1)$ & $-3\epsilon$\\
        $(0,0,0,0,0,0,0,0;1)$         & $(0,0,0,0,0,0,0,0;1)$ & $0$\\
    \end{tabular}
    \caption{Regge-limit expansion of graph $G_{\bullet \bullet}$, fourth integral dissection $\mathcal{I}_4$. The region $(-1,-1,-1,0,-1,-1,-1,0;1)$ in the original variables, is present in other integral dissections, e.g. $\mathcal{I}_{11}$. The region $(0,-1,-1,-1,0,-1,-1,-1;1)$ in the original polytope, is entirely absent after dissection.}
    \label{tab:regions_forward_split1} 
\end{table}

\subsection{On-shell expansion of \texorpdfstring{$G_{\bullet s}$}{TEXT}}
\label{appendix-examples_regions_gbs_onshell}

\begin{table}[H]
    \small
    \centering
    \begin{tabular}{l|l}
        \bf $\v_\mathrm{R}$ ($x_0, x_1, \ldots, x_8 ;\lambda$)      & \bf degree of divergence \\\hline
        $(-2,-1,-2,-1,-2,-1,-2,-1,-2;1)$ & $-2-6\epsilon$\\
        $(-2,-1,-2,-1,-1,-1,-1,0,-2,1;1)$ & $-1-5\epsilon$\\
        $(-2,-1,-2,-1,-1,0,-1,-1,-2;1)$ & $-1-5\epsilon$\\
        $(-1,-1,-1,0,-2,-1,-2,-1,-2;1)$ & $-1-5\epsilon$\\
        $(-1,-2,-1,-2,-1,-2,-1,-2,0;1)$ & $-6\epsilon$\\
        $(-1,-1,-1,0,-1,0,-1,0,-1;1)$ & $-3\epsilon$\\
        $(-1,-1,-1,0,0,0,0,0,-1;1)$ & $-2\epsilon$\\
        $(-1,-1,0,0,0,0,0,0,0;1)$ & $-\epsilon$\\
        $(-1,-1,0,-1,0,-1,0,-1,0;1)$ & $1-3\epsilon$\\
        $(0,0,0,0,0,0,0,0;1)$         & $0$\\
    \end{tabular}
    \caption{Regions obtained by directly applying the MoR to the on-shell expansion of graph $G_{\bullet s}$ with $p_1^2 \sim \lambda Q^2$ and $p_i^2=0\ (i=2,3,4)$.}
    \label{tab:regions_onshell_original_3902}
\end{table}

\begin{table}[H]
    \small
    \centering
    \begin{tabular}{l|l|l}
        \multirow{2}{6cm}{\bf $\v_\mathrm{R}$ ($y_0, x_1, y_2, x_3, y_4, x_5, y_6, x_7, x_8 ;\lambda$)} &  \multirow{2}{5cm}{\bf $\v_\mathrm{R}$ ($x_0, x_1, \ldots, x_8; \lambda$)} & \bf degree of \\
         & & \bf divergence \\\hline
        $(-1,-1,-1,-1,-1,-1,-1,-1,-2;1)$ & $(-2,-1,-2,-1,-2,-1,-2,-1,-2;1)$ & $-2-6\epsilon$ \\
        $(1,-2;1,-2;1,-2;0,-2;0;2)$ & $(-2,-2,-2,-2,-2,-2,-2,-2;0;2)$ & $-1-6\epsilon$ \\
        $(-1,-1,-1,-1,-1;0;0,-1,-2;1)$ & $(-2,-1,-2,-1,-1;0,-1,-1,-2;1)$ & $-1-5\epsilon$ \\
        $(1,-2;1,-2;1,-2;1,-2;0;1)$ & $(-1,-2,-1,-2,-1,-2,-1,-2;0;1)$ & $-6\epsilon$ \\
        $(0,-1,-1;0,-1;0;0,-1,-2;1)$ & $(-1,-1,-1;0,-1;0,-1,-1,-2;1)$ & $-4\epsilon$ \\
        $(0,-1;1,-1;1,-1;0,-1;0;1)$ & $(-1,-1,-1,-1,-1,-1,-1,-1;0;1)$ & $-3\epsilon$ \\
        $(1,-1;0,-1;1,-1;0,-1;0;1)$ & $(-1,-1,-1,-1,-1,-1,-1,-1;0;1)$ & $-3\epsilon$ \\
        $(0,-1,-1;0;0;0;0;0,-1;1)$ & $(-1,-1,-1;0;0;0;0;0,-1;1)$ & $-2\epsilon$ \\
        $(0,-1;0;0;0;0;0;0;0;1)$ & $(-1,-1;0;0;0;0;0;0;0;1)$ & $-\epsilon$ \\
        $(0,-1,-1;0,-1;1;0;0,-1;1)$ & $(-1,-1,-1;0;0;1;0;0,-1;1)$ & $1-2\epsilon$ \\
        $(0,0,0,0,0,0,0,0,0,1)$ & $(0,0,0,0,0,0,0,0,0,1)$ & $0$ \\
    \end{tabular}
    \caption{On-shell expansion of $G_{\bullet s}$, with $p_1^2 \sim \lambda Q^2$ and $p_i^2=0\ (i=2,3,4)$, first dissection~$\mathcal{I}_1$. The regions $(-1,-1,-1,0,-1,0,-1,0,-1;1)$ and $(-1,-1,0,-1,0,-1,0,-1,0;1)$ in the original variables, are absent after the dissection. The remaining regions appear in other dissected integrals.}
    \label{tab:regions_onshell_split1_3902}
\end{table}

\subsection{Regge-limit expansion of \texorpdfstring{$G_{\bullet s}$}{TEXT}}
\label{appendix-examples_regions_gbs_regge}

\begin{table}[H]
    \small
    \centering
    \begin{tabular}{l|l}
        \bf $\v_\mathrm{R}$ ($x_0, x_1, \ldots, x_8 ;\lambda$)      & \bf degree of divergence \\\hline
        $(-1,-1,-1,0,-1,-1,-1,0,-1;1)$ & $-1-3\epsilon$\\
        $(-1,0,-1,-1,-1,0,-1,-1,-1;1)$ & $-1-3\epsilon$\\
        $(-1,-1,0,-1,-1,-1,0,-1,0;1)$ & $-3\epsilon$\\
        $(0,-1,-1,-1,0,-1,-1,-1,0;1)$ & $-3\epsilon$\\
        $(0,0,0,0,0,0,0,0;1)$ & $0$\\
    \end{tabular}
    \caption{Regions obtained by directly applying the MoR to the forward-scattering expansion of graph $G_{\bullet s}$.}
    \label{tab:regions_forward_original_3902}
\end{table}

\begin{table}[H]
    \small
    \centering
    \begin{tabular}{l|l|l}
        \multirow{2}{6cm}{\bf $\v_\mathrm{R}$ ($y_0, x_1, y_2, x_3, y_4, x_5, y_6, x_7, x_8 ;\lambda$)} &  \multirow{2}{5cm}{\bf $\v_\mathrm{R}$ ($x_0, x_1, \ldots, x_8; \lambda$)} & \bf degree of \\
         & & \bf divergence \\\hline
        $(-1,0,0,-1,-1,0,0,-1,-1;1)$ & $(-1,0,-1,-1,-1,0,-1,-1,-1;1)$ & $-1-3\epsilon$\\
        $(0,-1,1,-1,0,-1,0,-1,0;1)$ & $(-1,-1,-1,-1,-1,-1,-1,-1,0;1)$ & $-1-3\epsilon$\\
        $(1,-1,0,-1,0,-1,0,-1,0;1)$ & $(-1,-1,-1,-1,-1,-1,-1,-1,0;1)$ & $-1-3\epsilon$\\
        $(0,0,0,0,0,0,0,0,0;1)$ & $(0,0,0,0,0,0,0,0,0;1)$ & $0$\\
    \end{tabular}
    \caption{Regge-limit expansion of graph $G_{\bullet s}$, third integral dissection $\mathcal{I}_3$. The region $(-1,-1,-1,0,-1,-1,-1,0,-1;1)$ in the original variables, is present in other dissections, e.g.~$\mathcal{I}_{11}$. The regions $(-1,-1,0,-1,-1,-1,0,-1,0;1)$ and $(0,-1,-1,-1,0,-1,-1,-1,0;1)$ in the original variables, are absent after dissecting the integral.}
    \label{tab:regions_forward_split3_3902}
\end{table}

\subsection{Regge-limit expansion of \texorpdfstring{$G_{tt}$}{TEXT}}
\label{appendix-examples_regions_gtt_regge}

\begin{table}[H]
    \small
    \centering
    \begin{tabular}{l|l}
        \bf $\v_R$ ($x_0, x_1, \ldots, x_9 ;\lambda$) & \bf degree of divergence \\\hline
        $(-3,-2,-1,-1,-3,-2,-1,-1,-3,-1;1)$ & $-4-7\epsilon$\\
        $(-2,-3,-1,-1,-2,-3,-1,-1,-1,-3;1)$ & $-4-7\epsilon$\\
        $(-1,-1,-3,-2,-1,-1,-3,-2,-3,-1;1)$ & $-4-7\epsilon$\\
        $(-1,-1,-2,-3,-1,-1,-2,-3,-1,-3;1)$ & $-4-7\epsilon$\\
        $(-1,-1,-1,-1,-1,-1,-1,-1,-2,-1;1)$ & $-3-4\epsilon$\\
        $(-1,-1,-1,-1,-1,-1,-1,-1,-1,-2;1)$ & $-3-4\epsilon$\\
        $(-1,-1,-1,0,-1,-1,-1,0,-1,-1;1))$ & $-2-3\epsilon$\\
        $(-1,-1,0,-1,-1,-1,0,-1,-1,-1;1))$ & $-2-3\epsilon$\\
        $(-1,0,-1,-1,-1,0,-1,-1,-1,-1;1)$ & $-2-3\epsilon$\\
        $(0,-1,-1,-1,0,-1,-1,-1,-1,-1;1)$ & $-2-3\epsilon$\\
        $(-1,-1,0,0,-1,-1,0,0,-1,-1;1))$ & $-2-2\epsilon$\\
        $(0,0,-1,-1,0,0,-1,-1,-1,-1;1)$ & $-2-2\epsilon$\\
        $(0,0,0,0,0,0,0,0;1)$ & $0$\\
    \end{tabular}
    \caption{The 13 Regions obtained by directly applying the MoR to the forward-scattering expansion of graph $G_{tt}$.}
\label{tab:regions_forward_original_Gtt}
\end{table}

\begin{table}[H]
    \small
    \centering
    \begin{tabular}{l|l|l}
    $\v_\mathrm{R}$ ($y_0, x_1, y_2, x_3, y_4, x_5, y_6, x_7, x_8, x_9 ;\lambda$) &  
    $\v_\mathrm{R}$ ($x_0, x_1, \ldots, x_9; \lambda$) & 
    $\mu$ \\\hline
$(0,-1,0,-1,0,-1,0,-1,-2,-1;1)$ & $(-1,-1,-1,-1,-1,-1,-1,-1,-2,-1;1)$ & $-3-4\epsilon$ \\
$(0,-1,0,-1,0,-1,0,-1,-1,-2;1)$ & $(-1,-1,-1,-1,-1,-1,-1,-1,-1,-2;1)$ & $-3-4\epsilon$ \\
$(\frac{1}{2},-1,\frac{1}{2},-1,0,-1,\frac{1}{2},-1,-1,-1;1)$ & $(-2,-2,-2,-2,-2,-2,-2,-2,-2,-2;2)$ & $-\frac{5}{2}-3\epsilon$ \\$(\frac{-1}{2},\frac{-1}{2},\frac{-1}{2},\frac{-1}{2},\frac{-1}{2},\frac{-1}{2},\frac{-1}{2},\frac{-1}{2},-1,-1;1)$ & $(-2,-1,-2,-1,-2,-1,-2,-1,-2,-2;2)$ & $-2-3\epsilon$ \\
$(1,-1,0,-1,0,-1,1,-1,-1,-1;1)$ & $(-1,-1,-1,-1,-1,-1,-1,-1,-1,-1;1)$ & $-2-3\epsilon$ \\
$(1,-1,1,-1,0,-1,0,-1,-1,-1;1)$ & $(-1,-1,-1,-1,-1,-1,-1,-1,-1,-1;1)$ & $-2-3\epsilon$ \\
$(0,-1,0,0,0,-1,0,0,-1,-1;1)$ & $(-1,-1,0,0,-1,-1,0,0,-1,-1;1)$ & $-2-2\epsilon$ \\
$(0,0,0,-1,0,0,0,-1,-1,-1;1)$ & $(0,0,-1,-1,0,0,-1,-1,-1,-1;1)$ & $-2-2\epsilon$ \\
$(-1,0,-1,0,0,-1,-1,0,-1,-1;1)$ & $(-1,0,-1,0,-1,-1,-1,0,-1,-1;1)$ & $-1-3\epsilon$ \\
$(0,0,0,0,0,0,0,0,0,0;1)$ & $(0,0,0,0,0,0,0,0,0,0;1)$ & $0$ \\
    \end{tabular}
    \caption{Regge-limit expansion of graph $G_{t t}$, $20$-th dissection, $\mathcal{I}_{20}$, in the new variables, original variables, and the degree of divergence of each region, $\mu$. All the new region vectors are characterised by non-analytic behaviour of the form $x^{-3\epsilon}$, consistently with the hidden (Glauber) region.
    The regions $(-1,-1,-1,0,-1,-1,-1,0,-1,-1;1)$, $(-1,-1,0,-1,-1,-1,0,-1,-1,-1;1)$ and $(0,-1,-1,-1,0,-1,-1,-1,-1,-1;1)$, in the original variables, are absent after dissecting; meanwhile all other regions obtained by directly applying the MoR to $G_{t t}$ appear in at least one other polytope dissection. Note that the vectors $(-2,-1,-2,-1,-2,-1,-2,-1,-2,-2;2)$ and $(-1,0,-1,0,-1,-1,-1,0,-1,-1;1)$ in the original variables are incompatible with cancellations for arbitrary values of $s_{12}$ and $s_{13}$, thus are not region vectors.}
\label{tab:regions_forward_split3_4437}
\end{table}

\bibliographystyle{JHEP}
\bibliography{refs}

\providecommand{\href}[2]{#2}\begingroup\raggedright\begin{thebibliography}{100}

\bibitem{Smirnov:2004ym}
V.~A. Smirnov, {\it {Evaluating Feynman integrals}},  {\em Springer Tracts Mod. Phys.} {\bf 211} (2004) 1--244.

\bibitem{Henn:2014qga}
J.~M. Henn, {\it {Lectures on differential equations for Feynman integrals}},  {\em J. Phys. A} {\bf 48} (2015) 153001, [\href{http://xxx.lanl.gov/abs/1412.2296}{{\tt arXiv:1412.2296}}].

\bibitem{Weinzierl:2022eaz}
S.~Weinzierl, {\it Feynman integrals},  {\em arXiv preprint arXiv:2201.03593} {\bf 5} (2022), no.~6.

\bibitem{SmnTtyk09FIESTA}
A.~Smirnov and M.~Tentyukov, {\it Feynman integral evaluation by a sector decomposition approach (fiesta)},  {\em Computer Physics Communications} {\bf 180} (2009), no.~5 735--746.

\bibitem{SmnSmnTtyk11FIESTA2}
A.~Smirnov, V.~Smirnov, and M.~Tentyukov, {\it Fiesta 2: parallelizeable multiloop numerical calculations},  {\em Computer Physics Communications} {\bf 182} (2011), no.~3 790--803.

\bibitem{Smn14FIESTA3}
A.~V. Smirnov, {\it Fiesta 3: cluster-parallelizable multiloop numerical calculations in physical regions},  {\em Computer Physics Communications} {\bf 185} (2014), no.~7 2090--2100.

\bibitem{Smn16FIESTA4}
A.~V. Smirnov, {\it Fiesta 4: Optimized feynman integral calculations with gpu support},  {\em Computer Physics Communications} {\bf 204} (2016) 189--199.

\bibitem{Smn22FIESTA5}
A.~Smirnov, N.~Shapurov, and L.~Vysotsky, {\it Fiesta5: numerical high-performance feynman integral evaluation},  {\em Computer Physics Communications} {\bf 277} (2022) 108386.

\bibitem{KnkUeda10}
T.~Kaneko and T.~Ueda, {\it A geometric method of sector decomposition},  {\em Computer Physics Communications} {\bf 181} (2010), no.~8 1352--1361.

\bibitem{Carter:2010hi}
J.~Carter and G.~Heinrich, {\it {SecDec: A general program for sector decomposition}},  {\em Comput. Phys. Commun.} {\bf 182} (2011) 1566--1581, [\href{http://xxx.lanl.gov/abs/1011.5493}{{\tt arXiv:1011.5493}}].

\bibitem{Borowka:2015mxa}
S.~Borowka, G.~Heinrich, S.~P. Jones, M.~Kerner, J.~Schlenk, and T.~Zirke, {\it {SecDec-3.0: numerical evaluation of multi-scale integrals beyond one loop}},  {\em Comput. Phys. Commun.} {\bf 196} (2015) 470--491, [\href{http://xxx.lanl.gov/abs/1502.06595}{{\tt arXiv:1502.06595}}].

\bibitem{Heinrich:2023til}
G.~Heinrich, S.~P. Jones, M.~Kerner, V.~Magerya, A.~Olsson, and J.~Schlenk, {\it {Numerical scattering amplitudes with pySecDec}},  {\em Comput. Phys. Commun.} {\bf 295} (2024) 108956, [\href{http://xxx.lanl.gov/abs/2305.19768}{{\tt arXiv:2305.19768}}].

\bibitem{pySecDec17}
S.~Borowka, G.~Heinrich, S.~Jahn, S.~P. Jones, M.~Kerner, J.~Schlenk, and T.~Zirke, {\it {pySecDec: a toolbox for the numerical evaluation of multi-scale integrals}},  {\em Comput. Phys. Commun.} {\bf 222} (2018) 313--326, [\href{http://xxx.lanl.gov/abs/1703.09692}{{\tt arXiv:1703.09692}}].

\bibitem{Brsk20}
M.~Borinsky, {\it {Tropical Monte Carlo quadrature for Feynman integrals}},  {\em Ann. Inst. H. Poincare D Comb. Phys. Interact.} {\bf 10} (2023), no.~4 635, [\href{http://xxx.lanl.gov/abs/2008.12310}{{\tt arXiv:2008.12310}}].

\bibitem{BrskMchTld23}
M.~Borinsky, H.~J. Munch, and F.~Tellander, {\it {Tropical Feynman integration in the Minkowski regime}},  {\em Comput. Phys. Commun.} {\bf 292} (2023) 108874, [\href{http://xxx.lanl.gov/abs/2302.08955}{{\tt arXiv:2302.08955}}].

\bibitem{LiuMa23AMFlow}
X.~Liu and Y.-Q. Ma, {\it {AMFlow: A Mathematica package for Feynman integrals computation via auxiliary mass flow}},  {\em Comput. Phys. Commun.} {\bf 283} (2023) 108565, [\href{http://xxx.lanl.gov/abs/2201.11669}{{\tt arXiv:2201.11669}}].

\bibitem{Hdg21DiffExp}
M.~Hidding, {\it {DiffExp, a Mathematica package for computing Feynman integrals in terms of one-dimensional series expansions}},  {\em Comput. Phys. Commun.} {\bf 269} (2021) 108125, [\href{http://xxx.lanl.gov/abs/2006.05510}{{\tt arXiv:2006.05510}}].

\bibitem{AmdlBcnDvtRanaVcn23SeaSyde}
T.~Armadillo, R.~Bonciani, S.~Devoto, N.~Rana, and A.~Vicini, {\it {Evaluation of Feynman integrals with arbitrary complex masses via series expansions}},  {\em Comput. Phys. Commun.} {\bf 282} (2023) 108545, [\href{http://xxx.lanl.gov/abs/2205.03345}{{\tt arXiv:2205.03345}}].

\bibitem{BnkSmn97}
M.~Beneke and V.~A. Smirnov, {\it {Asymptotic expansion of Feynman integrals near threshold}},  {\em Nucl. Phys.} {\bf B522} (1998) 321--344, [\href{http://xxx.lanl.gov/abs/hep-ph/9711391}{{\tt hep-ph/9711391}}].

\bibitem{Smirnov:2002pj}
V.~A. Smirnov, {\it {Applied asymptotic expansions in momenta and masses}},  {\em Springer Tracts Mod. Phys.} {\bf 177} (2002) 1--262.

\bibitem{PakSmn11}
A.~Pak and A.~Smirnov, {\it Geometric approach to asymptotic expansion of feynman integrals},  {\em The European Physical Journal C} {\bf 71} (2011), no.~4 1--6.

\bibitem{Jtz11}
B.~Jantzen, {\it {Foundation and generalization of the expansion by regions}},  {\em JHEP} {\bf 12} (2011) 076, [\href{http://xxx.lanl.gov/abs/1111.2589}{{\tt arXiv:1111.2589}}].

\bibitem{JtzSmnSmn12}
B.~Jantzen, A.~V. Smirnov, and V.~A. Smirnov, {\it Expansion by regions: revealing potential and glauber regions automatically},  {\em The European Physical Journal C} {\bf 72} (2012), no.~9 1--14.

\bibitem{SmnvSmnSmv19}
T.~Y. Semenova, A.~V. Smirnov, and V.~A. Smirnov, {\it On the status of expansion by regions},  {\em The European Physical Journal C} {\bf 79} (2019), no.~2 1--12.

\bibitem{GrdHzgJnsMaSchlk22}
E.~Gardi, F.~Herzog, S.~Jones, Y.~Ma, and J.~Schlenk, {\it {The on-shell expansion: from Landau equations to the Newton polytope}},  {\em JHEP} {\bf 07} (2023) 197, [\href{http://xxx.lanl.gov/abs/2211.14845}{{\tt arXiv:2211.14845}}].

\bibitem{Ma23}
Y.~Ma, {\it {Identifying regions in wide-angle scattering via graph-theoretical approaches}},  \href{http://xxx.lanl.gov/abs/2312.14012}{{\tt arXiv:2312.14012}}.

\bibitem{Plenter:2020lop}
J.~Plenter and G.~Rodrigo, {\it {Asymptotic expansions through the loop-tree duality}},  {\em Eur. Phys. J. C} {\bf 81} (2021), no.~4 320, [\href{http://xxx.lanl.gov/abs/2005.02119}{{\tt arXiv:2005.02119}}].

\bibitem{IsgWise89}
N.~Isgur and M.~B. Wise, {\it Weak decays of heavy mesons in the static quark approximation},  {\em Physics Letters B} {\bf 232} (1989), no.~1 113--117.

\bibitem{IsgWise90}
N.~Isgur and M.~B. Wise, {\it Weak transition form factors between heavy mesons},  {\em Physics Letters B} {\bf 237} (1990), no.~3-4 527--530.

\bibitem{Nbt98book}
M.~Neubert, {\it B decays and the heavy quark expansion},  {\em Adv. Ser. Dir. High Energy Phys} {\bf 15} (1998) 239.

\bibitem{Mnh00book}
A.~V. Manohar and M.~B. Wise, {\em Heavy quark physics}.
\newblock Cambridge university press, 2000.

\bibitem{Grz04book}
A.~G. Grozin, {\em Heavy quark effective theory}, vol.~201.
\newblock Springer Science \& Business Media, 2004.

\bibitem{BchBrgFrl15book}
T.~Becher, A.~Broggio, and A.~Ferroglia, {\it {Introduction to Soft-Collinear Effective Theory}},  {\em Lect. Notes Phys.} {\bf 896} (2015) pp.1--206, [\href{http://xxx.lanl.gov/abs/1410.1892}{{\tt arXiv:1410.1892}}].

\bibitem{BurFlmLk00}
C.~W. Bauer, S.~Fleming, and M.~E. Luke, {\it {Summing Sudakov logarithms in B ---> X(s gamma) in effective field theory}},  {\em Phys. Rev.} {\bf D63} (2000) 014006, [\href{http://xxx.lanl.gov/abs/hep-ph/0005275}{{\tt hep-ph/0005275}}].

\bibitem{BurPjlSwt02-1}
C.~W. Bauer, D.~Pirjol, and I.~W. Stewart, {\it {Soft collinear factorization in effective field theory}},  {\em Phys. Rev.} {\bf D65} (2002) 054022, [\href{http://xxx.lanl.gov/abs/hep-ph/0109045}{{\tt hep-ph/0109045}}].

\bibitem{BurPjlSwt02-2}
C.~W. Bauer, D.~Pirjol, and I.~W. Stewart, {\it {Power counting in the soft collinear effective theory}},  {\em Phys. Rev.} {\bf D66} (2002) 054005, [\href{http://xxx.lanl.gov/abs/hep-ph/0205289}{{\tt hep-ph/0205289}}].

\bibitem{BnkChpkDhlFdm02}
M.~Beneke, A.~P. Chapovsky, M.~Diehl, and T.~Feldmann, {\it {Soft collinear effective theory and heavy to light currents beyond leading power}},  {\em Nucl. Phys. B} {\bf 643} (2002) 431--476, [\href{http://xxx.lanl.gov/abs/hep-ph/0206152}{{\tt hep-ph/0206152}}].

\bibitem{RstStw16}
I.~Z. Rothstein and I.~W. Stewart, {\it {An Effective Field Theory for Forward Scattering and Factorization Violation}},  {\em JHEP} {\bf 08} (2016) 025, [\href{http://xxx.lanl.gov/abs/1601.04695}{{\tt arXiv:1601.04695}}].

\bibitem{EdenLdshfOlvPkhn02book}
R.~J. Eden, P.~V. Landshoff, D.~I. Olive, and J.~C. Polkinghorne, {\em The analytic S-matrix}.
\newblock Cambridge University Press, 2002.

\bibitem{Stm95book}
G.~F. Sterman, {\em {An Introduction to quantum field theory}}.
\newblock Cambridge University Press, 1993.

\bibitem{ClsSprStm04}
J.~C. Collins, D.~E. Soper, and G.~F. Sterman, {\it {Factorization of Hard Processes in QCD}},  {\em Adv. Ser. Direct. High Energy Phys.} {\bf 5} (1989) 1--91, [\href{http://xxx.lanl.gov/abs/hep-ph/0409313}{{\tt hep-ph/0409313}}].

\bibitem{Stmg18}
A.~Strominger, {\em Lectures on the infrared structure of gravity and gauge theory}.
\newblock Princeton University Press, 2018.

\bibitem{Agarwal:2021ais}
N.~Agarwal, L.~Magnea, C.~Signorile-Signorile, and A.~Tripathi, {\it {The infrared structure of perturbative gauge theories}},  {\em Phys. Rept.} {\bf 994} (2023) 1--120, [\href{http://xxx.lanl.gov/abs/2112.07099}{{\tt arXiv:2112.07099}}].

\bibitem{Lnd59}
L.~Landau, {\it On analytic properties of vertex parts in quantum field theory},  {\em Nuclear Physics} {\bf 13} (1959), no.~1 181--192.

\bibitem{Bjorken:1959fd}
J.~D. Bjorken, {\em {Experimental tests of Quantum electrodynamics and spectral representations of Green's functions in perturbation theory}}.
\newblock PhD thesis, Stanford U., 1959.

\bibitem{Nakanishi}
N.~Nakanishi, {\it {Ordinary and Anomalous Thresholds in Perturbation Theory}},  {\em Progress of Theoretical Physics} {\bf 22} (07, 1959) 128--144, [\href{http://xxx.lanl.gov/abs/https://academic.oup.com/ptp/article-pdf/22/1/128/5427385/22-1-128.pdf}{{\tt https://academic.oup.com/ptp/article-pdf/22/1/128/5427385/22-1-128.pdf}}].

\bibitem{ClmNtn65}
S.~Coleman and R.~E. Norton, {\it {Singularities in the physical region}},  {\em Nuovo Cim.} {\bf 38} (1965) 438--442.

\bibitem{Stm78I}
G.~F. Sterman, {\it {Mass Divergences in Annihilation Processes. 1. Origin and Nature of Divergences in Cut Vacuum Polarization Diagrams}},  {\em Phys. Rev.} {\bf D17} (1978) 2773.

\bibitem{ClsStm81}
J.~C. Collins and G.~F. Sterman, {\it {Soft Partons in {QCD}}},  {\em Nucl. Phys.} {\bf B185} (1981) 172--188.

\bibitem{Stm96lectures}
G.~F. Sterman, {\it {Partons, factorization and resummation, TASI 95}},  in {\em {QCD and beyond. Proceedings, Theoretical Advanced Study Institute in Elementary Particle Physics, TASI-95, Boulder, USA, June 4-30, 1995}}, pp.~327--408, 1995.
\newblock \href{http://xxx.lanl.gov/abs/hep-ph/9606312}{{\tt hep-ph/9606312}}.

\bibitem{LeePmrsk13}
R.~N. Lee and A.~A. Pomeransky, {\it {Critical points and number of master integrals}},  {\em JHEP} {\bf 11} (2013) 165, [\href{http://xxx.lanl.gov/abs/1308.6676}{{\tt arXiv:1308.6676}}].

\bibitem{MzrTln22}
S.~Mizera and S.~Telen, {\it {Landau discriminants}},  {\em JHEP} {\bf 08} (2022) 200, [\href{http://xxx.lanl.gov/abs/2109.08036}{{\tt arXiv:2109.08036}}].

\bibitem{AkHmHlmMzr22}
N.~Arkani-Hamed, A.~Hillman, and S.~Mizera, {\it {Feynman polytopes and the tropical geometry of UV and IR divergences}},  {\em Phys. Rev. D} {\bf 105} (2022), no.~12 125013, [\href{http://xxx.lanl.gov/abs/2202.12296}{{\tt arXiv:2202.12296}}].

\bibitem{FvlMzrTln23prl}
C.~Fevola, S.~Mizera, and S.~Telen, {\it {Landau Singularities Revisited: Computational Algebraic Geometry for Feynman Integrals}},  {\em Phys. Rev. Lett.} {\bf 132} (2024), no.~10 101601, [\href{http://xxx.lanl.gov/abs/2311.14669}{{\tt arXiv:2311.14669}}].

\bibitem{FvlMzrTln23pld}
C.~Fevola, S.~Mizera, and S.~Telen, {\it {Principal Landau determinants}},  {\em Comput. Phys. Commun.} {\bf 303} (2024) 109278, [\href{http://xxx.lanl.gov/abs/2311.16219}{{\tt arXiv:2311.16219}}].

\bibitem{Brown:2009ta}
F.~C.~S. Brown, {\it {On the periods of some Feynman integrals}},  \href{http://xxx.lanl.gov/abs/0910.0114}{{\tt arXiv:0910.0114}}.

\bibitem{Panzer:2014caa}
E.~Panzer, {\it {Algorithms for the symbolic integration of hyperlogarithms with applications to Feynman integrals}},  {\em Comput. Phys. Commun.} {\bf 188} (2015) 148--166, [\href{http://xxx.lanl.gov/abs/1403.3385}{{\tt arXiv:1403.3385}}].

\bibitem{Dlapa:2023cvx}
C.~Dlapa, M.~Helmer, G.~Papathanasiou, and F.~Tellander, {\it {Symbol alphabets from the Landau singular locus}},  {\em JHEP} {\bf 10} (2023) 161, [\href{http://xxx.lanl.gov/abs/2304.02629}{{\tt arXiv:2304.02629}}].

\bibitem{Helmer:2024wax}
M.~Helmer, G.~Papathanasiou, and F.~Tellander, {\it {Landau Singularities from Whitney Stratifications}},  \href{http://xxx.lanl.gov/abs/2402.14787}{{\tt arXiv:2402.14787}}.

\bibitem{AnthnrySkrRmn19}
B.~Ananthanarayan, A.~Pal, S.~Ramanan, and R.~Sarkar, {\it Unveiling regions in multi-scale feynman integrals using singularities and power geometry},  {\em The European Physical Journal C} {\bf 79} (2019), no.~1 1--20.

\bibitem{HrchJnsSlk22}
G.~Heinrich, S.~Jahn, S.~Jones, M.~Kerner, F.~Langer, V.~Magerya, A.~Poldaru, J.~Schlenk, and E.~Villa, {\it Expansion by regions with pysecdec},  {\em Computer Physics Communications} {\bf 273} (2022) 108267.

\bibitem{Herzog:2023sgb}
F.~Herzog, Y.~Ma, B.~Mistlberger, and A.~Suresh, {\it {Single-soft emissions for amplitudes with two colored particles at three loops}},  {\em JHEP} {\bf 12} (2023) 023, [\href{http://xxx.lanl.gov/abs/2309.07884}{{\tt arXiv:2309.07884}}].

\bibitem{Pkhn63I}
J.~Polkinghorne, {\it High-energy behavior in perturbation theory},  {\em Journal of Mathematical Physics} {\bf 4} (1963), no.~4 503--506.

\bibitem{Pkhn63II}
J.~Polkinghorne, {\it High-energy behavior in perturbation theory. ii},  {\em Journal of Mathematical Physics} {\bf 4} (1963), no.~11 1396--1400.

\bibitem{IslLdsfTlr63}
J.~Islam, P.~Landshoff, and J.~Taylor, {\it Singularity of the regge amplitude},  {\em Physical Review} {\bf 130} (1963), no.~6 2560.

\bibitem{Hld63}
I.~G. Halliday, {\it {High-energy behaviour in perturbation theory}},  {\em Nuovo Cim.} {\bf 30} (1963) 177--192.

\bibitem{Mdst63}
S.~Mandelstam, {\it {Cuts in the Angular Momentum Plane. 2}},  {\em Nuovo Cim.} {\bf 30} (1963) 1148--1162.

\bibitem{Tktpls63}
G.~Tiktopoulos, {\it {High-Energy Behavior of Feynman Amplitudes}},  {\em Phys. Rev.} {\bf 131} (1963) 480--490.

\bibitem{Tktpls63nonplanar}
G.~Tiktopoulos, {\it {High-Energy Behavior of Feynman Amplitudes. 2. Nonplanar Graphs}},  {\em Phys. Rev.} {\bf 131} (1963) 2373--2380.

\bibitem{Hld64}
I.~G. Halliday, {\it {High-energy behavior at fixed angle in perturbation theory}},  {\em Annals Phys.} {\bf 28} (1964) 320--345.

\bibitem{Lam69}
C.~S. Lam, {\it {High-energy behaviour of feynman diagrams via the electric-circuit analogy}},  {\em Nuovo Cim. A} {\bf 62} (1969) 97--112.

\bibitem{Ldsf74}
P.~V. Landshoff, {\it {Model for elastic scattering at wide angle}},  {\em Phys. Rev. D} {\bf 10} (1974) 1024--1030.

\bibitem{Cls77book}
P.~D.~B. Collins, {\em {An Introduction to Regge Theory and High Energy Physics}}.
\newblock Cambridge Monographs on Mathematical Physics. Cambridge University Press, 7, 2023.

\bibitem{Pkhn80book}
J.~C. Polkinghorne, {\em {Models of High Energy Processes}}.
\newblock Cambridge Monographs on Mathematical Physics. Cambridge Univ. Press, Cambridge, UK, 4, 2010.

\bibitem{Henn:2020lye}
J.~Henn, B.~Mistlberger, V.~A. Smirnov, and P.~Wasser, {\it {Constructing d-log integrands and computing master integrals for three-loop four-particle scattering}},  {\em JHEP} {\bf 04} (2020) 167, [\href{http://xxx.lanl.gov/abs/2002.09492}{{\tt arXiv:2002.09492}}].

\bibitem{Bargiela:2021wuy}
P.~Bargiela, F.~Caola, A.~von Manteuffel, and L.~Tancredi, {\it {Three-loop helicity amplitudes for diphoton production in gluon fusion}},  {\em JHEP} {\bf 02} (2022) 153, [\href{http://xxx.lanl.gov/abs/2111.13595}{{\tt arXiv:2111.13595}}].

\bibitem{BottsStm89}
J.~Botts and G.~F. Sterman, {\it {Hard Elastic Scattering in QCD: Leading Behavior}},  {\em Nucl. Phys. B} {\bf 325} (1989) 62--100.

\bibitem{BottsStm89SudakovEffects}
J.~Botts and G.~F. Sterman, {\it {Sudakov Effects in Hadron Hadron Elastic Scattering}},  {\em Phys. Lett. B} {\bf 224} (1989) 201. [Erratum: Phys.Lett.B 227, 501 (1989)].

\bibitem{LbtdStm85}
J.~M.~F. Labastida and G.~F. Sterman, {\it {Inclusive Hadron - Hadron Scattering in the Feynman Gauge}},  {\em Nucl. Phys.} {\bf B254} (1985) 425--440.

\bibitem{JiangLiuXuYang24}
X.~Jiang, J.~Liu, X.~Xu, and L.~L. Yang, {\it {Symbol letters of Feynman integrals from Gram determinants}},  \href{http://xxx.lanl.gov/abs/2401.07632}{{\tt arXiv:2401.07632}}.

\bibitem{CrHtCrrGrx24}
S.~Caron-Huot, M.~Correia, and M.~Giroux, {\it {Recursive Landau Analysis}},  \href{http://xxx.lanl.gov/abs/2406.05241}{{\tt arXiv:2406.05241}}.

\bibitem{Bogner:2007cr}
C.~Bogner and S.~Weinzierl, {\it {Resolution of singularities for multi-loop integrals}},  {\em Comput. Phys. Commun.} {\bf 178} (2008) 596--610, [\href{http://xxx.lanl.gov/abs/0709.4092}{{\tt arXiv:0709.4092}}].

\bibitem{Kaneko:2010kj}
T.~Kaneko and T.~Ueda, {\it {Sector Decomposition Via Computational Geometry}},  {\em PoS} {\bf ACAT2010} (2010) 082, [\href{http://xxx.lanl.gov/abs/1004.5490}{{\tt arXiv:1004.5490}}].

\bibitem{Schlenk:2016cwf}
J.~Schlenk and T.~Zirke, {\it {Calculation of Multi-Loop Integrals with SecDec-3.0}},  {\em PoS} {\bf RADCOR2015} (2016) 106, [\href{http://xxx.lanl.gov/abs/1601.03982}{{\tt arXiv:1601.03982}}].

\bibitem{Schlenk:2016epj}
J.~K. Schlenk, {\em {Techniques for higher order corrections and their application to LHC phenomenology}}.
\newblock PhD thesis, Munich, Tech. U., 8, 2016.

\bibitem{Svtr24}
G.~Salvatori, {\it {The Tropical Geometry of Subtraction Schemes}},  \href{http://xxx.lanl.gov/abs/2406.14606}{{\tt arXiv:2406.14606}}.

\bibitem{dlCrz24}
L.~de~la Cruz, {\it {Polytope symmetries of Feynman integrals}},  {\em Phys. Lett. B} {\bf 854} (2024) 138744, [\href{http://xxx.lanl.gov/abs/2404.03564}{{\tt arXiv:2404.03564}}].

\bibitem{Engel:2022kde}
T.~Engel, {\em {Muon-Electron Scattering at NNLO}}.
\newblock PhD thesis, Zurich U., 2022.
\newblock \href{http://xxx.lanl.gov/abs/2209.11110}{{\tt arXiv:2209.11110}}.

\bibitem{Maple}
{Maplesoft, a division of Waterloo Maple Inc..}, ``Maple.''

\bibitem{Soper:1999xk}
D.~E. Soper, {\it {Techniques for QCD calculations by numerical integration}},  {\em Phys. Rev. D} {\bf 62} (2000) 014009, [\href{http://xxx.lanl.gov/abs/hep-ph/9910292}{{\tt hep-ph/9910292}}].

\bibitem{Binoth:2005ff}
T.~Binoth, J.~P. Guillet, G.~Heinrich, E.~Pilon, and C.~Schubert, {\it {An Algebraic/numerical formalism for one-loop multi-leg amplitudes}},  {\em JHEP} {\bf 10} (2005) 015, [\href{http://xxx.lanl.gov/abs/hep-ph/0504267}{{\tt hep-ph/0504267}}].

\bibitem{Nagy:2006xy}
Z.~Nagy and D.~E. Soper, {\it {Numerical integration of one-loop Feynman diagrams for N-photon amplitudes}},  {\em Phys. Rev. D} {\bf 74} (2006) 093006, [\href{http://xxx.lanl.gov/abs/hep-ph/0610028}{{\tt hep-ph/0610028}}].

\bibitem{Anastasiou:2007qb}
C.~Anastasiou, S.~Beerli, and A.~Daleo, {\it {Evaluating multi-loop Feynman diagrams with infrared and threshold singularities numerically}},  {\em JHEP} {\bf 05} (2007) 071, [\href{http://xxx.lanl.gov/abs/hep-ph/0703282}{{\tt hep-ph/0703282}}].

\bibitem{Lazopoulos:2007ix}
A.~Lazopoulos, K.~Melnikov, and F.~Petriello, {\it {QCD corrections to tri-boson production}},  {\em Phys. Rev. D} {\bf 76} (2007) 014001, [\href{http://xxx.lanl.gov/abs/hep-ph/0703273}{{\tt hep-ph/0703273}}].

\bibitem{Beerli:2008zz}
S.~Beerli, {\it {A New method for evaluating two-loop Feynman integrals and its application to Higgs production}},  other thesis, 2008.

\bibitem{Borowka:2012yc}
S.~Borowka, J.~Carter, and G.~Heinrich, {\it {Numerical Evaluation of Multi-Loop Integrals for Arbitrary Kinematics with SecDec 2.0}},  {\em Comput. Phys. Commun.} {\bf 184} (2013) 396--408, [\href{http://xxx.lanl.gov/abs/1204.4152}{{\tt arXiv:1204.4152}}].

\bibitem{Borowka:2014aaa}
S.~C. Borowka, {\em {Evaluation of multi-loop multi-scale integrals and phenomenological two-loop applications}}.
\newblock PhD thesis, Munich, Tech. U., 2014.
\newblock \href{http://xxx.lanl.gov/abs/1410.7939}{{\tt arXiv:1410.7939}}.

\bibitem{TomLoopsLegsProc}
S.~Jones, A.~Olsson, and T.~Stone, {\it {Evaluating Parametric Integrals in the Minkowski Regime without Contour Deformation}},  in {\em {Loops and Legs in Quantum Field Theory}}, 7, 2024.
\newblock \href{http://xxx.lanl.gov/abs/2407.06973}{{\tt arXiv:2407.06973}}.

\bibitem{LbyStm78}
S.~B. Libby and G.~F. Sterman, {\it {Jet and Lepton Pair Production in High-Energy Lepton-Hadron and Hadron-Hadron Scattering}},  {\em Phys. Rev.} {\bf D18} (1978) 3252.

\bibitem{AmtFbnStgln62asymptotic}
D.~Amati, S.~Fubini, and A.~Stanghellini, {\it {Asymptotic Properties of Scattering and Multiple Production}},  {\em Phys. Lett.} {\bf 1} (1962) 29--32.

\bibitem{AmtFbnStgln62theory}
D.~Amati, A.~Stanghellini, and S.~Fubini, {\it {Theory of high-energy scattering and multiple production}},  {\em Nuovo Cim.} {\bf 26} (1962) 896--954.

\bibitem{GrbPmrch62}
V.~N. Gribov and I.~Y. Pomeranchuk, {\it {Regge poles and Landau singularities}},  {\em Zh. Eksp. Teor. Fiz.} {\bf 43} (1962) 1970.

\bibitem{LeeSyr62}
B.~W. Lee and R.~F. Sawyer, {\it {Regge Poles and High-energy Limits in Field Theory}},  {\em Phys. Rev.} {\bf 127} (1962) 2266--2273.

\bibitem{Mnk64}
M.~Menke, {\it High-energy behaviour of feynman integrals involving singular configurations},  {\em Il Nuovo Cimento (1955-1965)} {\bf 34} (1964) 351--370.

\bibitem{Hprcht65}
B.~Hamprecht, {\it High-energy behaviour of feynman amplitudes},  {\em Il Nuovo Cimento A (1965-1970)} {\bf 40} (1965) 542--559.

\bibitem{GrdHzgJnsMaprepare}
E.~Gardi, F.~Herzog, S.~Jones, and Y.~Ma, {\it {Regions in the Regge limit of two to two scattering}},  {\em -- in preparation}.

\bibitem{DiVita:2014pza}
S.~Di~Vita, P.~Mastrolia, U.~Schubert, and V.~Yundin, {\it {Three-loop master integrals for ladder-box diagrams with one massive leg}},  {\em JHEP} {\bf 09} (2014) 148, [\href{http://xxx.lanl.gov/abs/1408.3107}{{\tt arXiv:1408.3107}}].

\bibitem{Canko:2021xmn}
D.~D. Canko and N.~Syrrakos, {\it {Planar three-loop master integrals for 2 \textrightarrow{} 2 processes with one external massive particle}},  {\em JHEP} {\bf 04} (2022) 134, [\href{http://xxx.lanl.gov/abs/2112.14275}{{\tt arXiv:2112.14275}}].

\bibitem{Syrrakos:2023mor}
N.~Syrrakos and D.~D. Canko, {\it {Three-loop master integrals for H+jet production at N3LO: Towards the non-planar topologies}},  {\em PoS} {\bf RADCOR2023} (2024) 044, [\href{http://xxx.lanl.gov/abs/2307.08432}{{\tt arXiv:2307.08432}}].

\bibitem{Henn:2023vbd}
J.~M. Henn, J.~Lim, and W.~J. Torres~Bobadilla, {\it {First look at the evaluation of three-loop non-planar Feynman diagrams for Higgs plus jet production}},  {\em JHEP} {\bf 05} (2023) 026, [\href{http://xxx.lanl.gov/abs/2302.12776}{{\tt arXiv:2302.12776}}].

\bibitem{Gao:2024qsg}
A.~Gao, I.~Moult, S.~Raman, G.~Ridgway, and I.~W. Stewart, {\it {A collinear perspective on the Regge limit}},  {\em JHEP} {\bf 05} (2024) 328, [\href{http://xxx.lanl.gov/abs/2401.00931}{{\tt arXiv:2401.00931}}].

\bibitem{Moult:2022lfy}
I.~Moult, S.~Raman, G.~Ridgway, and I.~W. Stewart, {\it {Anomalous dimensions from soft Regge constants}},  {\em JHEP} {\bf 05} (2023) 025, [\href{http://xxx.lanl.gov/abs/2207.02859}{{\tt arXiv:2207.02859}}].

\bibitem{Milloy:2022fns}
C.~Milloy, G.~Falcioni, E.~Gardi, N.~Maher, and L.~Vernazza, {\it {High-energy limit of 2-\ensuremath{>}2 scattering amplitudes at NNLL}},  {\em PoS} {\bf LL2022} (2022) 044, [\href{http://xxx.lanl.gov/abs/2207.07441}{{\tt arXiv:2207.07441}}].

\bibitem{Falcioni:2021dgr}
G.~Falcioni, E.~Gardi, N.~Maher, C.~Milloy, and L.~Vernazza, {\it {Disentangling the Regge Cut and Regge Pole in Perturbative QCD}},  {\em Phys. Rev. Lett.} {\bf 128} (2022), no.~13 132001, [\href{http://xxx.lanl.gov/abs/2112.11098}{{\tt arXiv:2112.11098}}].

\bibitem{Caola:2021izf}
F.~Caola, A.~Chakraborty, G.~Gambuti, A.~von Manteuffel, and L.~Tancredi, {\it {Three-Loop Gluon Scattering in QCD and the Gluon Regge Trajectory}},  {\em Phys. Rev. Lett.} {\bf 128} (2022), no.~21 212001, [\href{http://xxx.lanl.gov/abs/2112.11097}{{\tt arXiv:2112.11097}}].

\bibitem{Fadin:2023aen}
V.~S. Fadin, {\it {Regge Cuts in QCD}},  {\em Phys. Part. Nucl. Lett.} {\bf 20} (2023), no.~3 341--346.

\bibitem{Fadin:2021csi}
V.~S. Fadin, {\it {Three-Reggeon Cuts in QCD Amplitudes}},  {\em Phys. Atom. Nucl.} {\bf 84} (2021), no.~1 100--104.

\end{thebibliography}\endgroup

\end{document}